\documentclass[12pt]{article}
\usepackage{cite}
\usepackage{amsmath, amsthm, amssymb, graphicx,slashed}
\parskip=\baselineskip
\def\Bbb{\mathbb}

\def\Tr{{\rm Tr}}
\def\T{{\cal T}}
\def\B{{\mathcal B}}
\def\M{{\mathcal M}}
\def\16{{\bf 16}}
\def\1{{\bf 1}}
\def\2{{\bf 2}}
\def\4{{\bf 4}}
\def\bar{\overline}
\def\tilde{\widetilde}
\def\R{{\Bbb{R}}}\def\Z{{\Bbb{Z}}}
\def\N{{\mathcal N}}
\def\hat{\widehat}
\numberwithin{equation}{section}
\def\frak{\mathfrak}
\font\teneurm=eurm10 \font\seveneurm=eurm7 \font\fiveeurm=eurm5
\newfam\eurmfam
\textfont\eurmfam=\teneurm \scriptfont\eurmfam=\seveneurm
\scriptscriptfont\eurmfam=\fiveeurm

 \font\teneusm=eusm10 \font\seveneusm=eusm7 \font\fiveeusm=eusm5
\newfam\eusmfam
\textfont\eusmfam=\teneusm \scriptfont\eusmfam=\seveneusm
\scriptscriptfont\eusmfam=\fiveeusm
\def\eusm#1{{\fam\eusmfam\relax#1}}
\font\tencmmib=cmmib10 \skewchar\tencmmib='177
\font\sevencmmib=cmmib7 \skewchar\sevencmmib='177
\font\fivecmmib=cmmib5 \skewchar\fivecmmib='177
\newfam\cmmibfam
\textfont\cmmibfam=\tencmmib \scriptfont\cmmibfam=\sevencmmib
\scriptscriptfont\cmmibfam=\fivecmmib

\parskip=\baselineskip
\def\Bbb{\mathbb}

\def\Bbb{\mathbb}
\def\Tr{{\rm Tr}}
\def\A{{\mathcal A}}
\def\C{\Bbb{C}}
\def\1{{\bf 1}}
\def\2{{\bf 2}}
\def\3{{\bf 3}}
\def\4{{\bf 4}}
\def\bar{\overline}
\def\tilde{\widetilde}

\def\R{{\Bbb{R}}}
\def\Z{{\Bbb{Z}}}
\def\N{{\mathcal N}}
\def\hat{\widehat}
\def\neg{\negthinspace}
\numberwithin{equation}{section}
\begin{document}

\begin{titlepage}
\begin{flushright}
CERN-TH-PH/2009-019\\
\end{flushright}
\vskip 1.5in
\begin{center}
{\bf\Large{Branes, Instantons, And Taub-NUT Spaces}}\vskip 0.5cm
{Edward Witten\footnote{On leave from Institute for Advanced Study,
Princeton NJ 08540 USA. Supported in part by NSF Grant
PHY-0503584.}} \vskip 0.3in {\small{ \textit{Theory Group,
CERN}\vskip 0cm {\textit{Geneva, Switzerland}}}}

\end{center}
\vskip 0.5in

\baselineskip 16pt

\begin{abstract}
ALE and Taub-NUT (or ALF) hyper-Kahler four-manifolds can be naturally constructed as hyper-Kahler
quotients.  In the ALE case, this construction has long been understood in terms of D-branes;
here we give a D-brane derivation in the Taub-NUT case.
Likewise, instantons on ALE spaces  and on Taub-NUT spaces  have ADHM-like constructions related to hyper-Kahler quotients.  Here we refine the analysis in the Taub-NUT case by making use
of a D-brane probe.
\end{abstract}
\end{titlepage}
\vfill\eject

\date{December, 2008}

\tableofcontents

\section{Introduction}\label{intro}
\def\ello{l}
\def\RR{{\mathcal R}}
\def\fB{\frak B}
Some of the simplest non-flat solutions of Einstein's equations are the ALE (asymptotically locally
Euclidean) hyper-Kahler four-manifolds, which can be obtained by
a hyper-Kahler resolution of a quotient singularity $\R^4/\Gamma$, where $\Gamma$ is a finite
subgroup of $SU(2)$.  Such resolutions can be naturally constructed \cite{Kronheimer} as a hyper-Kahler quotient
of a finite-dimensional Euclidean space.   This construction has a natural interpretation in terms of D-branes \cite{Douglas:1996sw} and  has served as an important example in string theory.

Close cousins of these spaces are the ALF (asymptotically locally flat) hyper-Kahler four-manifolds.
These are locally asymptotic at infinity to $\R^3\times S^1$.  The examples relevant for us
are the multi-centered Taub-NUT solutions of \cite{Hawking:1976jb}. They also can be constructed
\cite{Gibbons:1996nt}
as a hyper-Kahler quotient of a finite-dimensional flat space, but this fact has not yet been interpreted in terms
of branes.   The first goal of the present paper will be, in section 2, to accomplish this.
For that purpose, we will use
the $T$-duality \cite{Gregory:1997te}  between an NS5-brane localized at a point on $\R^3\times S^1$, and an ALF space.
(This $T$-duality has been further studied in \cite{Tong}. A related statement that symmetry enhancement by $k$ parallel
NS5-branes is dual to symmetry enhancement due to an ${\mathrm A}_{k-1}$ singularity was first argued in \cite{OV}.)
Probing this $T$-duality with a suitable D-brane, we arrive at the desired hyper-Kahler quotient
construction, as well as a new understanding of this somewhat subtle $T$-duality.
The construction that we use here is really a special case of a recent construction of instantons
on Taub-NUT spaces
\cite{Cherkis:2008ip}, as will become clear.

A generalization of the fact that an ALE space $X$ can be described by a hyper-Kahler quotient of
a flat space  is that moduli spaces of instantons on $X$ can be described by such quotients \cite{KN}.  Indeed, $X$ itself can be regarded as a degenerate case of a moduli space of instantons on $X$ -- it parametrizes D0-branes on $X$, which correspond to instantons of rank 0 and second Chern class 1.  The hyper-Kahler quotient construction of instantons on an ALE space is a natural generalization of the ADHM construction of instantons on $\R^4$. It is also natural in terms of D-branes
\cite{Douglas:1996sw}, and has served, again, as an important example in string theory.

Recently \cite{Cherkis:2008ip}, a brane construction has been used to describe an analog of the ADHM construction for
instantons on an ALF space.\footnote{In a forthcoming paper \cite{CherkisNew}, which I received after completing the
present one, the differential geometry of this construction is described in detail, and a calculation is performed similar to
what we do in section 2.}
  Our goal in section 3 is to use a suitable D-brane probe, analogous
to the one used in section 2,
to slightly extend this construction in several directions. Among other things, we determine
the relation between the ADHM data and the Chern classes of the instanton bundle, and
we describe a further relationship to $M$-theory at a product of
${\mathrm A}_{k-1}$ and $ {\mathrm A}_{p-1}$ orbifold singularities.

Certain classes of instanton solutions on ALF spaces have been
constructed directly, for example in \cite{EH,ES}.  For somewhat
related work on singular monopoles and Nahm's equations, see for
example \cite{CK,CD}.  For more general reviews of Nahm's equations,
see \cite{WY,Gaiotto:2008sa}.

\section{D3-Brane Probe Of $T$-Duality}\label{probe}

\subsection{An NS5-Brane On A Transverse Circle}\label{transverse}

\def\d{\mathrm d}
\def\S{{\widetilde S}}
\def\Z{{\widetilde Z}}
\def\TN{{\rm TN}}
\def\A{{\mathrm A}}

Our basic idea in finding a D-brane construction of Taub-NUT space as a hyper-Kahler quotient
is to use the fact that the Taub-NUT space can be generated \cite{Gregory:1997te}
by  $T$-duality from a configuration consisting of an NS5-brane localized at a point on $S^1\times \R^3$.  We will call this configuration
$Z$. The $T$-dual of $S^1\times \R^3 $ without the NS5-brane is simply
$\S^1\times \R^3 $ (where $\S^1$ is the circle dual to $S^1$, with the radius inverted).  However,
$T$-duality converts the charge of the
NS5-brane into an ``$H$-monopole'' charge that is encoded in the geometry.  As a result, the $T$-dual of $Z$ is a space $\Z$ that at infinity looks like a nontrivial $\S^1$ bundle
over $\R^3$.   The first Chern class of the $\S^1$ bundle is 1 (the original NS5-brane charge).
More precisely, the region near infinity in $\R^3$ is homotopic to a two-sphere $S^2$, and the
$\S^1$ fibration, restricted to this two-sphere, has first Chern class 1.

In fact, $\Z$ coincides with the Taub-NUT space, which we will call
$\TN$, but it is difficult to show this directly via $T$-duality.
The reason for this is that  the starting point $Z$  involves an
NS5-brane, which is described by a two-dimensional conformal field
theory, but not one which is elementary or known in any explicit
form.  Consequently, the standard arguments \cite{Gregory:1997te}
for determining $\Z$ are slightly abstract.  We recall these
arguments for completeness.  The first step is to analyze the
symmetries.   Because the original configuration $Z$ has a conserved
winding number symmetry, $\Z$ must have a conserved momentum
symmetry, that is a $U(1)$ symmetry that rotates the fibers of the
$\S^1$ fibration.   Supersymmetry implies that $\Z$ is hyper-Kahler
and that the $U(1)$ symmetry commutes with the hyper-Kahler
structure of $\Z$; a symmetry with this property is called
triholomorphic.  A triholomorphic $U(1)$ symmetry
has\footnote{\label{wurf} The hyper-Kahler moment map was first
defined in \cite{HKUR}. For a charged hypermultiplet $H$, regarded
as a complex doublet, the moment map is defined by
$\vec\mu=H^\dagger\vec\sigma H$.  In general, it is defined up to an
additive constant by $d\vec\mu=\iota_V\vec\omega$, where
$\vec\omega$ are the symplectic forms defining the hyper-Kahler
manifold, $V$ is the vector field that generates the $U(1)$ action,
$\iota_V$ is contraction with $V$, and $\vec\mu$ is the moment map.}
a hyper-Kahler ``moment map'' $\vec\mu:\Z\to \R^3$, whose fibers are
the $U(1)$ orbits. Near infinity, where $U(1)$
  acts freely, this map is a circle fibration, confirming that $\Z$ is asymptotically an $\S^1$ fibration.
  In fact,\footnote{The form of the metric (\ref{genf}) implies that there cannot be orbits left fixed
  by a non-trivial finite subgroup of $U(1)$.}  the moment map is an $\S^1$  fibration everywhere except at the $U(1)$ fixed points (where the
radius of  $\S^1$ shrinks to zero).  The first
Chern class of the fibration at infinity is the number of fixed points, and to get first Chern class 1,
we need precisely 1 fixed point.  The unique hyper-Kahler four-manifold with this property is  $\TN$, so this must be $\Z$.

Finally, as a check, one notes that in the sigma model\footnote{In the full string theory, as opposed to the sigma
model, there is a conserved quantity that generalizes the string winding number. This is shown in \cite{Gregory:1997te}.}
 with target $\TN$, there is no conserved quantity
that corresponds to the winding of strings around  $\S^1$.  The reason
for this is that a wrapped string can unwind at the fixed point, where  $\S^1$ collapses to a point.
The dual of this is that in the original sigma model that describes an NS5-brane localized at a point on
$S^1\times \R^3$, there is no conserved momentum along the $S^1$ because translation invariance
along the $S^1$ is broken by the presence of the NS5-brane.

\subsection{Explicit Form Of The Taub-NUT  Metric}\label{expform}

\def\W{{\mathcal W}}

The space $\TN$ can be described very simply.  We write $\vec X$ for coordinates on $\R^3$, and  $\theta$ for an angular parameter on $S^1$.
A four-dimensional hyper-Kahler metric with a triholomorphic $U(1)$ symmetry can be put in the general
form\footnote{\label{put}
For this metric to be smooth, $\theta$ has period $4\pi$.  This normalization is used in much of the literature to avoid some factors of 2, and will be followed in all similar formulas in this paper.}
  \cite{Lindstrom:1983rt}
\begin{equation}\label{genf} \d s^2=U\,\d\vec X\cdot \d\vec X +\frac{1}{U}(\d\theta+\vec \omega\cdot \d\vec X)^2,\end{equation}
where $U$ is a harmonic function on $\R^3$, and $w=\vec \omega\cdot \d\vec X$ is a $U(1)$ connection
on $\R^3$ such that $\d U=\star \d w$.  $\TN$  corresponds to the special case of
(\ref{genf}) with
\begin{equation}\label{homely}U=\frac{1}{|\vec X-\vec x|}+\frac{1}{{\lambda^2}}.\end{equation}
Here $\lambda$ is a constant, which determines the asymptotic radius of $\S^1$.
And $\vec x$ is a point in $\R^3$ that is given by  the position
of the NS5-brane in the original description on $S^1\times \R^3$, projected to the second
factor.  For $\vec X\to \vec x$,  $\S^1$ collapses to a point.
(Of course, we could remove the dependence on $\vec x$ by shifting the coordinates $\vec X$, but we retain it since
we will eventually be interested in the relative positions of several NS5-branes.)

As for the projection of the NS5-brane position to $S^1$, it corresponds to a mode of the $B$-field on $\TN$.  This may be deduced as follows.  The position of the NS5-brane
in $S^1$ can, of course, be changed by a rotation of $S^1$.  The $T$-dual of the rotation group
of $S^1$ is a group of gauge transformations of the $B$-field that are asymptotically constant
at infinity.  A general $B$-field gauge transformation is of the form $B\to B+\d\Lambda$ where
$\Lambda$ is a one-form.  By an asymptotically constant $B$-field gauge transformation, we mean
one such that at infinity $\Lambda\to f(\d\theta+\vec \omega\cdot\d \vec X)$, for constant $f$.  We only care about
the asymptotic behavior of $\Lambda$,
because in general two gauge transformations that coincide at infinity act identically on physical states.
Moreover, we only care about the value of $f$ modulo an integer, because again, a $B$-field gauge transformation by a one-form $\Lambda$ that is closed and whose periods are integer multiples of
$2\pi$ acts trivially on physical states.  (Differently put, the gauge transformation corresponding
to $\Lambda$ is trivial if, when regarded as
an abelian gauge field, $\Lambda$ is pure gauge.)  Hence the asymptotic value of $2\pi f$ is an
angle.

In the present case, because the circle $\tilde S^1$ shrinks to zero size in the interior of $\TN$, a one-form  $\Lambda$  that is asymptotic to $f(\d\theta+\vec \omega\cdot \d\vec X)$ (and so has a nonzero
integral over $\tilde S^1$)
cannot be closed, and therefore a gauge transformation $B\to B+\d \Lambda$ generates a non-trivial
shift of the $B$-field.  This is dual to the fact that a rotation of the original $S^1$ shifts the position
of the NS5-brane.
The  $T$-dual of the angular position of the NS5-brane is encoded in a $B$-field of the form
  $B=\d\Lambda$, where at infinity
$\Lambda\to f(d\theta+\vec \omega\cdot \d\vec X)$. The angular position of the original
NS5-brane is $2\pi f$.

Apart from being of the pure gauge form $B= \d\Lambda$, the two-form $B$ that is dual to the fivebrane position should be anti-selfdual
(and in particular harmonic) for supersymmetry.  It should also be invariant under rotations of
$\R^3$ around the point $\vec X=\vec x$, since the original NS5-brane had this property.
In fact, there is a unique anti-selfdual harmonic two-form on $\TN$ with all the right behavior (apparently first
constructed in \cite{EgH,P}; see also \cite{Gi}), namely
 $B=\d\Lambda$ with
\begin{equation}\label{kk}\Lambda=\frac{r}{r+\lambda^2}\left(\d\theta+\vec \omega\cdot\vec \d X\right),~~
r=|\vec X-\vec x|.\end{equation}
This formula and the previous ones of this subsection have a natural explanation by constructing $\TN$ as a hyper-Kahler quotient \cite{Gibbons:1996nt}, as we will recall in section \ref{review}.

 \subsection{A More Concrete Approach}\label{concrete}

 \def\xx{\vec {x}}
The argument of section \ref{transverse} for finding the $T$-dual of the space $Z$ is clear but somewhat abstract.  We would like to find a more concrete way to
analyze the $T$-dual of $Z$.  To do this, we will consider
a D-brane probe.  The D-brane will be an  E1-brane (a Euclidean D-brane of one-dimensional
worldvolume) wrapped on $L=S^1\times \xx$, where $\xx$ is a point
in $\R^3$.  The moduli space $W$ of supersymmetric configurations of the E1-brane is four-dimensional; the choice of $\xx$ depends on three parameters, and (at least when the E1-brane is far away from the NS5-brane) the fourth parameter is  the holonomy around
$L$ of the $U(1)$ gauge field of the E1-brane.  $T$-duality on $S^1$, while leaving
$W$ unchanged,  converts the E1-brane to
an E0-brane on $\Z$.  But the moduli space that parametrizes E0-branes on $\Z$ is simply
a copy of $\Z$ -- an E0-brane is supported at a point, which may be any point in $\Z$.
So $\Z$ is the same as $W$,
and in other words, to determine the $T$-dual of $Z$, it suffices to determine the E1-brane moduli
space $W$.  This can be done using arguments of a standard type, and will lead us to an explicit
description of  $W$ or  $\Z$ that will coincide with the
hyper-Kahler quotient construction \cite{Gibbons:1996nt} of the Taub-NUT space $\TN$.

\def\N{{\mathcal N}}
\def\X{{{\vec X}}}
\def\Y{{{\vec Z}}}
\def\Y{{\vec Y}}
Though  determining $W$ is just a question about the sigma model with target $Z$, the steps required to
answer it are probably more familiar if we
embed the problem in Type IIB superstring theory.
 We take the ten-dimensional spacetime $M$ to be
\begin{equation}\label{gurf}M=\R^3\times S^1\times \R^3_{\vec X}\times \R^3_{\vec Y}.\end{equation}
All branes we consider will have a worldvolume that includes the first factor $\R^3$, but they
will in general have no other factors in common.  Such branes will generate at low energies
an effective  gauge theory on $\R^3$; we will choose the branes so that this theory has
half of the possible supersymmetry (eight supercharges).   The last two factors $
\R^3_\X$ and $\R^3_\Y$ are parametrized by triplets of real coordinates $\X$ and $\Y$;
the corresponding rotation groups $SO(3)_\X$ and $SO(3)_\Y$ are $R$-symmetry groups in the three-dimensional gauge theory.
To embed  in this context our problem of finding the $T$-dual of $Z$,
we introduce an NS5-brane whose
worldvolume is $\R^3\times \R^3_\Y$ times a point in $S^1\times \R^3_\X$.  Thus, $S^1\times \R^3_\X$,
with the embedded NS5-brane, is what we earlier called $Z$.  Up to a translation symmetry, we can take
the NS5-brane to be located at $0=y=\vec X$, where $S^1$, whose circumference we call $2\pi R$, is parametrized by a variable
$y$ with $y\cong y+2\pi R$.  Similarly, we promote the E1-brane
wrapped on a circle in $Z$ to a D3-brane whose worldvolume is $\R^3\times S^1$ times a
point in $\R^3_\X\times \R^3_\Y$.  Since the NS5-brane is invariant under translations in $\vec Y$,
there is no essential loss in taking the D3-brane to be at $\vec Y=0$.  The value of $\vec X$ for
the D3-brane parametrizes three of its four moduli.

The effective action on the worldvolume of a stack of $N$ D3-branes is four-dimensional $\N=4$ super Yang-Mills theory,
with gauge group $U(N)$.   (In our present discussion, we will take a single D3-brane,
but the generalization to any number is important later.)    When a
D3-brane intersects an NS5-brane, the gauge theory is ``cut.'' We get separate $U(N)$ gauge theories
on either side of the NS5-brane, coupled via the existence of a bifundamental hypermultiplet supported
at the D3-NS5 intersection.

In the case of D3-branes wrapped on $\R^3\times S^1$ (the first two factors in $M$),
their motion in the normal space $\R^3_\X\times \R^3_\Y$ is described by adjoint-valued
scalar
fields that we call $\vec X$ and $\vec Y$. Incorporation of an NS5-brane (embedded as above) reduces
four-dimensional $\N=4$ supersymmetry to what we might call three-dimensional $\N=4$ supersymmetry
(with half as many supercharges).
Invariance under this remaining supersymmetry algebra
requires that $\vec Y$ should be a constant (which we will take to be
zero), while $\vec X$ should obey Nahm's equations.  In their gauge-invariant form, these equations
read
\begin{equation}\label{cuffy} \frac{D\X}{D y}+\vec X\times \vec X=0. \end{equation}
Here $D/Dy=d/dy+A_y$, with $A_y$ the component of the gauge field in
the $S^1$ direction.  Henceforth we denote $A_y$ simply as $A$.
Locally, $A$ can be gauged away, but it is more useful to write the
equations in a gauge-invariant way.    Also, $\vec X\times \vec X$
is defined by $(\vec X\times \vec
X)_a=\frac{1}{2}\epsilon_{abc}[\X_b,\X_c]$, for $a,b,c=1,2,3$. (For
more detail on Nahm's equations, see, for example, \cite{WY} or
section 2 of \cite{Gaiotto:2008sa}.)

\def\NN{ N}
The effect of introducing an  NS5-brane at $y=y_0$ (eventually we
will set $y_0=0$) is that, locally,  the D3-branes support separate
$U(N)$ gauge theories for $y\leq y_0 $ and $y\geq y_0$. (If, as in
our problem, there is only one NS5-brane and $y$ parametrizes a
circle,  then these two gauge theories are connected by going the
long way around the circle.) We write $(\X^-,A^-)$ and $(\X^+,A^+)$
for the fields to the left and right of $y=y_0$.  At $y=y_0$, there
is a bifundamental hypermultiplet field $H$, transforming as
$(\NN,\bar\NN)\oplus (\bar\NN,\NN)$ under $U(N)\times U(N)$.   It
has hyper-Kahler moment maps\footnote{See footnote \ref{wurf} for
the definition of the hyper-Kahler moment map.} $\vec\mu^-$ and
$\vec \mu^+$ for the left and right actions of $U(N)$. Supersymmetry
requires that $(\X^-,A^-)$ and $(\X^+,A^+)$ obey Nahm's equations
for $y\not= y_0$, and additionally leads to boundary
conditions\footnote{For brevity, we set the gauge coupling to 1
except in section \ref{review}. Otherwise, $\vec X^\pm$ should be
multiplied here by $1/e^2$.}
\begin{equation}\label{zelf} \X^-(y_0)=\vec\mu^-,~~  \X^+(y_0)=-\vec\mu^+.\end{equation}

Now let us specialize this to our problem, in which there is only a single D3-brane, with gauge group  $U(1)$, and there is just a single NS5-brane, which we place at $y=0$.
Because the gauge group is abelian, Nahm's equations reduce to
\begin{equation}\label{jura}\frac{\d\X}{\d y}=0.\end{equation}
The bifundamental hypermultiplet $H$ reduces to a single
hypermultiplet that transforms with charge $1$ or $-1$ under gauge
transformations acting on the left or right. In particular, this
means that $\vec\mu^-=-\vec\mu^+$; we henceforth write $\vec \mu_H$
for $\vec\mu^+$. Setting $y_0=0$, it is convenient to unwrap the
circle $S^1$ to an interval $I:0\leq y\leq 2\pi R$; after doing so,
we set $ \X^+(y_0)=\vec X(0)$ and $\X^-(y_0)=\vec X(2\pi R)$.
Approaching $y_0$ from right or left now corresponds to $y=0$ or
$y=2 \pi R$, so the boundary conditions (\ref{zelf}) become
\begin{equation}\label{telf} \vec X(0)=-\vec \mu_H =\vec X(2\pi R).\end{equation}
If we also impose Nahm's equations (\ref{jura}), then $\X(0)=\X(2\pi R)$ and the second condition
in (\ref{telf}) can be dropped.

At this point, we can rather trivially describe topologically
the moduli space $W$ of supersymmetric states of the
D3-brane.  (It will take more work to determine its hyper-Kahler metric.)
The hypermultiplet $H$ parametrizes a copy of $\R^4$.
  After picking $H$, we compute $-\vec\mu_H$ and use (\ref{telf}) to determine $ \X(0)$.  Then according
  to (\ref{jura}), we set $ \X(y)=\X(0)$ for all $y$.  Finally, we gauge  away the $U(1)$
   gauge field
  $A$, which does not appear in any of the above equations, and contains no gauge-invariant
  information.  The upshot is that there is a unique supersymmetric configuration for each choice of
  $H$, so $W\cong \R^4$.

  \def\W{{\mathcal W}}
  \def\G{{\mathcal G}}
  \def\d{\mathrm d}
  We recall that the expected answer is that $W$ should be the Taub-NUT hyper-Kahler four-manifold
  $\TN$.
  Topologically, this is equivalent to $\R^4$, so we are on the right track so far.
  To determine the hyper-Kahler metric of $W$, we repeat the above analysis with more care.
  Nahm's equations and the boundary condition (\ref{telf}) can be interpreted in terms of
  an infinite-dimensional hyper-Kahler quotient.  We write $\W$ for the space of triples $(\vec X,A,H)$.
   $\W$ carries a natural flat hyper-Kahler metric that we will write down shortly, and
   $W$ can be interpreted, roughly speaking,
  as the hyper-Kahler quotient of $\W$ by the group of gauge transformations.

  However, constant
  gauge transformations act trivially on $\W$, since {\it (i)} the gauge
  group is abelian (so constant gauge transformations do not act on $\vec X$ or $A$), and {\it (ii)} $H$ transforms with equal and opposite charges under gauge transformations
  at $y=0$ and $y=2\pi R$ (hence $H$ is invariant under gauge transformations that are actually    constant).
  So we really want to remove constant gauge transformations from the discussion.  A convenient
  way to do this is to allow only gauge transformations that are trivial at $y=0$.  We let $\G$
  be the group of such gauge transformations, that is, the group of maps $g(y):I\to U(1)$ such that
  $g(0)=1$.   With this definition, $W$ is the hyper-Kahler quotient of $\W$ by $\G$.
  Here $\W$ is endowed with a natural flat hyper-Kahler metric:
 \begin{equation}\label{urolf}\d s^2=|\d H|^2+\frac{1}{2}\int_0^{2\pi R}\d y \left( \d \X(y)\cdot \d \X(y)+
 \d A(y)^2\right);\end{equation}
 $|\d H|^2$ is the flat hyper-Kahler metric of the space $\R^4$ that is parametrized by $H$.
 The hyper-Kahler moment map for the action of $\G$ on $\W$ is
 \begin{equation}
 \label{hypmom} \vec\mu_\W(y)=\frac{\d\X}{\d y}+\delta(y-2\pi R)(\vec X(2\pi R)+\vec\mu_H).\end{equation}
 (This computation is explained in section  2.3.2 of \cite{Gaiotto:2008sa}.)  So the condition $\vec\mu_\W=0$ gives
 Nahm's equations plus the boundary condition (\ref{telf}).

 \def\neg{\negthinspace}
 Thus, the space $W$ of supersymmetric vacua of this system can be interpreted as the hyper-Kahler
 quotient of $\W$ by $\G$. Such a hyper-Kahler quotient is often denoted $\W/\neg /\neg /\G$.   We are
 getting close to the result of \cite{Gibbons:1996nt}, but we are not there yet. So far
 we have exhibited $W$ as
 an infinite-dimensional hyper-Kahler quotient, while in \cite{Gibbons:1996nt}, $\TN$  is realized as  a
 hyper-Kahler quotient of a finite-dimensional flat manifold.

We can establish the equivalence of these two constructions by using the fact that the infinite-dimensional group $\G$ has a codimension
 1 normal subgroup $\G_\star$, consisting of gauge transformations that are trivial at $y=2\pi R$
 (as well as at $y=0$).  Thus, an element of $\G_\star$ is a map $g(y):I\to U(1)$ with $g(0)=g(2\pi R)=1$.
The group $\G$ maps to $U(1)$ by evaluating a map $g(y)$ at $y=2\pi R$, and this gives $\G$ as
an extension of $\G_\star$ by $U(1)$:
\begin{equation}\label{hobo} 1\to \G_\star\to \G\to U(1)\to 1. \end{equation}
We can compute the hyper-Kahler quotient of $\W$ by $\G$ by first taking its hyper-Kahler
quotient by $\G_\star$, to get a space $W_\star$ that as we will see shortly is finite-dimensional and flat.
Then $W$ is the hyper-Kahler quotient of $W_\star$ by $U(1)$.  When we make this
explicit,  we will recover the description of $\TN$ given in  \cite{Gibbons:1996nt}, confirming finally
that $W$ is the same as $\TN$.  (The relevant part of \cite{Gibbons:1996nt} is reviewed in section
\ref{review}.)

The hyper-Kahler moment map for $\G_\star$ is the same as (\ref{hypmom}), except that we should
omit the delta function term (since a generator of $\G_\star$ vanishes at $y=2\pi R$).  So one step
in determining $W_\star=\W/\neg/\neg/\G_\star$ is to impose Nahm's equations and set $\X(y)$ to a constant, which we just call $\X$.
This gives a subspace of $\W$ that we will call $\W'$.  Then $W_\star$ is the quotient of $\W'$ by
$\G_\star$. Here $\G_\star$ acts only on $A$; in its action on $A$,
the only invariant is
\begin{equation}\label{korn}\alpha=\int_0^{2\pi R}\d y\, A~~~~~{\rm mod}~2\pi.\end{equation}
$\alpha$ is a $\G_\star$-invariant mod $2\pi$ since a gauge transformation in $\G_\star$ is trivial at both endpoints (gauge transformations in $\G_\star$
that have a non-trivial
winding number around $U(1)$ can shift $\alpha$ by an integer multiple of $2\pi$).

To find the induced hyper-Kahler metric on $W_\star=\W'/\G_\star$, we take a slice of the action of
$\G_\star$ on $\W'$ that is orthogonal to the $\G_\star$ orbits (with respect to the metric (\ref{urolf})), and evaluate the metric (\ref{urolf}) on this slice.  The appropriate slice is obtained simply by taking $A(y)$
to be constant:
\begin{equation}\label{zorn}A(y)=\frac{\alpha}{2\pi R}.\end{equation}
(To see that this slice is orthogonal to the orbits, we note that the change in $A(y)$ under an infinitesimal gauge transformation
in $\G_\star$ is of the form $A(y)\to A(y)+\d\epsilon/\d y$, where $\epsilon(0)=\epsilon(2\pi R)=0$;
orthogonality means that $\int_0^{2\pi R}\d y\,\alpha\,(\d\epsilon/\d y)=0$ for constant $\alpha$.)
With this slice, we compute the metric on $W_\star$ by simply evaluating (\ref{urolf}), and get
\begin{equation}\label{zurolf} \d s^2=|\d H|^2+\pi R\,\d\X\cdot \d\X +\frac{1}{4\pi R}\d\alpha^2.
\end{equation}

In particular, this is a flat metric on $W_\star=\R^7\times S^1$.  The space $M$ is the finite-dimensional
hyper-Kahler quotient $M=W_\star/\neg/\neg/U(1)$.  (The moment map for this $U(1)$ is simply the
coefficient of the delta function in   (\ref{hypmom}).)  This is precisely the description of the Taub-NUT
manifold given in \cite{Gibbons:1996nt}, so we have accomplished our task of identifying the D-brane moduli
space -- and hence the $T$-dual of the original configuration $Z$ -- with that space.

\subsection{The Multi-Centered Case}\label{mulcit}

\def\k{{\bf k}}
The problem we have considered so far has a natural generalization.  We consider
$k$ NS5-branes in $S^1\times \R^3$, at locations $p_\sigma=y_\sigma\times \vec x_\sigma$,
$\sigma=1,\dots,k$.  We denote $S^1\times \R^3$ with these fivebranes as $Z_{\k}$.  We now want to determine what happens
when we apply to $Z_\k$ the usual $T$-duality of $S^1$.

The answer is known to be a multicentered Taub-NUT or ALF manifold, as first constructed in \cite{Hawking:1976jb}.
This can be shown by arguments similar to those that were summarized in section \ref{transverse}.
The $T$-dual of $Z_{\k}$ must be a hyper-Kahler four-manifold with a triholomorphic $U(1)$ symmetry.  It therefore can be put in the general form
of eqn. (\ref{genf}), for some harmonic function $U$.  $U$ must approach a constant at infinity
(since the radius of the dual circle $\tilde S^1$ must be asymptotically constant) and must have
a singularity as in (\ref{homely}) for each fivebrane.  So we must have
\begin{equation}\label{gomely}U=\sum_{\sigma=1}^k\frac{1}{|\vec X-\vec x_\sigma|}+\frac{1}{{\lambda^2}}.\end{equation}
This is the multi-centered Taub-NUT or  ALF geometry \cite{Hawking:1976jb}; we denote this space as $\TN_\k$.

As in section \ref{expform}, our goal is to understand this result in a more explicit way, using
a D-brane probe.   The strategy will be the same. The $T$-dual of $Z_{\k}$ is the same as the moduli
space of supersymmetric states of an E1-brane wrapped on the product of  $S^1$ with a point in $\R^3$.
In the Type IIB interpretation, we consider the spacetime
\begin{equation}\label{gurft}M=\R^3\times S^1\times \R^3_{\vec X}\times \R^3_{\vec Y}\end{equation}
with NS5-branes supported on $\R^3\times y_\sigma\times \vec x_\sigma\times \R^3_\Y$, $\sigma=1,\dots,k$.  In this spacetime, we consider a D3-brane probe supported on
$\R^3\times S^1\times \vec q\times \{0\}$, where $\vec q$ is a point in $\R^3_\X$.    The $T$-dual of
$Z_\k$ is
the moduli space $W$ of supersymmetric vacua in the probe theory. (More exactly, we want the subspace consisting
of such vacua  that are invariant under $SO(3)_\Y$ --  that is, those that are supported at $\Y=0$.  This
is why in what follows we ignore the motion of the probe in $\Y$.)

It is convenient to begin with the case that the NS5-branes are all at the same location in
$\R^3_\X$, say $\vec x_\sigma=0$, $\sigma=1,\dots,k$.  The D3-brane can ``break'' when it crosses
an NS5-brane.  This results in general in $k$  D3-brane slabs, supported respectively on
$\R^3\times [y_\sigma,y_{\sigma+1}]$, where  $  [y_\sigma,y_{\sigma+1}]$ is the closed interval in $S^1$
with the indicated endpoints.  Each slab supports a $U(1)$ gauge theory with $\N=4$
supersymmetry; this theory has its own fields $\vec X$ and $A$.  To simplify the notation,
we group these all together as a single set of fields $\vec X(y)$, $A(y)$; but when we do this,
 $\vec X(y)$ may
be discontinuous at $y=y_{\sigma},\,\sigma=1,\dots ,k $, and the gauge transformations acting on $A(y)$ may likewise
be discontinuous.  (In the notation, we suppress the dependence of the fields on the $\R^3$ directions of $\R^3\times S^1$,
since supersymmetric vacua are described by fields that are constant in those directions.)  Supported on each common boundary  $\R^3\times \{y_\sigma\}$ of two adjacent slabs is a bifundamental hypermultiplet $H_\sigma$ that
transforms with equal and opposite charges under gauge transformations of the left or right slab restricted to $y=y_\sigma$.

Supersymmetry requires that $\vec X$ obeys Nahm's equations
\begin{equation}\label{longnahm}\frac{\d \vec X}{\d y}=0,\end{equation}
with possible jumps across slab boundaries at $y=y_\sigma$.  At $y=y_\sigma$, we need
boundary conditions analogous to (\ref{zelf}).  We write $\X^-(y_\sigma)$ and $\X^+(y_\sigma)$ for
the limits of $\X(y)$ for $y\to y_\sigma$ from the left or right.  Similarly, we write $\vec\mu_{L,\sigma}$
and $\vec\mu_{R,\sigma}$ for the hyper-Kahler moment map of $H_\sigma$ under gauge transformations acting on the left or right.\footnote{We fix an additive constant in $\vec \mu$ by
requiring it to vanish when $H_\sigma=0$.  This is the unique choice that makes $H_\sigma$
$SO(3)_\X$-invariant.  That invariance will be relaxed shortly.}
As in (\ref{zelf}), the boundary condition is then
\begin{equation}\label{bcon} \X^-(y_\sigma)=\vec\mu_{L,\sigma},~~\X^+(y_\sigma)=-\vec\mu_{R,\sigma}.\end{equation}
In this form, the boundary condition holds in the nonabelian case with any number of probe
D3-branes.  However, if there is only a single D3-brane, the left and right hypermultiplet moment
maps are equal and opposite as before, and we write $\vec\mu_{L,\sigma}=-\vec\mu_{R,\sigma}
=-\vec\mu_{H_\sigma}$.  The boundary condition is then
\begin{equation}\label{zcon}\X^-(y_\sigma)=-\vec\mu_{H_\sigma}=\X^+(y_\sigma).\end{equation}

The $T$-dual of $Z_\k$ is simply the space of solutions of (\ref{longnahm}) and (\ref{bcon})
for a triple $(X(y),A(y),H_\sigma)$, modulo
gauge transformations acting on $A$ and $H_\sigma$.  ($A(y)$ actually does not appear in the equations.) Before determining the moduli
space, let us generalize slightly to the case that the NS5-branes have arbitrary transverse positions
$\vec x_\sigma\in \R^3_\X$.

For any one value of $\sigma$, the effect of this can be eliminated by shifting $\vec X$ by
$\vec x_\sigma$.   The resulting generalization of (\ref{zcon}) is
\begin{equation}\label{zocon}\X^-(y_\sigma)=-\vec\mu_{H_\sigma}+\vec x_\sigma=\X^+(y_\sigma).\end{equation}
  Of course, if there
are several NS5-branes with different values of $\vec x_\sigma$, we cannot shift $\vec X$ so
as to remove these constants for all values of $\sigma$.

Equation (\ref{zocon}) has a simple interpretation.  Assuming it exists (which is so under a mild topological
condition), the hyper-Kahler moment map $\vec\mu$ of a hyper-Kahler manifold with $U(1)$ symmetry
-- in this case the flat hyper-Kahler manifold parametrized by $H_\sigma$ --   is unique only up to an additive constant.  Precisely such a constant is visible in  eqn.  (\ref{zocon}), which says that the limit
of $\X(y)$ for $y\to y_\sigma$ (from left or right) is equal to $-\vec\mu'_{H_\sigma}$, where
we define a shifted moment map
\begin{equation}\label{shifted}\vec \mu\,'_{H_\sigma}
=\vec\mu_{H_\sigma}-\vec x_\sigma \end{equation} with an additive constant $-\vec x_\sigma$.  Adding such a constant to the moment map preserves the hyper-Kahler
nature of the hyper-Kahler quotient.
This is also clear in the present example from the fact that these constants result from shifts in
NS5-brane positions that preserve supersymmetry.

The explicit determination of the moduli space is similar to what we have already seen in section
\ref{concrete}.  We let $\W$ be the space of data $(\vec X,A,H_1,\dots,H_k)$.  $\W$ has a flat
hyper-Kahler metric that is the obvious generalization of (\ref{urolf}):
 \begin{equation}\label{zzurolf}\d s^2=\sum_{\sigma=1}^k|\d H_\sigma|^2+\frac{1}{2}\int_0^{2\pi R}\d y \left( \d \X(y)\cdot \d \X(y)+
 \d A(y)^2\right).\end{equation}
We want to take the hyper-Kahler quotient of $\W$ by the group of gauge transformations,
but in doing so, we again take into account the fact that constant gauge transformations act trivially.
It is therefore convenient to pick a point, say $y=0$, and consider only gauge transformations that are trivial at that point.
Also, a gauge transformation may be discontinuous at the points $y=y_\sigma$, $\sigma=1,\dots,k$.
It therefore has separate limits $g^-(y_\sigma)$ and $g^+(y_\sigma)$ as $y\to y_\sigma$ from the left or
right.  We set
\begin{equation}\label{crf} u_\sigma=g^-(y_\sigma)^{-1}g^+(y_\sigma),~\sigma=1,\dots,k.\end{equation}
We may as well pick coordinates so that $0=y_0<y_1<\dots <y_{k-1}$.  (We consider the index $\sigma$ to have period $k$, so that $y_0$ is the same as $y_k$.  At the end of the analysis,
one can permit some of the $y_\sigma$ to coincide.)  We let $\G$ be the  group of maps
$g(y):S^1\to U(1)$, which are continuous except possibly at the points $y_\sigma$,
and which obey $g^+(0)=1$.  The moduli space we want is the hyper-Kahler quotient $W=\W/\neg/\neg/\G$.

This describes the $T$-dual of $Z_\k$ as an infinite-dimensional hyper-Kahler quotient, but as in the
example with a single fivebrane, it is possible to reduce this to a finite-dimensional hyper-Kahler quotient.  We let $\G_\star$ be the normal subgroup of $\G$ consisting of continuous gauge
transformations, in other words those for which $u_\sigma=1$, $\sigma=1,\dots,p$.
Thus, we have an exact  sequence
\begin{equation}\label{ont}1\to \G_\star\to \G\to U(1)^p\to 1,\end{equation}
mapping an element of $\G$ to the corresponding collection of $u_\sigma$'s.  We can reduce
to finite dimensions by first taking a hyper-Kahler quotient by $\G_\star$, after which we
take a finite-dimensional hyper-Kahler quotient by $F=U(1)^p$.  So we define $W_\star=\W/\neg/\neg/\G_\star$
and the desired moduli space is then $W=W_\star/\neg/\neg/F$.

To make this explicit, we need to construct $W_\star$.   $\G_\star$ acts trivially on the hypermultiplets $H_\sigma$, which therefore are unaffected by the hyper-Kahler quotient by $\G_\star$.  The vanishing
of the $\G_\star$ moment map simply gives Nahm's equations $\d\X/\d y=0$.  Because $\G_\star$ consists
of continuous gauge transformations on the circle, nothing special happens at the points
$y=y_\sigma$.  Therefore, when the $\G_\star$ moment map vanishes,  $\X$ is constant, independent of $y$.

A longer way to obtain the same statement is to observe that away from the special points $y_\sigma$,
vanishing of the $\G_\star$ moment map certainly gives Nahm's equations $\d\X/\d y=0$.  At
$y=y_\sigma$, because of requiring a gauge transformation to be continuous, we only get
one linear combination of the two conditions $\X^-(y_\sigma)
=-\vec\mu_{H_\sigma}+\vec x_\sigma$ and  $\X^+(y_\sigma)
=-\vec\mu_{H_\sigma}+\vec x_\sigma$ in (\ref{zocon}).  This linear combination is independent of $H_\sigma$ (which does not appear in the $\G_\star$ moment map, since $\G_\star$ acts trivially on $H_\sigma$),
so even without computation, it must be the difference of the two conditions, $\X^-(y_\sigma)
=\X^+(y_\sigma)$.

To complete the hyper-Kahler quotient, we must also divide by $\G_\star$, which acts only on $A$.
As in (\ref{korn}), the only invariant is the global holonomy $\alpha=\int_0^{2\pi R}\d y\,A$, valued in
$\R/2\pi {\Bbb Z}$.  $W_\star$ is therefore a simple product $\prod_{\sigma=1}^p\R^4_\sigma\times (\R^3\times
S^1)$, where $\R^4_\sigma$ is a copy of $\R^4$ parametrized by $H_\sigma$, $\R^3$ is
parametrized by $\vec X$, and $S^1$ is parametrized by $\alpha$.   The hyper-Kahler metric
of $W_\star$ is the obvious analog of (\ref{zurolf}):
\begin{equation}\label{turolf} \d s^2=\sum_{\sigma=1}^p|\d H_\sigma|^2+{\pi R}\,\d\X\cdot \d\X +\frac{1}{4\pi R}\d\alpha^2.
\end{equation}
This can be shown by the same reasoning as before.

Finally, we have to take the hyper-Kahler quotient of $W_\star$ by $F=U(1)^p$.
We write $F=\prod_{\sigma=1}^p U(1)_\sigma$, where $U(1)_\sigma$ is a copy of $U(1)$ parametrized by $u_\sigma$ (defined in (\ref{crf})).  Thus, $U(1)_\sigma$ acts in the usual way on $H_{\sigma}$
and acts trivially on the other hypermultiplets.
On the other fields $(\X,\alpha)$, all of the $U(1)_\sigma$'s act in the same way, independent of $\sigma$: they act
trivially on $\X$, and by rotation of the angle $\alpha$.
The $U(1)_\sigma$ moment map for this action
is precisely $-\X+\vec\mu_{H_\sigma}-\vec x_\sigma$, whose
vanishing gives the remaining conditions in (\ref{zocon}).

So the moduli space we want -- the $T$-dual of $Z_\k$ --  is the hyper-Kahler quotient
$W_\star/\neg/\neg/F$.  On the other hand, according to \cite{Gibbons:1996nt} (and as reviewed
in section \ref{review}), precisely this finite-dimensional
hyper-Kahler quotient is equal to the multi-centered Taub-NUT space $\TN_\k$
(with $U$ given in eqn.  (\ref{gomely})).   So we have obtained a more concrete, although perhaps
longer, explanation of why the $T$-dual of $Z_\k$ is $\TN_\k$.

The moduli of $Z_\k$, apart from the radius of the circle, are the positions $\vec x_\sigma\times y_\sigma$ of the NS5-branes.  The $\vec x_\sigma$ enter  the geometry of the $T$-dual -- because of their
appearance in the moment map, and consequently in the function $U$ of eqn. (\ref{gomely}) -- while as in section \ref{expform}, the $y_\sigma$ are modes
of the $B$-field on $\TN_\k$.

\subsubsection{Topology}\label{topology}

\begin{figure}
  \begin{center}
  \includegraphics[width=3in]{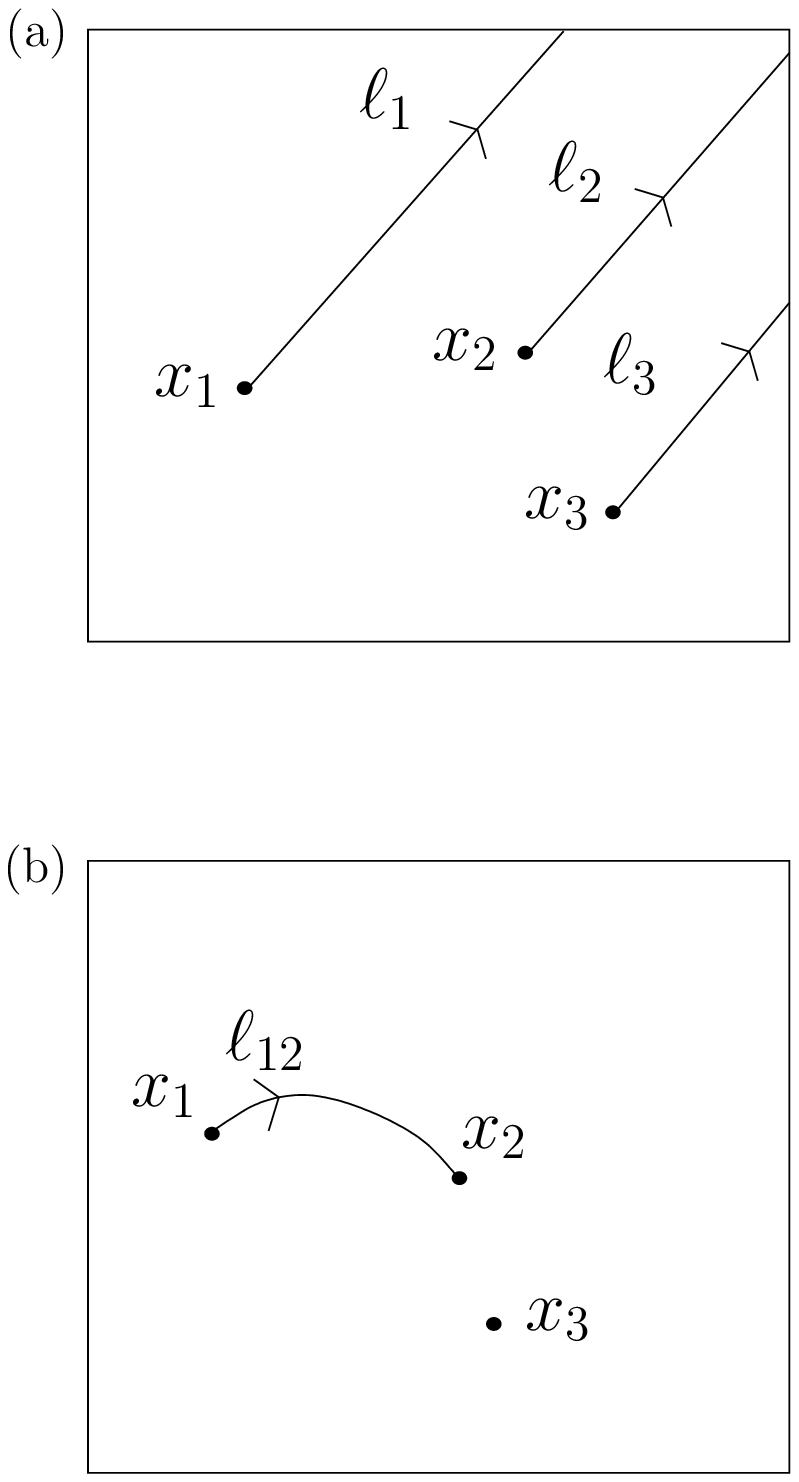}
  \end{center}

\caption{\small   (a) Nonintersecting curves $\ello_\sigma$, $\sigma=1,\dots,k$, connecting the points $x_\sigma\in\R^3$ to infinity
(in the figure, $k=3$).  Their inverse images in $\TN_\k$ are the noncompact cycles $C_\sigma$.  (b) A curve $\ello_{\sigma
\sigma'}$ connecting two of the points $x_\sigma,x_{\sigma'}$, and not meeting any of the others (sketched here for $\sigma,\sigma'=1,2$).  Its inverse image is a two-cycle $C_{\sigma\sigma'}\subset\TN_\k$
that is topologically $S^2$.  $\ello_{\sigma\sigma'}$ is homologous to the difference $\ello_\sigma-\ello_{\sigma'}$ (which represents a curve from
$x_\sigma$ to infinity and back to $x_{\sigma'}$).}
  \label{Curves}\end{figure}

The relevant modes of the $B$-field can be
described topologically as follows.   It is convenient to refer back to the multi-Taub-NUT metric
\begin{equation}\label{ogenf} \d s^2={U}\,\d\vec X\cdot \d\vec X +\frac{1}{U}(\d\theta+\vec \omega\cdot \d\vec X)^2,~~U=\sum_{\sigma=1}^k\frac{1}{|\vec X-\vec x_\sigma|}+\frac{1}{{\lambda^2}}.\end{equation}
The moment map $\vec\mu:\TN\to \R^3_\X$ is the map that forgets
$\theta$ and remembers $\vec X$. The inverse image of a generic
point in $\R^3_\X$ is a circle $\S^1$ (parametrized by $\theta$),
but the inverse image of one of the points $\vec x_\sigma$,
$\sigma=1,\dots,k$ (at which $\S^1$ shrinks to zero radius) is a
point. As in fig. \ref{Curves}(a), for $\sigma=1,\dots,k$, let
$\ello_\sigma$ be a path in $\R^3_\X$ from $\vec x_\sigma$ to
infinity, and not passing through any of the other points $\vec
x_{\sigma'}$.  We can take the $\ello_\sigma$ to be parallel rays in
some generic common direction. Then for each $\sigma$,
$C_\sigma=\vec\mu^{-1}(\ello_\sigma)$ is topologically an open disc
(and metrically a sort of  semi-infinite cigar). We define
\begin{equation}\label{kids}\theta_\sigma=\int_{C_\sigma}B.\end{equation}
$\theta_\sigma$ does not depend on the precise path $\ello_\sigma$ (as long as $B$ is flat
and vanishes at infinity) and is invariant mod $2\pi {\Bbb Z}$ under $B$-field gauge transformations
that are trivial at infinity.  Thus, we can regard the $\theta_\sigma$ as angles; they are dual
to the angular positions $y_\sigma/R$  of the original NS5-branes.   As in section
\ref{expform}, a rotation of the original circle in $Z_\k$ is dual to a $B$-field gauge transformation
that is nonzero but constant at infinity ($B\to B+\d\Lambda$ where $\Lambda$ is asymptotically
 $f(\d\theta+\vec \omega\cdot \d\vec X)$, with constant $f$).  A gauge transformation of this type shifts
all $\theta_\sigma$ by $2\pi f$, as one can readily calculate using the definition (\ref{kids}),
so the differences $\theta_\sigma-\theta_{\sigma'}$ are invariant.
The fact that these
 are completely gauge-invariant (regardless of the behavior of the
gauge parameter at infinity) is clear from the following.
The difference $\ello_\sigma-\ello_\sigma'$ is homologous to a path $\ello_{\sigma\sigma'}$ from
$\vec x_\sigma$ to $\vec x_{\sigma'}$ (fig. \ref{Curves}(b)).   The inverse image $C_{\sigma\sigma'}=\vec\mu^{-1}(
\ello_{\sigma\sigma'})$ is a compact two-cycle, topologically a copy of $S^2$. So
\begin{equation}\label{horse}\theta_\sigma-\theta_{\sigma'}=\int_{C_{\sigma\sigma'}}B\end{equation}
is completely gauge-invariant.

\begin{figure}
  \begin{center}
    \includegraphics[width=3in]{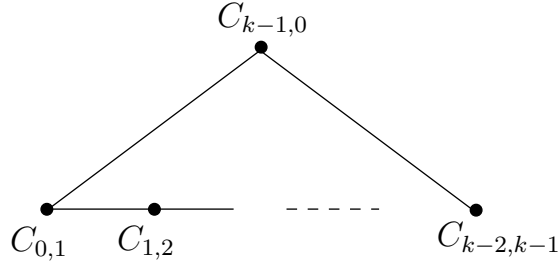}
  \end{center}
\caption{\small   The intersection pairings of the cycles
$C_{\sigma,\sigma+1 }$ have a natural interpretation in terms of the
Dynkin diagram of the group ${\mathrm A}_{k-1}$.  More precisely,
the cycles $C_{01},C_{12},\dots,C_{k-2,k-1}$, which are arranged
here in the horizontal row at the bottom, correspond to the nodes of
the ordinary Dynkin diagram of ${\mathrm A}_{k-1}$.  The diagonal elements of the intersection
matrix are The cycle
$C_{k-1,0}$, depicted here at the top, then represents the
additional node of the extended
 Dynkin diagram of this group.}
  \label{Dynkin}\end{figure}

\def\ZZ{{\Bbb Z}}
The compact cycles $C_{\sigma\sigma'}$ generate the second homology
group of $\TN_\k$. Indeed, that group is isomorphic to $\ZZ^{k-1}$,
generated (for example) by the cycles
$D_\sigma=C_{\sigma-1,\sigma}$, $\sigma=1,\dots,k-1$.  These cycles
intersect like the simple roots of the group ${\mathrm A}_{k-1}\cong
SU(k)$.  To prove this, we represent each cycle $D_\sigma$ by an
oriented path $\ello_{\sigma-1,\sigma}$, and we count intersections
of cycles, which come from intersections of paths.  For example,
$D_{\sigma}$ has one point of intersection with $D_{\sigma\pm 1}$,
coming from the endpoints of the paths, and this contributes $1$ to
the intersection number. On the other hand, for $|\sigma-\tau| \geq
2$, $D_{\sigma}$ is disjoint from $D_{\tau}$ (if the paths are
suitably chosen), so the intersection number vanishes.  Finally the
self-intersection number of each $D_{\sigma}$ is $-2$; for this, we
observe that $\ello_{\sigma\sigma'}$ can be deformed to a second path
$\tilde\ello_{\sigma\sigma'}$ between the same two points;
generically the two paths intersect precisely at the two endpoints,
and allowing for orientations, each contributes $-1$ to the
intersection number.  Putting all this together, the matrix $\eusm
I$  of intersections of the cycles $D_\sigma$  is
\begin{equation}\label{cartan}{\eusm I}_{\sigma\tau}=\begin{cases} -2 & {\mathrm{if}}~ \sigma=\tau\\
                                                                                                     1 & {\mathrm {if}} ~\sigma=\tau\pm 1 \\
                                                                                                      0 & {\mathrm{if}}~|\sigma-\tau|\geq 2.\end{cases}\end{equation}
$\eusm I$ is the negative of the Cartan matrix of the group ${\mathrm A}_{k-1}$.                                                                                                        If we include one more cycle
$D_0=C_{k-1,0}$ (which in homology is minus the sum of the others), we
get the negative of the extended Cartan matrix.  (The cycles can be naturally
arranged as the nodes of a Dynkin diagram, as in  fig. \ref{Dynkin};
we interpret $D_0$ as the extended node.) This is related to
the fact that $\TN_\k$ generates an ${\mathrm A}_{k-1}$ singularity
when all points $\vec x_\sigma$ coincide; this leads in $M$-theory to ${\mathrm A}_{k-1}=SU(k)$
gauge symmetry.

The results described in the previous paragraph reflect the fact that $H_2(\TN_\k,\ZZ)$ is
isomorphic to the weight lattice of ${\mathrm A}_{k-1}$, and in particular is isomorphic to $\ZZ^{k-1}$.
  The first
Chern class of an instanton bundle over $\TN_\k$ takes values in
$H^2(\TN_\k,\ZZ)$, so for our later study of instantons, it will
help to have an explicit description of that group.   We simply use
the fact that $H^2(\TN_ \k,\ZZ)$ is dual to $H_2(\TN_\k,\ZZ)$ (and
in particular is also isomorphic to $\ZZ^{k-1}$). The duality means
that there is a natural pairing in which an element $b\in
H^2(\TN_\k,\ZZ)$ assigns an integer $b_{\sigma\sigma'}$ to each
compact cycle $C_{\sigma\sigma'}$, and we must have
$b_{\sigma\sigma'}+b_{\sigma'\sigma''}+b_{\sigma''\sigma}=0$, since
the sum $C_{\sigma\sigma'}+C_{\sigma'\sigma''}+C_{\sigma''\sigma}$
vanishes in homology.  So $b_{\sigma\sigma'}=
b_\sigma'-b_{\sigma'}$, for some integers $b_\sigma$, which are
uniquely determined up to $b_\sigma\to b_\sigma+b$ for some integer
$b$.  (One choice of the $b_\sigma$ is  $b_1=0$,
$b_{\sigma}=-b_{1,\sigma}$ for $\sigma>1$.) Thus $H^2(\TN_\k,\ZZ)$
is spanned by integer-valued sequences
$\{b_\sigma|\sigma=1,\dots,k\}$, modulo $b_\sigma\to b_\sigma+b$.

Since $\TN_\k$ is not compact, we should distinguish the second cohomology group from the second cohomology with
compact support, which is its dual.  We denote the second cohomology with compact support
as $H^2_{\mathrm {cpct}}(\TN_\k,\ZZ)$; this is naturally isomorphic  to $H_2(\TN_\k,\ZZ)$, by the map
which takes a homology cycle (which is compact by definition) to its Poincar\'e dual.  They are
both dual to $H^2(\TN_\k,\ZZ)$.
  A cohomology class $f$ with compact support is represented
by an integer-valued sequence $\{f_\sigma|\sigma=1,\dots,k\}$ with $\sum_\sigma f_\sigma=0$.
This is the dual of the cohomology, as there
is a natural pairing between a cohomology class $b=\{b_\sigma\}$ and a cohomology class with
compact support $f=\{f_\sigma\}$:
\begin{equation}\label{zumor}(b,f)=\sum_\sigma b_\sigma f_\sigma.\end{equation}
In this way of representing the dual of $H^2(\TN_\k,\ZZ)$, the homology cycle $C_{\rho\rho'}$ is related to the sequence $f_\sigma=\delta_{\sigma\rho}
-\delta_{\sigma\rho'}$, which indeed obeys $\sum_\sigma f_\sigma=0$.

There is also a ``geometrical'' (rather than topological)
version of $H_2(\TN_\k,\ZZ)$, generated by the noncompact cycles $C_\sigma$, $\sigma=1,\dots,k$.
Reasoning as above, the intersection matrix of these cycles is
simply $\langle C_\sigma,C_{\sigma'}\rangle=\delta_{\sigma\sigma'}$, corresponding to the weight
lattice of the group $U(k)$.
(To make this rigorous, one must count intersections for a prescribed behavior of the cycles at infinity,
i.e. a prescribed asymptotic behavior of the paths $\ello_\sigma$.)
This is related to the fact that $k$ NS5-branes $T$-dual to
$\TN_\k$ generate $U(k)$  gauge symmetry (rather than $SU(k)$).  The corresponding geometrical
version of $H^2(\TN_\k,\ZZ)$ is $\ZZ^k$, labeled by sequences $\{b_\sigma|\sigma=1,\dots,k\}$ with
no equivalences.

\subsection{Review Of Hyper-Kahler Quotient Construction}\label{review}

To tie up various loose ends, we will now  briefly review  the
hyper-Kahler quotient construction \cite{Gibbons:1996nt} of the multi-Taub-NUT spaces $\TN_\k$,
which we have recovered as the output of our analysis.
We begin with the basic case $k=1$. (We largely follow the notation of   \cite{Gibbons:1996nt} with
some modifications to be consistent
with the present paper as well as \cite{Gaiotto:2008sa}.)

\def\rr{{\bf r}}
Let $H$ be a hypermultiplet parametrizing a flat hyper-Kahler manifold $\R^4$,
and pick a triholomorphic $U(1)$ symmetry that acts linearly on $H$.  (This $U(1)$ is simply
a subgroup of the $SU(2)$ rotation group that acts linearly on
$\R^4$ preserving its hyper-Kahler structure.)
Writing $\vec\mu_H=\vec\rr$ for the moment map,
and $r=|\vec\rr|$ for its magnitude, the metric of $\R^4$ can be put in the form
\begin{equation}\label{lovely} \d s^2=\frac{1}{r}\d\vec\rr^2+r(\d\psi+\vec\omega\cdot \d
\vec\rr)^2.\end{equation}  This is a special case of (\ref{genf}) with $U=1/r$.
The triholomorphic  $U(1)$ acts by shifts of $\psi$.

Now we consider the flat hyper-Kahler manifold $\R^3\times S^1$, with a triholomorphic
$U(1)$ symmetry that acts by rotation of the second factor.  We parametrize $\R^3$ by a triple
of coordinates $\vec X$, and $S^1$ by an angular variable\footnote{\label{hut} We take $\theta$ to have period $4\pi$, like
$\psi$ (recall footnote \ref{put}), to minimize factors of 2.}  $\theta$, so the metric is
\begin{equation}\label{trusty} \d s^2=\d\X^2+\lambda^2\d\theta^2,\end{equation}
with a constant $\lambda$ that controls the radius of the circle.  With a natural orientation
of the hyper-Kahler structure,
the $U(1)$  moment map is  $\lambda\X$.

The combined metric on $\R^4\times \R^3\times S^1$ is
\begin{equation}\label{lop}\d s^2=\frac{1}{r}\d\vec\rr^2+r(\d\psi+\vec\omega\cdot \d
\vec\rr)^2+\d\X^2+\lambda^2\d\theta^2.\end{equation}
The $U(1)$ action is $(\psi,\theta)\to (\psi+t,\theta+t)$, so $\chi=\psi-\theta$ is invariant.
The moment map of the combined system is $\vec\mu=\vec\rr+\lambda \X$.

To construct the hyper-Kahler quotient $W=(\R^4\times\R^3\times S^1)/\neg/\neg/U(1)$, we must
restrict to $\vec\mu^{-1}(0)$ and then divide by $U(1)$.  Restricting to $\vec\mu^{-1}(0)$ is
accomplished by simply setting $\X=-\vec\rr/\lambda$.  We write the resulting metric on $\vec\mu^{-1}(0)$ in terms of $\chi$ and $\theta$ (rather than $\psi$ and $\theta$):
\begin{equation}\label{onox}\d s^2=\left(\frac{1}{r}+\frac{1}{\lambda^2}\right)\d\vec\rr^2
+r(\d\chi+d\theta+\vec\omega\cdot \d
\vec\rr)^2+\lambda^2\d\theta^2.
\end{equation}
Completing the square for the terms involving $\d\theta$, the metric on $\vec\mu^{-1}(0)$ is equivalently
\begin{equation}\label{nox}\d s^2=\left(\frac{1}{r}+\frac{1}{\lambda^2}\right)\d\vec\rr^2
+\left(\frac{1}{r}+\frac{1}{\lambda^2}\right)^{-1}\left(\d\chi+\vec\omega\cdot \d\vec\rr\right)^2
+(r+\lambda^2)\left(\d\theta+\frac{r}{r+\lambda^2}(\d\chi+\vec\omega\cdot \d\vec\rr)\right)^2.
\end{equation}

In these coordinates, the triholomorphic $U(1)$ symmetry acts by shifts of $\theta$, so to
construct $W=\vec\mu^{-1}(0)/U(1)$ as a space, we simply drop $\theta$ from the description.
However, to construct the hyper-Kahler metric on $W$, we are supposed to identify the tangent
space to $W$ with the subspace of the tangent space to $\vec\mu^{-1}(0)$ that is orthogonal
to the Killing vector field $\partial/\partial\theta$.  We do this simply by setting
$\d\theta+\frac{r}{r+\lambda^2}(\d\chi+\vec\omega\cdot \d\vec\rr)=0$.  The resulting metric on $W$ is
\begin{equation}\label{znox}\d s^2=\left(\frac{1}{r}+\frac{1}{\lambda^2}\right)\d\vec\rr^2
+\left(\frac{1}{r}+\frac{1}{\lambda^2}\right)^{-1}\left(\d\chi+\vec\omega\cdot \d\vec\rr\right)^2,\end{equation}
and now we recognize $W$ as the Taub-NUT manifold $\TN$.

As a bonus, we see that $\vec\mu^{-1}(0)$ is a circle bundle over $\TN$.  Going back to (\ref{nox}), the part of the
metric of $\vec\mu^{-1}(0)$ that involves the fiber coordinate $\theta$ has the Kaluza-Klein
form $A(\d\theta+\Lambda)^2$, where $A=r+\lambda^2$ is a function on the base space $\TN$, and
$\Lambda=\frac{r}{r+\lambda^2}(\d\chi+\vec\omega\cdot \d\vec\rr)$ is locally a one-form on the
base space.  Globally, $\Lambda$ is best understood as a connection on the $U(1)$ bundle
$\vec\mu^{-1}(0)\to\TN$.  On general grounds, its curvature form $B=\d\Lambda$ is of type $(1,1)$
in each of the complex structures of the hyper-Kahler manifold $\TN$, and hence (as $\TN$ is
four-dimensional) it is anti-selfdual.  Thus, we have accounted for the earlier formula (\ref{kk}).

\subsubsection{The  Multi-Centered Case}\label{multicc}

Now we want to extend this to the multi-centered case.  We begin with the flat metric on
$\left(\prod_{\sigma=1}^k \R^4_\sigma \right)\times \R^3\times S^1$, which we describe by the obvious analog of (\ref{lop}):
\begin{equation}\label{ilop}\d s^2=\sum_{\sigma=1}^k\left(\frac{1}{r_\sigma}\d\vec\rr_\sigma^2+r_\sigma(\d\psi_\sigma+\vec\omega_\sigma\cdot \d
\vec\rr_\sigma)^2\right)+\d\X^2+\lambda^2\d\theta^2.\end{equation}
Here $\vec\rr_\sigma$ is the moment map for the action of $U(1)_\sigma$ on $\R^4_\sigma$, and
$r_\sigma=|\vec\rr_\sigma|$.
The moment map for the $U(1)_\sigma$ action on $\R^3\times S^1$ is $\lambda\vec X$, and
we add constants $\vec x_\sigma$ so that the moment  map of the combined system is $\vec\rr_\sigma+\lambda\vec X
+\vec x_\sigma$.  We set the moment map to zero by setting $-\lambda\vec X=\vec\rr_\sigma+\vec x_\sigma$, and we write $\vec\rr$ for $-\lambda\vec X$.    The metric on $\vec\mu^{-1}(0)$ comes out to be
\begin{equation}\label{thatmet} \d s^2= \left(\sum_\sigma\frac{1}{|\vec\rr-\vec x_\sigma|}+\frac{1}{\lambda^2}
\right) \d\vec\rr^2+\sum_\sigma |\vec\rr-\vec x_\sigma|(\d\psi_\sigma+\vec\omega_\sigma\cdot \d
\vec\rr)^2+\lambda^2\d\theta^2.\end{equation}  (Here $\vec\omega_\sigma$ is evaluated at
 $\vec \rr_\sigma=\vec\rr-\vec x_\sigma$.)

To avoid complicated algebra, we organize the remaining steps as follows.
The tangent space to $\TN_\k$ is the subspace of the tangent space to
$\vec\mu^{-1}(0)$ that is orthogonal to the orbits of the group $F$.  The condition for
orthogonality gives
\begin{equation}\label{olp}|\vec\rr-\vec x_\sigma|(\d \psi_\sigma+\omega_\sigma\cdot \d\rr)
+\lambda^2\,\d\theta=0.\end{equation}
By these conditions, one can eliminate $\d\psi_\sigma$ and $\d\theta$ in favor of $\d\chi$,
where $\chi$ is the invariant $\chi=\sum_\sigma\psi_\sigma-\theta$.  This leads to the metric
\begin{equation}\label{lp} \d s^2=\left(\sum_\sigma\frac{1}{|\vec\rr-\vec x_\sigma|}+\frac{1}{\lambda^2}
\right) \d\vec\rr^2+\left(\sum_\sigma\frac{1}{|\vec\rr-\vec x_\sigma|}+\frac{1}{\lambda^2}
\right)^{-1}\left(\d\chi+\sum_\sigma\vec\omega_\sigma\cdot\d\vec\rr\right)^2,\end{equation}
which describes the expected multi-Taub-NUT space $\TN_\k$.

\def\L{\mathcal L}
\subsubsection{Line Bundles}\label{lb}

The space
$\vec\mu^{-1}(0)$, which is described explicitly in (\ref{thatmet}), is a fiber bundle over $\TN_\k=\vec\mu^{-1}(0)/F$ with fiber $F\cong U(1)^k$.
This fibration, moreover, comes with a natural connection, coming from the Riemannian
connection on $\vec\mu^{-1}(0)$.
So the construction automatically gives us a $U(1)^k$ gauge field over $\TN_k$, or equivalently
$k$ independent $U(1)$ gauge fields $\TN_k$; equivalently,
we get $k$ complex line bundles
 $\L_\sigma\to \TN_k$, $\sigma=1,\dots,k$.  Each of these has a curvature that is
of type $(1,1)$ in each complex structure, and is therefore anti-selfdual.  This gives us the $k$
$B$-field modes that, as indicated at the end of section \ref{mulcit},  describe the duals to the angular
positions of the $k$ NS5-branes in the original description via branes on $\R^3\times S^1$.

\def\D{{\mathrm D}}
To describe these line bundles explicitly, we pick a particular value of $\sigma$, say $\sigma=\rho$,
and we impose the equations (\ref{olp}) for all other values of $\sigma$.  This gives enough
to eliminate the $\d\psi_\sigma$, $\sigma=1,\dots,k$, in favor of the invariants $\d\chi$ and $\d\psi_\rho$.
We then get a metric on a five-manifold $M_\rho$ that is a circle bundle over $\TN_k$, with
$\psi_\rho$ as the fiber coordinate.  The metric on $M_\rho$ turns out to be the sum of
(\ref{lp}) plus
\begin{equation}\label{songo}A\left(\D\psi_\rho-\frac{\D\chi}{|\rr-\vec x_\rho|\left(\lambda^{-2}+\sum_\sigma 1/|\rr-\vec x_\sigma|\right)}\right)^2
\end{equation}
with
\begin{align}\label{elk}A=&|\rr-\vec x_\rho| \frac{\left(1+\lambda^2\sum_\sigma 1/|\rr-\vec x_\sigma|\right)}{\left(1+\lambda^2
\sum_{\sigma\not=\rho} 1/|\rr-\vec x_\sigma|\right)} \\ \notag \D\psi_\rho=&\d\psi_\rho+\vec\omega_\rho
\cdot \d\vec\rr, ~~ \D\chi=\d\chi+\sum_\sigma \vec\omega_\sigma\cdot \d\vec\rr.\end{align}
We can also introduce conventional angular variables $\tilde\psi_\rho=\psi_\rho/2$, $\tilde \chi=\chi/2$, with
periods $2\pi$ (recall footnotes \ref{put} and \ref{hut}), and rewrite (\ref{songo}) as
\begin{equation}\label{songox}4 A\left(\D\tilde \psi_\rho-\frac{\D\tilde \chi}{|\rr-\vec x_\rho|\left(\lambda^{-2}+\sum_\sigma 1/|\rr-\vec x_\sigma|\right)}\right)^2
\end{equation}
with $\D\tilde\psi_\rho=\D\psi_\rho/2$, $\D\tilde\chi=\D\chi/2$.
The $U(1)$ connection associated with the circle bundle $M_\rho\to \TN_\k$ can thus
be described by the connection
\begin{align}\label{pij}\Lambda_\rho &= f_\rho \,\D\tilde \chi+\frac{1}{2}\vec\omega_\rho\cdot \d\rr\end{align}
where
\begin{align}\label{hope} f_\rho&=-\frac{1}{|\rr-\vec x_\rho|\left(\lambda^{-2}+\sum_\sigma 1/
|\rr-\vec x_\sigma|\right)}.\end{align}

The curvature $B_\rho=\d\Lambda_\rho$ of this connection is of type $(1,1)$ for each complex structure on $\TN_\k$ and
hence is anti-selfdual.   From the fact that the function $f_\rho$ vanishes for $\vec\rr$ approaching
$\infty$ or $\vec x_\sigma$, $\sigma\not=\rho$ and equals $-1$ for  $\rr=\vec x_\rho$, it follows that
if $C_\sigma$ are the two-cycles introduced at the end of section \ref{mulcit}, then
\begin{equation}\label{lofty}\int_{C_\sigma}\frac{B_\rho}{2\pi}=\delta_{\rho\sigma}.\end{equation}
(In evaluating the integral, a factor of $2\pi$ comes from
integrating over $\tilde\chi$.)   Thus, comparing to the description of
$H^2(\TN_\k,\ZZ)$ given in section \ref{topology}, $c_1(\L_\rho)$ is
associated with the sequence
\begin{equation}\label{orno}b^{(\rho)}_\sigma=\delta_{\rho\sigma},~~
\sigma=1,\dots,k.\end{equation}

For $k>1$, the line bundles $\L_\rho$ are topologically non-trivial, as their first Chern classes
are nontrivial.
 However, the tensor
product $\L_*=\L_1\otimes \L_2\otimes \cdots\otimes \L_k$ is
topologically trivial; by virtue of (\ref{lofty}), its curvature
$B=\sum_\sigma B_\sigma$ has a vanishing integral over each compact
cycle $C_{\sigma \sigma'}$.  The line bundle $\L_*$ can be represented
by the connection form $\sum_\sigma\Lambda_\sigma$.  A short
computation shows that after adding the exact form $\d\tilde\chi$, we can
take the connection form of $\L_*  $ to be
\begin{equation}\label{elfk}\Lambda=\frac{\D\tilde\chi}{1+\lambda^2\sum_\sigma\frac{1}{|\rr-\vec x_\sigma|}}.
\end{equation}
$\Lambda$ is actually a globally defined one-form (the singularities at $\rr=\vec x_\sigma$ are only
apparent), so we can multiply $\Lambda$ by an arbitrary real number $t$ to get
another one-form $\Lambda_t=t\Lambda$.  We denote as $\L_*^t$ a trivial complex line bundle with
connection form $\Lambda_t$.
The circle $\tilde S^1$ at infinity in the Taub-NUT space
$\TN_\k$ is parametrized by $\tilde\chi$; the integral of $\Lambda_t$ over this circle is $2\pi t$.
So the holonomy of $\L_*^t$ over the circle at infinity is $\exp(2\pi i t)$.

On the other hand,
the connection forms $\Lambda_\rho$ vanish at infinity, so the line bundles $\L_\sigma$ have
trivial holonomy at infinity.

A general unitary line bundle  ${\mathcal T}\to\TN_\k$ with
anti-selfdual curvature is of the form
\begin{equation}\label{dono}{\mathcal
T}=\L_*^t\otimes\left(\otimes_{\sigma=1}^k\L_\sigma^{n_\sigma}\right),\end{equation}
where $t$ is real and the $n_\sigma$ are integers. The holonomy of
this line bundle over the circle at infinity is $\exp(2\pi i t)$.
Its first Chern class is associated with the sequence $(n_1,n_2,
\dots,n_k)$.  The
representation (\ref{dono}) is not unique, as one may add 1 to $t$
and subtract 1 from each $n_\sigma$.

For future reference, let us summarize the construction of the $\L_\rho$.  First we divide $\W$
by the group $\G_\star$ of continuous gauge transformations.  Then,  imposing (\ref{olp}) for $\sigma\not=\rho$, we divide by $\prod_{\sigma\not=\rho}U(1)_\sigma$.  The net effect is to divide $\W$ by
the subgroup of $\G$ consisting of gauge transformations that are continuous at $y_\rho$.  This
subgroup, which we will call $\G^\rho$, is of codimension 1 in $\G$ and fits in an exact sequence
\begin{equation}\label{remember}1\to \G^\rho\to \G\to U(1)\to 1.\end{equation}
The map $\G\to U(1)$ maps a gauge transformation $g(y)$ to the discontinuity $u_\rho$ at $y=y_\rho$.
By dividing $\W$ by the codimension 1 subgroup $\G^\rho$, we have obtained a $U(1)$ bundle
 $\W/\G^\rho\to \TN_\k$.  $\L_\rho$ is the associated complex line bundle  $(\W/\G^\rho\times \C)/U(1)$,
 where $U(1)$ acts in the natural way on $\C$.  A completely equivalent way
to define $\L_\rho$ is to divide by the action of $\G$ not on $\W$ but on $\W\times \C$, with the
usual $\G$ action on $\W$, and the action of $\G$ on $\C$ chosen so that a gauge transformation
$g(y)$ acts by multiplication by
\begin{equation}\label{gormy}u_\rho=g^-(y_\rho)^{-1}g^+(y_\rho),\end{equation}
 ensuring that $\G^\rho\subset \G$ acts trivially.  Since $\G^\rho$ acts trivially on $\C$, we have
 $(\W\times\C)/\G=\left(\W/\G^\rho\times\C\right)/U(1)$.

\subsubsection{ALE Limit}\label{el}

The space $\TN_\k$ looks near infinity like a circle bundle over $\R^3$.  The fiber $\tilde S^1$
 has radius $\lambda$,
according to the metric (\ref{lp}).  For $\lambda\to\infty$, the ALF space $\TN_\k$ becomes
an ALE space that is asymptotic at infinity to $\R^4/{\Bbb Z}_k$.  It is a hyper-Kahler resolution of the
${\mathrm A}_{k-1}$ singularity, as first described in \cite{Kronheimer}.  The metric on the ALE
space is obtained by just dropping the $1/\lambda^2$ term in (\ref{gomely}).

The ALE space is simply-connected (like its ALF precursor), but at infinity it has a fundamental
group ${\Bbb Z}_k$.  Consequently, a line bundle that is flat at infinity can have a global monodromy
at infinity, which must be a $k^{th}$ root of 1. The monodromy of the line bundles $\L_\rho$ is easily determined.
After setting $\lambda=\infty$,
 the connection form $\Lambda_\rho$ of eqn. (\ref{pij}) is asymptotic for
$\rr\to\infty$ to $-k^{-1}\D \tilde\chi$. The holonomy of the line bundle
$\L_\rho$ is therefore $\exp(-2\pi i/k)$, independent of $\rho$.

The tensor product $\L_*=\otimes_{\sigma=1}^k\L_\sigma$ therefore has trivial monodromy at
infinity.  This is also clear from eqn. (\ref{elfk}); the connection form $\Lambda$ vanishes uniformly
for $\lambda\to\infty$.

The fiber $\tilde S^1$ in the asymptotic fibration $\TN_\k\to\R^3$  is dual to the circle $S^1$ in the
original description via NS5-branes on $\R^3\times S^1$.  So the ALE limit that we have just
analyzed arises when the radius of the original circle goes to zero.

In the limit $\lambda\to\infty$, the construction of $\TN_\k$ as a finite-dimensional hyper-Kahler quotient
simplifies slightly.  From the starting point $(\R^4)^k\times (\R^3\times S^1)$, one can omit the factor
of $\R^3\times S^1$, while also restricting the gauge group from $F=U(1)^k$ to its subgroup $F'\cong U(1)^{k-1}$
that acts trivially on $\R^3\times S^1$.  This truncation is valid for $\lambda\to\infty$ because in this limit, the hyper-Kahler
quotient by the ``extra'' $U(1)$ (in a decomposition $F=F'\times U(1)$) serves just to eliminate the factor $\R^3\times S^1$.
The truncated theory is simply the familiar \cite{Kronheimer} construction of the ALE space as a finite-dimensional
hyper-Kahler quotient.

\subsubsection{Other Examples}\label{othe}

In \cite{Gibbons:1996nt}, several other examples are given of complete hyper-Kahler metrics
that can be constructed as hyper-Kahler quotients of flat spaces.  We will here briefly mention
how these may be treated along lines similar to the foregoing.

One example is the Lee-Weinberg-Yee metric \cite{LWY,M}.  It is a
moduli space of solutions of the Bogomolny equations on $\R^3$, and
therefore \cite{Dia} it is the moduli space of supersymmetric vacua
of a D3-D5 system.  Because only one D3-brane is required, the gauge
group in Nahm's equations is $U(1)$ and hence a treatment analogous
to the above is  possible.  To get the LWY metric, we place
D5-branes at points $y_0<y_1<\dots < y_{m+1}$, with a single
D3-brane supported on the interval $I=[y_0,y_{m+1}]$.  This leads to
Nahm's equations for $U(1)$ fields $\vec X,A$  interacting with
charge 1 hypermultiplets $H_\lambda$ that are supported at the
points $y_\lambda$, $\lambda=1,\dots,m$.  (The hypermultiplets make
delta function contributions in Nahm's equations, as in eqn.
(\ref{gommo}).) The resulting moduli space $W$ is locally the
product of the LWY metric with a copy of $\R^3\times S^1$ that
describes the center of mass  motion of the monopoles.   To
construct $W$ as a hyper-Kahler quotient, let $\W$ be the space of
tuples $(\vec X,A,H_\lambda)$, and $\G$ the group of maps $g:I\to
U(1)$ such that $g(y_0)=g(y_{m+1})=1$. Then
$W=\W/\negthinspace/\negthinspace/ \G$.  To get instead a
construction of $W$ by a finite-dimensional hyper-Kahler quotient,
we let $\G_\star$ be the subgroup of $\G$ characterized by the
condition $g(y_\lambda)=1$, $\lambda=1,\dots,m$, so $\G_\star$ is a
normal subgroup fitting in an exact sequence
\begin{equation}\label{kelp}1\to \G_\star\to \G\to U(1)^m\to 1.\end{equation}
Then setting $F=U(1)^m$ and $W_\star = \W/\negthinspace/\negthinspace/\G$, $W$ can be constructed
as a finite-dimensional hyper-Kahler quotient $W=W_\star/\negthinspace/\negthinspace/F$.  Explicitly, one finds that
$W_\star = (\R^4)^m\times
(\R^3\times S^1)^{m+1}$, and apart from an extra factor of $\R^3\times S^1$ that is locally decoupled,
this gives the construction in \cite{Gibbons:1996nt} of the LWY metric as a hyper-Kahler quotient.

The Taubian-Calabi metric, also treated in \cite{Gibbons:1996nt}, arises in the
limit $y_1=\dots=y_m$.  In this limit,
all D5-branes are coincident, so there is a global $U(m)$ symmetry.  Also, as
all D5-branes are at the same value of $y$,
one can take $\G_\star$ to be of codimension one in $\G$, leading to a construction
of the Taubian-Calabi metric
as a hyper-Kahler quotient of a flat space by a single $U(1)$.

As in these examples, a major simplification occurs whenever the
gauge group in Nahm's equations is abelian.  In
\cite{Cherkis:2008ip}, this fact is used to explicitly analyze the
one-instanton solution in $SU(2)$ gauge theory on $\TN_\k$.

\section{Instantons On A Taub-NUT Space}\label{insttn}
\subsection{Overview}

The construction that we have analyzed in section \ref{probe} is really a special case of a more
general brane construction \cite{Cherkis:2008ip} of instantons on $\TN_\k$.    In effect, we have been
studying the special case of instantons of zero rank and second Chern class equal
to 1.  An instanton with this property is a point instanton, represented by an E0-brane supported
at a point in $\TN_\k$.  The moduli space of such E0-branes, which is isomorphic to $\TN_\k$ itself,
is precisely what we have been studying, in a $T$-dual formulation.

To describe more general instantons on $\TN_\k$, one simply incorporates additional branes
in a supersymmetric fashion.  In the Type IIB language in terms of branes on
\begin{equation}\label{kelkoo} M=\R^3\times S^1\times \R^3_\X\times \R^3_\Y,\end{equation}
so far we have considered NS5-branes wrapped on $\R^3\times \R^3_\Y$
(and localized in $S^1\times\R^3_\X$), along with a probe D3-brane
wrapped on $\R^3\times S^1$.  Without breaking supersymmetry, it is
possible to also include D5-branes wrapped on $\R^3\times \R^3_\X$
(and localized in the other factors), and additional D3-branes
wrapped on $\R^3\times S^1$. (This is a familiar supersymmetric
brane configuration \cite{G,HW}, frequently studied in a limit in
which the $S^1$ is decompactified and replaced by $\R$.) This also
has one very important further refinement: instead of $\R^3\times
S^1$, the support of a D3-brane might be $\R^3\times I$, where $I$
is an interval in $S^1$, bounded by two fivebranes.   We will call a
general configuration of this kind an NS5-D5-D3 configuration.

Regardless of the details of such a configuration, what happens to it when we perform $T$-duality
along $S^1$?  The $T$-dual geometry depends on the positions of the NS5-branes but not of the
D-branes; the reason for this is that $T$-duality can be performed in the limit of weak string coupling,
where the effects of any finite set of D-branes are infinitesimal.  So the $T$-dual geometry is
simply the one that we have explored in section \ref{probe}: the $T$-dual of $Z_\k$, that is
of $S^1\times \R^3_\X$ with $k$ embedded NS5-branes, is the multi-Taub-NUT space $\TN_\k$.

$T$-duality maps the D5-branes and D3-branes to other D-branes.  In
particular, the $T$-dual of a D5-brane is a D6-brane wrapped on
$\R^3\times \TN_\k$.  If there are $p$ D5-branes, we get $p$
D6-branes supporting a $U(p)$ gauge symmetry.

In section \ref{probe}, the $T$-dual of a D3-brane on $\R^3\times
S^1$ was a D2-brane wrapped on $\R^3$ and supported at a point $q\in
\TN_\k$.  However, once we introduce D5-branes, so that the $T$-dual
description has D6-branes that fill $\R^3\times \TN_\k$, a D2-brane
on $\R^3\times q$ can dissolve into an instanton on $\TN_\k$.

Superficially, it seems that the instanton number may be simply the
number of D3-branes in the original description via Type IIB theory
on $M$.  This is actually correct if one considers only ``complete''
D3-branes that wrap all the way around the $S^1$, rather than
``fractional'' D3-branes that are suspended between two fivebranes.
The topological description becomes more complicated when there are
fractional D3-branes, since the instanton bundle may have a first
Chern class.   The topology allows this, as explained in section
\ref{topology}.  One of the main reasons for re-examining in this
section the brane construction of instantons on  $\TN_\k$ is  to
compute the first Chern class of the instanton bundle.

We begin in section \ref{lineb} with the simplest case that there
are D5-branes (as well as NS5-branes) but no D3-branes.  Roughly
speaking, each D5-brane has for its $T$-dual a D6-brane that
supports a Chan-Paton line bundle.  (A somewhat more precise
statement depends on the $B$-field, as described in section
\ref{lineb}.) The line bundle depends on the position of the
D5-brane.

The Chan-Paton line bundle supported on the $T$-dual to any given D5-brane is flat at infinity
on $\TN_\k$, and has anti-selfdual curvature everywhere by supersymmetry.  Hence, this line
bundle is one of the line bundles described in section \ref{lb}.
In section \ref{lineb}, we determine precisely which line bundle over $\TN_\k$ is associated
with any given D5-brane.  In particular, we determine both the first Chern class and the asymptotic
monodromy of this line bundle.   This then automatically determines the Chern classes and
asymptotic monodromy of the vector bundle -- a direct sum of line bundles -- associated to
any collection of D5-branes, as long as there are no D3-branes.

Incorporating D3- branes does not change the monodromy at infinity,
since the D3-branes are localized on $\R^3_\X$.  However, the
inclusion of D3-branes does change the first Chern class. Analyzing
this is the goal of section \ref{moveb}.  This analysis is based on the ability
to reduce any configuration to a more convenient one, without changing the topology,
by moving fivebranes along $S^1$ in a judicious fashion.

Finally, we conclude in section \ref{all} by considering the ALE
limit in which the radius of the circle in (\ref{kelkoo}) goes to
zero, and therefore the radius of the dual circle in the asymptotic
Taub-NUT fibration $\TN_\k\to\R^3$ goes to infinity.   In this
limit, some of our results have a natural $M$-theory interpretation.

\subsubsection{D3-Brane Probe}\label{morep}

One obvious question is whether the description of instantons on $\TN_\k$ via branes
on the dual geometry $M$ of eqn. (\ref{kelkoo}) is useful.

There are actually two questions here.  (1) Can one use this approach
to explicitly describe the moduli space $\M$  of instantons on $\TN_\k$?  (2) Can one use it
 to describe the instanton bundles themselves?

The second question may seem more basic, but it turns out that it is
better to start with the first. The answer to that question is that
$\M$ is the moduli space of supersymmetric vacuum states of the
relevant NS5-D5-D3 brane configuration on $M$.   Such vacua
correspond to solutions of Nahm's equations along $S^1$.  One must
solve Nahm's equations for the group $U(n)$ where $n$ is the number
of D3-branes.  (If some D3-branes end on fivebranes, $n$ may jump as
one moves around the circle.)  Fivebranes correspond to defects,
discontinuities,  poles, and jumps in rank in the solution of Nahm's
equations. These facts are used in \cite{Cherkis:2008ip}; the
relevant facts about Nahm's equations are also reviewed in detail in
\cite{Gaiotto:2008sa}.

In general, one cannot explicitly solve Nahm's equations, but the
description of $\M$ via Nahm's equations is very powerful
nonetheless.  For example, one can effectively describe the space of
solutions of Nahm's equations as a complex manifold in any of its
complex structures (though it is difficult to find the hyper-Kahler
metric).  This fact is exploited in \cite{Cherkis:2008ip} to show
that, as  a complex manifold in any of its complex structures, the
moduli space of instantons on $\TN_\k$ is independent of the radius
of the circle at infinity (and of the monodromy at infinity) and
coincides with the moduli space of instantons on a corresponding ALE
space. For example, as a complex manifold, the moduli space of
instantons on the basic Taub-NUT manifold $\TN$ coincides with the
moduli space of instantons on $\R^4$.  The reasoning is briefly
sketched in Appendix A.

Now let us consider\footnote{For a somewhat similar discussion in a more elaborate
context, see \cite{MRD}.}  question (2).  Let $V\to \TN_\k$ be an instanton
bundle.  For $q$ a point in $\TN_\k$, we would like to find a
natural way to use branes to extract $V_q$, the fiber of $V$ at $q$.
Moreover, the construction should have the property that as $q$
varies, we can extract the connection on $V$.

\def\B{{\mathcal B}}
To do this, we let $\B$ be the brane on $\TN_\k$ associated to the
instanton bundle $V$.  Let $\B'$ be an E0-brane supported at the
point $q\in \TN_\k$.  Actually, in the string theory description,
the supports of $\B$ and $\B'$ are $\R^3\times \TN_\k$ and
$\R^3\times q$, respectively. Then the space of $(\B',\B)$ string
ground states is\footnote{More precisely, this space is the zero
momentum part of a  hypermultiplet that is the tensor product with
$E_q$ of a standard hypermultiplet. In topological string theory on
$\TN_\k$, one would get just $E_q$.} $E_q$, and the instanton
connection on $E$ can be extracted from the natural connection
on the space of $(\B',\B)$ string ground states, as $q$ varies.

To compute the space of $(\B',\B)$ strings, we use a $T$-dual
description via branes on $M$. $\B$ corresponds to a supersymmetric
vacuum of an NS5-D5-D3 system.  This supersymmetric vacuum
corresponds to a solution of Nahm's equations; we schematically
denote this solution as $\diamondsuit$.  Its rank depends on the
chosen configuration. $\B'$ corresponds to a probe D3-brane, wrapped
on $\R^3\times S^1$, of the type that was extensively discussed in
section \ref{probe}. As in sections \ref{concrete} and \ref{mulcit},
the point $q\in \TN_\k$ corresponds to a supersymmetric state of the
probe D3-brane which can be described by a rank 1 solution of Nahm's
equations. We schematically denote this solution as $\star$.

The direct (or disjoint) sum of branes $\B'\oplus \B$ can be represented then by a reducible solution
of Nahm's equations, schematically of the form
\begin{equation}\label{pelm}\begin{pmatrix} \star & 0 \\ 0 & \diamondsuit \end{pmatrix}.\end{equation}
Now to describe the space of $(\B',\B)$ strings, we consider solutions of Nahm's equations in which
the diagonal blocks are kept fixed but the lower left block is allowed to vary.  Thus, the space of
$(\B',\B)$ strings is the space of solutions of Nahm's equations of the form
\begin{equation}\label{pelmo}\begin{pmatrix} \star & 0 \\ \Delta & \diamondsuit \end{pmatrix},\end{equation}
where only the block $\Delta$ is allowed to vary.  When we let the
block $\star$ vary (so as to vary the choice of a point $q\in
\TN_\k$) keeping the block $\diamondsuit$ fixed, the space of
possible $\Delta$'s varies as the fiber of a vector bundle
$V\to\TN_\k$.  The hyper-Kahler structure of the moduli space of
solutions of Nahm's equations gives a connection on this fibration,
and this is the instanton connection on $V$.

Applied to instantons on $\R^4$, constructed from E0-E4 brane systems, the same argument shows
that the usual ADHM description of the instanton bundles is a corollary of the ADHM description of
the moduli spaces of instantons.

Another application is as follows. As shown in \cite{Cherkis:2008ip}
(and as sketched in Appendix A), the description by Nahm's equations
implies that, when viewed as a complex symplectic manifold in any one complex
structure, certain components of the moduli space of instantons on $\TN_\k$ are
independent of the radius of the circle at infinity and the
monodromy around this circle. Let $\M$ be such a component. Adding a probe D3-brane and applying
the same reasoning to solutions of Nahm's equations of the form
(\ref{pelmo}), we deduce that if $m$ is a point in   $\M$, and
$V\to\TN_\k$ is the corresponding instanton bundle, then, as a
holomorphic vector bundle in any of the complex structures on
$\TN_\k$, $V$ is independent of the radius of the circle (and the
monodromy around the circle). As explained in \cite{Cherkis:2008ip},
the hyper-Kahler metric on $\M$ does depend on the radius of the
circle; the same is certainly true for the instanton connection on
$V$, since the self-duality condition depends on the metric of
$\TN_\k$.

\subsection{Line Bundles}\label{lineb}

\def\RR{\mathcal R}
\def\M{{\mathcal M}}
In the Type IIB spacetime
\begin{equation}\label{kelkp} M=\R^3\times S^1\times \R^3_\X\times \R^3_\Y,\end{equation}
we suppose that $S^1$ has radius $R$, and we parametrize it by a
variable $y$, $0\leq y\leq 2\pi R$. We suppose that there are $k$
NS5-branes localized at points $y_\sigma\times \vec x_\sigma \in
S^1\times \R^3_\X$.  And we include a D5-brane localized at $y=s$
(times the origin in $\R^3_\Y$).  For supersymmetry, we take the
Chan-Paton line bundle of this D5-brane to be trivial. The $T$-dual
of this D5-brane will be a D6-brane supported on $\R^3\times \TN_\k$
(times the origin in $\R^3_\Y$).  Roughly speaking, this D6-brane is
endowed with a Chan-Paton line bundle, and we want to know which
one; that is, we want to find the map $s\to \RR_s$ from a D5-brane
position to a line bundle $\RR_s\to\TN_\k$.

\def\T{\mathcal T}
This formulation is oversimplified because of the  role of the
$B$-field.  A $B$-field gauge transformation $B\to B+\d\Lambda$ acts
on the Chan-Paton gauge field $A$ of a D-brane by $A\to A+\Lambda$.
Here $\Lambda$ can be regarded as an abelian gauge field.  Accordingly, a $B$-field gauge transformation acts by
tensoring the Chan-Paton bundle of any D-brane by a line bundle
${\T}$ (with connection $\Lambda$).  Thus, under a $B$-field gauge transformation, we have
$\RR_s\to \T\otimes \RR_s$, where $\T $ is independent of $s$.

Hence the ratio of any two Chan-Paton line
bundles of branes, say $\RR_s\otimes \RR_{s'}^{-1}$, is invariant
under $B$-field gauge transformation. But a $B$-field gauge
transformation can be chosen to adjust $\RR_s$ for any one chosen
value of $s$ in an arbitrary fashion.

For our purposes, we will simply pick one value of $s$, say $s=0$,
and make a $B$-field gauge transformation to trivialize $\RR_s$.
Having done so, the $s$-dependence of $\RR_s$ is well-defined, and
we aim to compute it.

At infinity on $\TN_\k$, the curvature of $\RR_s$ vanishes, for any
$s$. This can be understood as follows. In the asymptotic fibration
$\TN_\k\to \R^3_\X$, the region at infinity in $\TN_\k$ lies over
the region at infinity in $\R^3_\X$.  Here one is far from any
NS5-brane (as those branes are localized in $\vec X$).
Asymptotically, the NS5-branes can be ignored.  In the absence of
any NS5-branes, we are just performing $T$-duality on the circle in
$M=\R^9\times S^1$, and this maps a D5-brane on $\R^6$ with trivial
Chan-Paton bundle to a D6-brane on $\R^6\times S^1$ with flat
Chan-Paton bundle.

We can carry this reasoning slightly farther to determine the
monodromy at infinity of $\RR_s$. This is a standard problem
\cite{DLP}, since the NS5-branes can be ignored; the monodromy on
the D6-brane is dual to the position of the D5-brane and is
$\exp(is/R)$. (The monodromy is trivial for $s=0$, since we made a
$B$-field gauge transformation to trivialize $\RR_s$ for that value
of $s$.)

In addition to being flat at infinity, $\RR_s$ has anti-selfdual
curvature everywhere by supersymmetry. It therefore must have the
form (\ref{dono})
\begin{equation}\label{incmon}\RR_s=\L_*^{s/2\pi R}\otimes\left(
\otimes_{\sigma=1}^k\L_\sigma^{n_\sigma}\right) ,\end{equation}
where we have determined the exponent of $\L_*$  from the monodromy
at infinity, and the $n_\sigma$ are integers that remain to be
determined.

Since $\RR_s$ is trivial for $s=0$, the $n_\sigma$ vanish if we set
$s=0$ in (\ref{incmon}).  At first sight, one might think that,
being integers, the $n_\sigma$ would have to vanish for all $s$.
This is fallacious, because the topological type of $\RR_s$ can jump
when $s$ is equal to the position $y_\sigma$ of one of the
NS5-branes. At that point, the D5-brane intersects an NS5-brane. If
one wants to move a D5-brane past an NS5-brane with the physics
varying smoothly, one must allow for the production of a D3-brane
that connects the two fivebranes.  (This fact played an important
role in \cite{HW}.) In the present discussion, we are considering
fivebrane configurations that do not have any D3-branes, so the
physics will jump when $s$ crosses the value of one of the
$y_\sigma$'s.

To verify that jumping must occur, we need only note that if $\RR_s$
varies continuously with $s$, then (\ref{incmon}) implies that
$\RR_s$ transforms to $\RR_s\otimes \L_*$ if $s$ increases by $2\pi
R$. But actually, the definition of $\RR_s$ makes clear that it is
invariant under $s\to s+2\pi R$.
  Recalling that
$\L_*=\otimes_{\sigma=1}^k\L_\sigma$, there is a natural guess for a
jumping behavior that will solve the problem: $\RR_s$ must jump by
\begin{equation}\label{zonk}\RR_s\to \RR_s\otimes\L_\sigma^{-1}\end{equation}
in crossing the point $s=y_\sigma$
from left to right. If so, then since $\RR_0$ is trivial, the
general form of $\RR_s$ is
\begin{equation}\label{genve}\RR_s =\L^{s/2\pi
R}\otimes\left(\otimes_{\sigma|s>y_\sigma}\L_\sigma^{-1}\right).
\end{equation}
In other words, starting at $s=0$ and increasing $s$, we start with
$\RR_s=\L^s$ for small $s$, and then include a jumping factor
$\L_\sigma^{-1}$ whenever $s$ crosses $y_\sigma$ for some $\sigma$.

\subsubsection{Determination Of The Line Bundle}\label{deter}

\def\U{{\mathcal U}}
 Now we will explain some preliminaries that will help us see
microscopically how the jumping comes about.

As usual, we introduce a probe D3-brane wrapped on $\R^3\times S^1$.
It interacts with $k$ NS5-branes that are supported at $y=y_\sigma$,
$\sigma=1,\dots,k$. Omitting for the moment the D5-brane, the
problem of the D3-brane interacting with the $k$ NS5-branes was
treated in section \ref{mulcit}. Supersymmetric states of this
system correspond to rank 1 solutions of Nahm's equations for data
$(\vec X,A)$ interacting with hypermultiplets $H_\sigma$ at the
points $y=y_\sigma$.  The fields and the gauge transformations to
which they are subject are allowed {\it a priori} to be
discontinuous at $y=y_\sigma$.  However, for rank 1, Nahm's
equations ultimately imply that $\vec X$ is continuous.

Let $W$ be the moduli space of such solutions. $W$ is defined as
$\vec\mu^{-1}(0)/\G$, where $\G$ is the group of gauge
transformations, and is a copy of $\TN_\k$, as we verified in
section \ref{multicc}. As in section \ref{mulcit}, we do not want to
divide by constant gauge transformations, so we define $\G$ to
consist of gauge transformations that equal 1 at some chosen
basepoint in $S^1$.  In what follows, it is convenient to take this
to be the point $s=0$ such that $\RR_s$ is trivial.  (Otherwise, we
get an alternative description that differs by a $B$-field gauge
transformation.)

Many codimension 1 normal subgroups of $\G$ can be constructed as
follows. We pick a point $s\in S^1$ and, if it is not one of the
special points $y_\sigma$, we define $\G_s$ to be the subgroup of
gauge transformations such that $g(s)=1$.  If $s$ is one of the
special points $y_\sigma$, we get two subgroups $\G_s^\pm$,
according to whether the limit of $g(y)$ as $y\to s$ from the right
or left is required to equal 1. The group $\G_s$ appears in an exact
sequence
\begin{equation}\label{turk}1\to \G_s\to \G\to U(1)\to
1,\end{equation} where the map $\G\to U(1)$ maps a gauge
transformation $g(y)$ to its value $g(s)$ (or its limiting value on
the left or right if $s$ is a special point).

\def\W{{\mathcal W}}
Now instead of dividing $\vec\mu^{-1}(0)$ by $\G$ to get $W$, we
can divide $\vec\mu^{-1}(0)$ by $\G_s$ to get a space $\W_s$ whose
dimension is 1 greater.  After dividing by $\G_s$, we can still
divide by the quotient group $U(1)$, giving $W=\W_s/U(1)$.  So
$\W_s$ is a $U(1)$ bundle over $W$.

We write $\U_s$ for the associated complex line bundle (that is,
$\U_s=(\W_s\times\C)/U(1)$, with standard action of $U(1)$ on $\C$).
An equivalent definition of $\U_s$ is the following.  The group $\G$
has an action on $\C$ defined by evaluating a gauge transformation
$g(y)$ at $s$ and then letting it act on $\C$ in the usual way. We
call this the action by evaluation at $s$.  Now start with the
product $\vec\mu^{-1}(0)\times \C$, and divide by $\G$ acting in the
usual way on $\vec\mu^{-1}(0)$, and acting on $\C$ by evaluation at
$s$.  The quotient is a complex line bundle over $W$ that is none
other than $\U_s$, since, as $\G_s$ acts trivially on $\C$, we have $(\vec\mu^{-1}(0)\times\C)/\G
=(\vec\mu^{-1}(0)/\G_s\times\C)/U(1)=(\W_s\times\C)/U(1)$.

If $s$ is one of the special points $y_\rho$, then we write $g^+(s)$
and $g^-(s)$ for the limiting values of $g(y)$ as $y\to s$ from
right or left.  Letting $g(y)$ act on $\C$ via $g^+(y_\rho)$ or
$g^-(y_\rho)$, we get two different lines bundles $\U^\pm_{\rho}$.

In section \ref{lb}, we made a variant of this construction to
define line bundles $\L_\rho\to W$ (see the last paragraph of
section \ref{lb}). Instead of $\G_s$, we used the codimension 1
subgroup $\G^\rho$. The action of $g(y)$ on $\C$ was multiplication
by $g^-(y_\rho)^{-1}g^+(y_\rho)$.

Comparing the last two paragraphs, we see that the relation between
$\L_\rho$ and $\U^\pm_\rho$ is
\begin{equation}\label{zinc}\L_\rho=\U^+_\rho\otimes
(\U^-_\rho{})^{-1}.\end{equation} Since $\U^\pm_\rho$ is just the
limit of $\U_s$ as $s$ approaches $y_\rho$ from right or left, we
conclude that $\U_s$ behaves as $\U_s\to \U_s\otimes \L_\rho$ when
$s$ crosses $y_\rho$ from left to right.   Looking back at
(\ref{zonk}), this suggests that the line bundle $\RR_s$  associated
with a D5-brane at $y=s$ is simply
\begin{equation}\label{orbox}\RR_s=\U_s^{-1}.\end{equation}

Once this statement is formulated, it is not difficult to see why it
is true.  We introduce a D5-brane located at $y=s$ and endowed with
a trivial Chan-Paton line bundle.  The fiber of $\RR_s$ at a point
$z\in W$ corresponding to a given D3-brane state is defined to be
the space of D3-D5 string ground states.   In $M=\R^3\times
S^1\times \R^3_\X\times \R^3_\Y$, the D3-brane and D5-brane
intersect at $\R^3\times \{s\}$ (times a point in $\R^3_\X$ which
depends on the D3-brane state, times the origin in $\R^3_\Y$).  A
D3-brane gauge transformation $g(y)$ acts on the space of D3-D5
string ground states by multiplication by\footnote{The exponent $-1$
results from the orientation of the string.  On the space of D5-D3
string ground states, the action is by $g(s)$.} $g(s)^{-1}$. Hence
the space of D3-D5 string ground states corresponds to the line
bundle $\U_s^{-1}$ over $W$.

Eqns.  (\ref{zinc}) and (\ref{orbox}) together imply that (\ref{genve})
gives the correct jumping behavior of $\RR_s$.  We already know that
(\ref{genve}) gives the monodromy at infinity correctly, so this is
sufficient to justify (\ref{genve}).

An interesting special case is that $\U_s$ is trivial at the
basepoint $s=0$, just like $\RR_s$.  In fact, since $\G$ consists of
gauge transformations $g(y)$ that are trivial at the basepoint, the
action of $\G$ on $\C$ by evaluation at the basepoint is trivial.
This accounts for the triviality of $\U_s$ for $s=0$.

The energetic reader may wish to explicitly construct the line
bundles $\U_s\to W$.  This can be done by explicitly dividing the
space of solutions of Nahm's equations by the group of gauge
transformations that are trivial at $y=s$ (as well as $y=0$),
somewhat as in section \ref{lb}, we explicitly divided by gauge
transformations that are continuous at $y=y_\rho$.

A final comment (not needed in this paper) is that for some purposes
it is useful to consider the dependence of $\U_s$ on $s$. By a
piecewise line bundle on $S^1$, we mean a collection of line bundles
on the intervals $y_\sigma\leq y\leq y_{\sigma+1}$.  A solution of
Nahm's equations is a triple $(\X,A,H_\sigma)$, where $A$ is a
connection on a piecewise line bundle, and $\X$ obeys Nahm's equations on
each interval with boundary conditions set by $H_\sigma$. As $s$ varies, the piecewise
line bundles $\U_s\to W$ fit together to a piecewise line bundle
$\hat \U\to W\times S^1$, while the other fields become appropriate objects on $M\times S^1$. It is natural to call $\hat \U$ a
universal piecewise line bundle. ``Universality'' means that if we
pick a point $z\in W$ and restrict $\hat\U$ to $\{z\}\times S^1$,
then $\hat U$ restricts to the piecewise line bundle of the solution
of Nahm's equations corresponding to $z$. We call $\hat \U$ ``a''
universal piecewise line bundle rather than ``the'' universal
piecewise line bundle because it depends on the choice of a
basepoint in the definition of the group $\G$.

\subsection{First Chern Class Of An Instanton Bundle}\label{moveb}
\subsubsection{Overview}\label{overv}

Our next goal is to understand the topology of the instanton bundle on $\TN_\k$
determined by a generic NS5-D5-D3 system.

One important detail concerns what we mean by topology.
In gauge theory, there is generally no way to interpolate from an instanton bundle $V\to\TN_\k $ to $V\otimes {\mathcal S}$,
where $\mathcal S$ is a non-trivial line bundle.  However, in string theory, there is a $B$-field, and in the presence of
the $B$-field one can smoothly interpolate from $V$ to $V\otimes {\mathcal S}$.  (Letting $F_{\mathcal S}$ be
the curvature of a connection on $\mathcal S$, one considers the one-parameter family of $B$-fields $B_t=t F_{\mathcal S}$,
$0\leq t\leq 1$.  One continuously interpolates from $t=0$ to $t=1$, and then one maps back from $t=1$ back to $t=0$ by a $B$-field gauge transformation that transforms $V$
to $V\otimes {\mathcal S}$.)  So there are two reasonable notions of topology and each is more relevant for some purposes.

An important tool is  the fact that by moving fivebranes in the $S^1$
direction, that is in the $y$ direction,
we can simplify a brane configuration without changing the topology.
Two NS5-branes can be displaced in $\vec X$ so that they can be moved past each other
in $y$ without ever meeting.  So there is no topological information in the ordering of NS5-branes
in $y$.  Similarly D5-branes can be displaced in $\vec Y$ and then reordered in $y$ without
meeting.   So again, the ordering of the D5-branes in $y$ contains no invariant information.

This procedure will be used shortly to put brane configurations in a standard form, but it has a drawback. There is no problem moving NS5-branes past each other, since
to construct $\TN_\k$ with generic metric parameters, the NS5-branes are located at generic points $\vec x_\sigma\in \R^3_\X$.
But to construct instanton bundles on $\TN_\k$, we place all D5-branes at the origin in $\R^3_\Y$, and moving
them past each other by displacing them in $\R^3_\Y$ can break supersymmetry.
So this is an operation that is useful for analyzing topology but possibly not for classifying the components
of the moduli space of
supersymmetric configurations.\footnote{A related remark was explained by S. Cherkis, correcting a claim in an earlier version of this paper.} Indeed, components of supersymmetric moduli space that cannot be put in the canonical form of section \ref{reduction}
have been studied in
\cite{Cherkis:2008ip,CherkisNew}.

A different type of problem arises when  a D5-brane is moved in $y$ so as to cross an NS5-brane. They inevitably
intersect, and  a D3-brane connecting them appears or disappears \cite{HW}.
So we cannot change the ordering of D5-branes relative to NS5-branes without changing the D3-brane
configuration at the same time.

\subsubsection{Relative Linking Number}\label{relative}

\begin{figure}
  \begin{center}
    \includegraphics[width=3in]{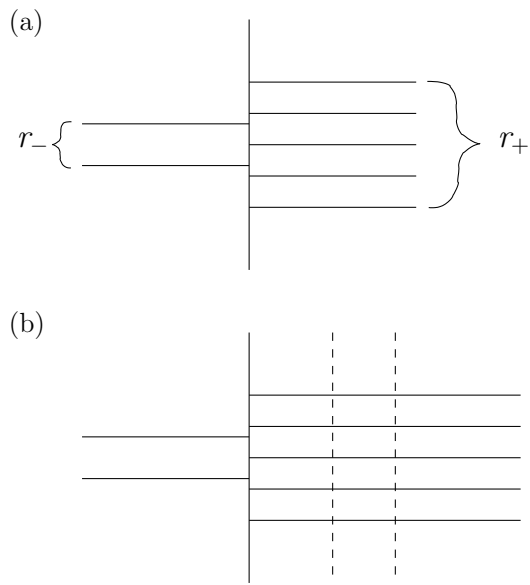}
  \end{center}

\caption{\small  The horizontal direction in this and later figures represents
$S^1$, and the vertical direction symbolically represents the six-dimensions of $\R^3_\X\times \R^3_\Y$.  Horizontal solid lines represent D3-branes, vertical solid lines
represent NS5-branes, and vertical dotted lines represent D5-branes.
(a) An NS5-brane with $r_+$ D3-branes
ending on its right and $r_-$ on its left.  We set $\delta r=r_+-r_-$. (b) If there are additionally $m$ D5-branes to the
right of the given NS5-brane, then its linking number is $\ell=\delta r - m$.  In the figure, $m=2$.    }
  \label{linking}\end{figure}

An essential concept in analyzing what happens when branes of different types cross
is the linking number of a fivebrane
\cite{HW,Gaiotto:2008ak}. First we recall the definition in the case
that $S^1$ is decompactified to $\R$, so that there is a
well-defined notion of whether one fivebrane is to the left or right
of another.  One contribution to the linking number of a fivebrane
is the net number of D3-branes ending on the given fivebrane.  This
is defined as the number $r_+$ of D3-branes to the right of the
given fivebrane, minus the number $r_-$ to its left (see fig.
\ref{linking}(a)). We set $\delta r=r_+-r_-$.

The linking number of a fivebrane is defined  as the sum of $\delta
r$ plus a second contribution from fivebranes of the opposite
type\footnote{In the past, definitions have been used that differ
from what follows by  inessential additive constants. For instance,
in \cite{Gaiotto:2008ak}, the linking number of an NS5-brane is
defined as the sum of $\delta r$ plus the number of D5-branes to its
left.  This differs from (1) below by a constant, the total number
of D5-branes.  The choice made here was suggested by D. Gaiotto.}:

(1) The linking number  $\ell$  of an NS5-brane equals
$\delta r$ minus the number of D5-branes to its right.  (For example, see fig. \ref{linking}(b).)

\def\tell{{\tilde \ell}}
(2) The linking number $\tell$ of a D5-brane equals the sum of $\delta r$ plus the number of NS5-branes
to its left.

The importance of the linking number is that it is unchanged when branes are reordered.
If a D5-brane is moved past an NS5-brane, the number of NS5-branes to its left changes,
but $\delta r$ changes in a way that compensates for this.

\begin{figure}
  \begin{center}
    \includegraphics[width=3in]{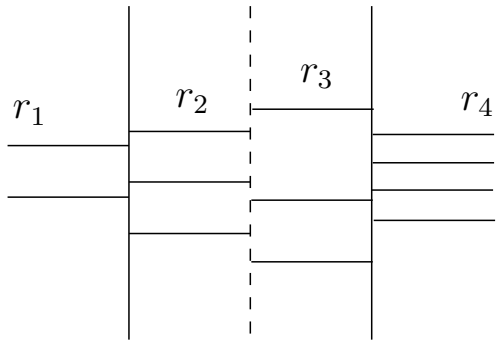}
  \end{center}

\caption{\small  This is part of a larger picture, possibly with additional fivebranes to the left and the right of those that are drawn.
Let $\ell$ and $\ell'$ be the linking numbers of the two NS5-branes on the leftmost and rightmost fivebrane in the
picture.  There is one D5-brane between them; it contributes $-1$ to $\ell$ and 0 to $\ell'$.  Any additional D5-branes
to the right of the picture  contribute $-1$ to both $\ell$ and $\ell'$.  So the difference $\Delta\ell=\ell'-\ell$
depends only on the branes that are depicted; its value is $\Delta\ell=1+\delta r'-\delta r$, where $\delta r'=r_4-r_3$
and $\delta r=r_2-r_1$ are the net numbers of D3-branes ending on the two D5-branes. }
  \label{example}\end{figure}

When we compactify $\R$ to $S^1$, we must be more careful,
since there is no invariant notion of ``left'' or ``right.''     Here
we simply take a pragmatic point of view.  We cannot define the
linking number of a single fivebrane, but there is no problem in
comparing the linking numbers of consecutive fivebranes of the same type. Going
back to fivebranes on $\R$, consider two   NS5-branes that are
consecutive in $y$. Suppose that a net  of $\delta r$ D3-branes end
on the first (the one of smaller $y$) and $\delta r'$ on the second.
And suppose that there are $w$ D5-branes between the two NS5-branes
(for an example with $w=1$, see fig. \ref{example}). Then, irrespective of any other D5-branes that are
to the left or the right of both NS5-branes, the difference between
their linking numbers is
\begin{equation}\label{polyo} \ell' - \ell = \delta r'-\delta r + w.\end{equation}

Now if we are on $S^1$, there is no natural way to define $\ell'$ or $\ell$ separately, but
there is no problem with the definition (\ref{polyo}) of the difference $\ell'-\ell$.  Thus,
we cannot define a natural integer-valued linking number for each fivebrane, but we can
define a relative linking number for two consecutive fivebranes of the same kind.

Labeling the consecutive NS5-branes on $S^1$ by $\sigma=1,\dots,k$, we write $\Delta \ell_\sigma$ for the difference
$\ell_{\sigma+1}-\ell_\sigma$.  Clearly
\begin{equation}\label{golyo}\sum_{\sigma=1}^k\Delta\ell_\sigma = p,\end{equation}
where $p$ is the total number of D5-branes.  Each D5-brane contributes to one term in the sum,
and D3-brane contributions cancel out.   This suggests that in some sense, as discussed further below, each NS5-brane linking number is well-defined mod $p$.

Similarly, the difference of linking numbers for consecutive D5-branes is $\tell'-\tell=\delta r'-\delta r
+m$, where $m$ is the number of NS5-branes between the two D5-branes.  This definition
makes perfect sense on $S^1$.  So labeling the consecutive D5-branes on $S^1$ by
$\lambda_1,\dots,\lambda_p$, we define $\Delta\tell_\lambda$ to be the difference $\tell_{\lambda+1}
-\tell_\lambda$.   Now we have
\begin{equation}\label{olyo} \sum_{\lambda=1}^p\Delta\tell_\lambda=k,\end{equation}
suggesting that in some sense each $\tell_\lambda$ is well-defined mod $k$.

\begin{figure}
  \begin{center}
    \includegraphics[width=3in]{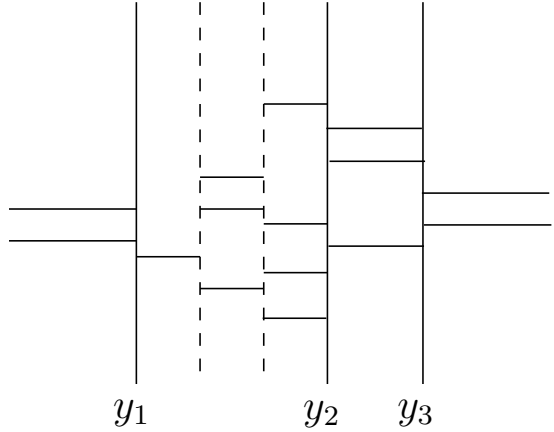}
  \end{center}

\caption{\small   After we remove from $S^1$ a point -- called the basepoint -- which does not coincide
with the position of any fivebrane, the brane configuration can be mapped to $\R$. (We have done this
without discussion in all of the previous figures!)  In this paper, we choose the basepoint
to be just to the left of the lefftmost NS5-brane, so that when we map to $\R$, the leftmost fivebrane is of NS type.
We label the positions of the NS5-branes from left to right as $y_\sigma$, $\sigma=1,\dots,k$.
}
  \label{basepoint}\end{figure}

A way of approaching the linking numbers that is less invariant but
is convenient in practice is to simply pick a point $y_0$ on $S^1$
that does not coincide with the position of any fivebrane. The
complement of $y_0$ is equivalent topologically to $\R$, so once
$y_0$ is chosen, we can make sense of whether a given fivebrane is
to the left or right of another.  This enables us to define the
individual $\ell_\sigma$ and $\tell_\lambda$, not just their
differences. Though not invariant, this approach is more convenient
than one might think because we have already essentially had to pick
a basepoint on $S^1$ in defining the line bundles $\RR_s$.    We
defined the $\RR_s$ so that just one of them is trivial, namely for
$s=0$. We define the linking numbers using the same basepoint.  We
pick the basepoint (as in fig. \ref{basepoint}) just to the left of
the first NS5-brane, so that that brane is the leftmost of all
fivebranes of either type.

The linking numbers change when the basepoint is moved across a
fivebrane.  Suppose that the basepoint crosses the $\sigma^{th}$ NS5-brane.
Then $\ell_\sigma$ changes by $\pm p$,  since all D5-branes go from being on the right
of the given NS5-brane to being on its left.  In the same process, all $\tell_\lambda$ change by $\mp 1$.
Conversely, if the basepoint moves across the $\lambda^{th}$ D5-brane, then $\tell_\lambda$ changes
by $\pm k$ and all $\ell_\sigma$ change by $\mp 1$.

If a fivebrane is moved clockwise all the way around the circle, it crosses from left to right all
fivebranes of the opposite type. So its linking number changes by $k$ in the case of a D5-brane, or by $p$ for
an NS5-brane.

\subsubsection{Reduction To Gauge Theory}\label{reduction}

Next, following \cite{Gaiotto:2008ak}, we want to describe a convenient ordering of the fivebranes.
First of all, we reorder the NS5-branes so that all relative linking
numbers $\Delta \ell$ are nonnegative as we move  toward increasing
$y$, or equivalently so that the linking numbers are
nondecreasing.\footnote{In \cite{Gaiotto:2008ak}, this condition and
the analogous one for D5-branes, which we impose momentarily, were
regarded as constraints that lead to more straightforward infrared
limits.  Here, the same conditions simplify the analysis of the
topology.}

We rearrange the D5-branes as follows.  First, moving them around
the $S^1$ as mentioned at the end of section \ref{relative}, we put
all their linking numbers in the range from 1 to $k$. Then we move
each fivebrane of linking number $\tell$ to the interval between the
$\tell^{th}$ and $\tell+1^{th}$ NS5-branes.  This in particular
means that the linking numbers of D5-branes are also nondecreasing
from left to right (and there are no D5-branes to the left of all
NS5-branes, in accord with our choice of basepoint).

\begin{figure}
  \begin{center}
    \includegraphics[width=3in]{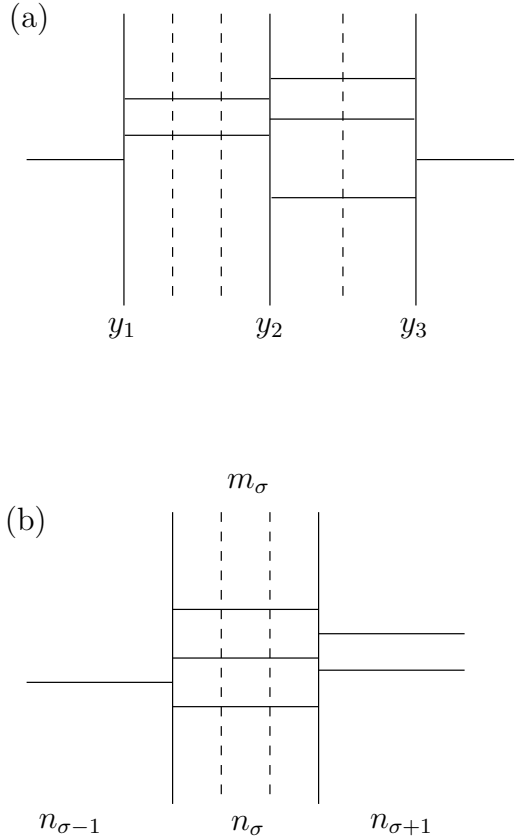}
  \end{center}

\caption{\small  (a)   Fivebranes can be ordered so that D3-branes end on NS5-branes only.  Here is
such an arrangement, with three NS5-branes at positions $y_\sigma$, $\sigma=1,2,3$.  (b) We write $n_\sigma$
for the number of D3-branes between the $\sigma^{th}$ and $\sigma+1^{th}$ NS5-brane, and $m_\sigma$ for the
number of D5-branes in that interval.  So in this figure, for example, we have $n_\sigma=3$ and $m_\sigma=2$.
The difference in linking numbers beween the two NS5-branes in this figure is
$\Delta\ell=n_{\sigma-1}+n_{\sigma+1}-2n_\sigma
+m_\sigma$.  This is similar to the case $r_2=r_3$ of fig. \ref{example}.
 }
  \label{gaugarrang}\end{figure}

At this stage, therefore, each D5-brane of linking number $\tell$
has precisely $\tell$ NS5-branes to its left.  Consequently, the net
number of D3-branes ending on any D5-brane is zero.  Accordingly,
D3-branes can be reconnected so as to end on NS5-branes only (as in
fig. \ref{gaugarrang}).    What we have gained this way is a canonical arrangement of branes in each topological
class.    For a reason mentioned in section \ref{overv}  (displacing fivebranes to move them past each other
may break supersymmetry), it does not give a classification of components of the supersymmetric moduli space,
only a convenient way to study the topology of the instanton bundle associated to a brane configuration.

We write $n_\sigma$, $\sigma=1,\dots,k$
for the number of D3-branes between the $\sigma^{th}$ and
$\sigma+1^{th}$ NS5-brane.  (This number is well-defined, since no
jumping occurs in crossing D5-branes.  The number of D3-branes to
the left, or equivalently to the right, of all fivebranes is denoted
as $n_0$ or $n_k$.) We also write $m_\sigma$ for the number of
D5-branes between the $\sigma^{th}$ and $\sigma+1^{th}$ NS5-branes,
that is, the number of D5-branes whose linking number is $\sigma$.
Then the formula (\ref{polyo}) for
$\Delta\ell_\sigma=\ell_{\sigma+1} -\ell_{\sigma}$ becomes
\begin{equation}\label{betterref}\Delta\ell_\sigma=n_{\sigma+1}+n_{\sigma-1}-2n_\sigma+m_\sigma \end{equation}
(see fig. \ref{gaugarrang}). We have ordered the NS5-branes so that these
numbers are all nonnegative.

A collection of canonically arranged branes leads to a quiver gauge theory in $3+1$ dimensions, with impurities
and discontinuities.
 The $n_\sigma$ D3-branes in the slab $y_\sigma\leq y\leq y_{\sigma+1}$ generate a $U(n_\sigma)$
gauge theory in this slab, with ${\cal N}=4$ supersymmetry. Each
slab is isomorphic to $\R^3\times I_\sigma$, where $I_\sigma$ is the
interval $[y_\sigma,y_{\sigma+1}]$.
 The
$m_\sigma$ D5-branes in the $\sigma^{th}$ slab generate matter
fields in the corresponding gauge theory.  Each D5-brane is
localized at some value of $y$, so its worldvolume intersects the
slab in a three-dimensional subspace, a copy of $\R^3$. This
subspace supports a hypermultiplet in the fundamental representation
of $U(n_\sigma)$. Finally, each boundary between two slabs, such as
the one at $y=y_\sigma$, supports a bifundamental hypermultiplet,
interacting with the $U(n_{\sigma-1})$ gauge theory to the left and
the $U(n_\sigma)$ gauge theory to the right.

\begin{figure}
  \begin{center}
    \includegraphics[width=3in]{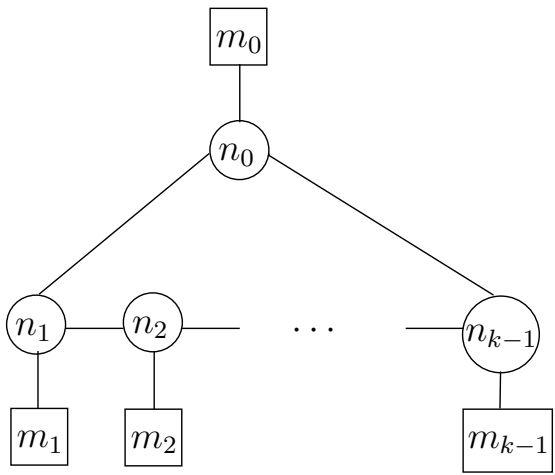}
  \end{center}

\caption{\small   A quiver of type ${\mathrm A}_{k-1}$ (analogous to the Dynkin diagram of fig. \ref{Dynkin}).
A node of the quiver -- that is a circle labeled by an integer $n_\sigma$, $\sigma=1,\dots,k$ -- represents a $U(n_\sigma)$
gauge theory that originates in one of the slabs, as described in the text. Hence the overall gauge group
is $\prod_{\sigma=1}^k U(n_\sigma)$.   The nodes are arranged
in a ring (which originates in the $S^1$ of the brane configuration).
Lines between neighboring nodes
represent bifundamental hypermultiplets of $U(n_\sigma)\times U(n_{\sigma+1})$, for various $\sigma$.  Finally,
a square labeled by an integer $m_\sigma$ represents $m_\sigma$ fundamental hypermultiplets of $U(n_\sigma)$.}
  \label{quiver}\end{figure}

What we have just described  is a slightly exotic four-dimensional
cousin of a three-dimensional quiver gauge theory. At low energies,
it reduces to an ordinary three-dimensional quiver gauge theory.
Indeed, at distances much greater than the width $r_\sigma$ of
$I_\sigma$, gauge theory on $\R^3\times I_\sigma$ reduces to purely
three-dimensional gauge theory on $\R^3$ (the three-dimensional
gauge coupling $g_3$ obeys $1/g_3^2=r_\sigma/g_4^2$, where  $g_4$ is
the four-dimensional gauge coupling).  The low energy limit is a
three-dimensional quiver gauge theory on what is usually called the
${\mathrm A}_{k-1}$ quiver. This quiver is a ring with $k$ nodes, as
in  fig. \ref{quiver}; there is  a $U(n_\sigma)$ gauge group at the
$\sigma^{th}$ node. Each such group interacts with $m_\sigma$
flavors of fundamental matter (indicated in the adjacent square) and
with bifundamental hypermultiplets associated with its links to the
two neighboring nodes.

This quiver gauge theory is more commonly used
\cite{KN,Douglas:1996sw} to describe instantons on an ALE space (the
hyper-Kahler resolution of the ${\mathrm A}_{k-1}$ singularity),
rather than on the ALF space $\TN_\k$. However, the two spaces are
topologically the same, and are holomorphically equivalent in any
one complex structure, though their hyper-Kahler metrics are
different. (To prove this, recall that the ALF space reduces to ALE if we set
$\lambda\to\infty$, as discussed in section \ref{el}.  From the
point of view of any one complex structure, $\lambda$ is a Kahler
parameter, which can be varied without changing the complex
structure.) It has been shown \cite{Cherkis:2008ip} using Nahm's
equations on $S^1$ that the components of instanton moduli space on the ALF
and ALE spaces that can be derived from canonical brane arrangements
are related in the same way -- equivalent in any one
complex structure but with different hyper-Kahler metrics. This is reviewed in Appendix A.  As we
explained in section \ref{morep}, the instanton bundles themselves
have the same property.

Our goal in this subsection is to compute the Chern classes of the instanton bundles corresponding
to a given NS5-D5-D3 configuration.  To this aim, in view of the observations in the last paragraph,
we could simply borrow standard results \cite{KN} about the ALE case.  We prefer, however,
to proceed with an independent derivation.

\subsubsection{The Number Of D3-Branes}

By a ``complete'' D3-brane, we mean a D3-brane that stretches all the way around $S^1$.
By a ``fractional'' D3-brane, we mean one that connects two fivebranes.
A complete D3-brane does not contribute to any linking numbers, but a fractional D3-brane does.

In the above construction, we have chosen to have all D3-branes end on NS5-branes,
so there are $k$  basic types of fractional D3-brane, connecting the successive NS5-branes.
 We will compute presently how the
topology of an instanton bundle depends on the fractional D3-branes.  However, one special
case is immediate.  A complete D3-brane can be moved far away from all fivebranes.
Its $T$-dual is then a D2-brane, supported at a point in $\TN_\k$.  This represents a point
instanton; it contributes 1  to the second Chern class and 0 to the first Chern class.

Adding a constant to all  $n_\sigma$, or subtracting such a
constant, means adding or removing a complete D3-brane.  So it
shifts the second Chern class without changing the first Chern
class. For specified first Chern class, there is a minimum possible
instanton number in a supersymmetric configuration: the bound comes
from requiring all $n_\sigma$ to be nonnegative.   (Negative
$n_\sigma$ means that some D3-branes are replaced by anti-branes,
breaking supersymmetry.) An ${\mathrm A}_{k-1}$ quiver with all
$n_\sigma$ nonnegative has a Higgs branch if and only if all
$\Delta\ell_\sigma$ are nonnegative \cite{Nak} (when the Higgs
branch exists, a generic point on it is smooth and corresponds to a
smooth instanton connection on $\TN_\k$).

We have, of course, broken the symmetry between NS5-branes and
D5-branes by requiring all D3-branes to end on NS5-branes.  We can
find a dual description by reversing the roles of the two types of
brane.  This leads to a ``mirror'' gauge theory associated with an
${\mathrm A}_{p-1}$ quiver and related to instantons on another ALF
space $\TN_{\mathrm p}$.

It is shown in section 2.2 of \cite{Gaiotto:2008ak} that, if all
fivebrane linking numbers are nondecreasing, all $n_\sigma$ are
nonnegative in one description if and only if they are nonnegative
in the dual description.  This analysis is made for the case that
$S^1$ is decompactified to $\R$. However, since a complete D3-brane
does not contribute to the linking numbers, the analysis can be made
after removing some D3-branes so that the minimum value of $n_\sigma
$ is zero.  At this point the brane configuration does not cover all
of $S^1$ and can be embedded in $\R$; hence the analysis in
\cite{Gaiotto:2008ak} applies.

\subsubsection{The First Chern Class: D5-Brane Contribution}

We aim to describe the topology of the instanton bundle $V\to
\TN_\k$ associated to a general NS5-D5-D3 configuration. Our main effort will be to determine the first Chern class $c_1(V)$.

According to section \ref{topology}, $c_1(V)$, which takes values in
$H^2(\TN_\k,\ZZ)$, can be represented by an integer-valued sequence
$\{b_\sigma|\sigma=1,\dots,k\}$, modulo $b_\sigma\to b_\sigma+b$ for
some fixed integer $b$.    We claim that $c_1(V)$ is associated with the
 sequence of
NS5-brane linking numbers $\{\ell_\sigma|\sigma=1,\dots,k\}$.

First let us check that this statement makes sense -- that the
linking number sequence has just the right sort of indeterminacy.
Suppose that we move the basepoint across a D5-brane.  Then all
NS5-brane linking numbers $\ell_\sigma$ change by $\pm 1$, an
operation that does not affect the element of $H^2(\TN_\k,\ZZ)$
determined by the linking number sequence.  On the other hand,
suppose that we move the basepoint across an NS5-brane.  The linking
number of this particular NS5-brane changes by $\pm p$.  Likewise,
if an NS5-brane is moved all the way around the circle, its linking
number changes by $\pm p$.  Either way, this corresponds to a
topologically non-trivial $B$-field gauge transformation, which acts
on $V$ by $V\to V\otimes \T$, for some line bundle $\T$.  As $V$ has
rank $p$, its first Chern class transforms by $c_1(V)\to c_1(V)+p\,
c_1(\T)$, consistent with the fact that the NS5-brane linking number
sequence changes by a multiple of $p$.

In our analysis, a choice of basepoint is used both to fix the
$B$-field gauge in defining $V$ and to define the NS5-brane linking
numbers. Having made these choices, we claim that the linking number
sequence determines $c_1(V)$ exactly (not just mod $p$). We first
establish this relation for the case that there are no D3-branes.
D3-branes are included in section \ref{threec}.

Suppose that there are $p$ D5-branes located at $y=s_i$, $i=1,\dots, p$. In this
case, $V$ is simply a direct sum of line bundles, $V=\oplus_{i=1}^p\RR_{s_i}$. Suppose
that $s_i$ is located between the $\sigma_i^{th}$ and
$\sigma_{i+1}^{th}$ NS5-brane.  Then according to (\ref{genve}),
from a topological point of view (recalling that $\L_*$ is
topologically trivial)
\begin{equation}\label{enve}
\RR_{s_i}=\otimes_{\rho=1}^{\sigma_i}\L_{\rho}^{-1}.\end{equation}
We also know from (\ref{orno}) that $c_1(\L_\rho)$ is associated to
the sequence $b^{(\rho)}_\sigma=\delta_{\rho\sigma}$.   Combining
these facts, the sequence associated to $c_1(\RR_{s_i})$ is
\begin{equation}\label{exon}b^{(i)}_\sigma=\begin{cases} -1 & {\mathrm {if}} ~ y_\sigma<s_i\\
  0 & {\mathrm {if}} ~y_\sigma>s_i. \end{cases}\end{equation}

The sequence associated to $c_1(V)=\sum_i c_1(\RR_{s_i})$ is
therefore given by
\begin{equation}\label{ytn}b_\sigma=\sum_i b^{(i)}_\sigma.\end{equation}
  The $\sigma^{th}$ entry in this
sequence is simply minus the number of D5-branes to the right of
$y_\sigma$, since each such fivebrane contributes $-1$ to
$b_\sigma$, and others do not contribute.  But this is precisely the
contribution of the D5-branes to the linking number $\ell_\sigma$.

\subsubsection{The D3-Brane Contribution}\label{threec}

\def\B{{\mathcal B}}

Now we include D3-branes.  A generic state of an NS5-D5-D3 system
has D5-D3 strings turned on, and leads to a sheaf on $\TN_\k$ that
is not simply a direct sum of contributions from the D5-branes and
the D3-branes.  (Generically, if $n_\sigma, \,\Delta\ell_\sigma\geq
0$, such a sheaf is a smooth instanton bundle.)  However, to compute
Chern classes, we can ignore the D5-D3 strings and treat the sheaf
on $\TN_\k$ as a direct sum.

Our goal is to compute the contribution of D3-branes to the first
Chern class, and show that this contribution does not spoil the
equality between the first Chern class and the linking number
sequence.  In the process, we will also learn how to compute the
second Chern class.

We denote as $\B_\tau$ a fractional D3-brane that stretches between
the $\tau^{th}$ and $\tau+1^{th}$ NS5-branes.  It is $T$-dual to a
sheaf  ${\mathcal V}_\tau\to \TN_\k$; we want to determine the brane
charges of this sheaf.

In general, we map a brane $\B$ to cohomology by taking the Chern
character of the corresponding sheaf $\mathcal V$.  The Chern
character ${\mathrm {ch}} \mathcal V$ takes values in
$H^0(\TN_\k,\ZZ)\oplus H^2(\TN_\k,\ZZ)\oplus H^4(\TN_\k,{\Bbb Q})$,
where the three summands  measure what we will call the E4-brane
charge, E2-brane charge, and E0-brane charge.\footnote{The reason to
call these brane charges E-brane charges (rather than D-brane
charges) is to emphasize that for now we are considering branes in
the $\TN_\k$ sigma model, rather than in the full ten-dimensional
string theory. The E0-brane charge can be fractional, and so takes
values in $H^4(\TN_\k,{\Bbb Q})$, rather than $H^4(\TN_\k,\ZZ)$.}
The rationale for the terminology is that the smallest possible
dimension of the support of a brane in $\TN_\k$ carrying E$n$-brane
charge, for $n=0,2,4$, is $n$.

\def\ch{{\mathrm {ch}}}
\def\V{{\mathcal V}}
The sheaf $\V_\tau\to \TN_\k$ that corresponds to a fractional
D3-brane $\B_\tau$ is supported on a proper
subspace\footnote{Intuitively, one suspects that the support would
be the compact two-cycle $C_{\tau ,\tau+1}$.  We do not have a
direct way to demonstrate this, but the determination below of the
E2-brane charge supports this idea.} of $\TN_\k$, since $\B_\tau$ is
localized in $\vec X$. $\V_\tau$ hence has vanishing E4-brane charge
(equivalently, the zero-dimensional part ${\mathrm{ch}}_0(\V_\tau)$
of its Chern character vanishes), so its Chern character is of the
form ${\mathrm {ch}}(\V_\tau)=0\oplus {\rm ch}_2(\V_\tau)\oplus
{\mathrm {ch}}_4(\V_\tau)$.  Here ${\mathrm {ch}}_2$ and ${\mathrm
{ch}}_4$ are the two-dimensional and four-dimensional parts of the
Chern character, which can be expressed in terms of Chern classes
$c_i,\,i=1,2$ by ${\mathrm {ch}}_2=c_1$, ${\mathrm
{ch}}_4=c_1^2/2-c_2$. Therefore an alternative expression of the
brane charges of $\V_\tau$ is
\begin{equation}\label{plyo}{\mathrm {ch}}(\V_\tau)=0\oplus
c_1(\V_\tau)\oplus
\left(\frac{c_1^2(\V_\tau)}{2}-c_2(\V_\tau)\right).\end{equation} As
in section \ref{lb}, we must consider the action of $B$-field gauge
transformations on $\V_\tau$.  A $B$-field gauge transformation
modifies the sheaf corresponding to any brane $\B$ by $\V\to
\V\otimes{\mathcal S}$, where  $\mathcal S$ is a line bundle that is
independent of $\B$. This modifies the Chern character by
$\ch(\V)\to \ch(\V)\exp(c_1({\mathcal S}))$.  In the case of
$\V_\tau$, because  $\ch_0(\V_\tau)$ vanishes, $\ch_2(\V_\tau)$
 is invariant under this operation, but
$\ch_4(\V_\tau)$ is not.

To compute ${\mathrm {ch}}(\V_\tau)$, we will compute the
``intersection number'' of the brane $\B_\tau$ with a D5-brane
supported at $y=s$, which we call $\tilde\B_s$.  The intersection
number is the supersymmetric index $\Tr\,(-1)^F$ in the space of
$\tilde\B_s-\B_\tau$ ground states.   The index is one if
$y_\tau<s<y_{\tau+1}$, because in that case the branes $\tilde\B_s$
and $\B_\tau$  have one positively oriented transverse point of
intersection. And it is zero otherwise, since if $s$ is outside the
indicated range, there are no intersections at all.   If
$y_\rho<s<y_{\rho+1}$, the sheaf on $\TN_\k$ corresponding to
$\tilde\B_s$ is the line bundle $\L_\rho$, and according to the
index theorem, the intersection number is $\int_{\TN_\k}\hat
A(\TN_\k) \ch(\RR_s^{-1})\ch(\V_\tau)$. (Because of the
noncompactness of $\TN_\k$, one should be careful in general in
using such a formula, but in this case there is no problem as
$\V_\tau$ has compact support.)  We have
$\ch(\RR_s^{-1})=\exp(-c_1(\RR_s))$.  Also $\hat A(\TN_\k)=1+\dots$
where the omitted term is a four-dimensional class that can be
dropped becuase $\V_\tau$ has no E4-brane charge. So the index
formula reduces to $\int\left({
\ch}_4(\V_\tau)-c_1(\RR_s)c_1(\V_\tau)\right)$.  This must equal the
expected intersection number:
\begin{equation}\label{celme}\int{\mathrm
{ch}}_4(\V_\tau)-\int
c_1(\RR_s)c_1(\V_\tau)=\delta_{\rho\,\tau}.\end{equation} Here
$\rho$ is such that $y_\rho<s<y_{\rho+1}$.

Let us first evaluate this formula for $\rho=0$ (which means that $s<y_1$ or $s>y_k$; $\rho=0$ is the same
as $\rho=k$, since the NS5-brane labels $\sigma,\rho,\tau$ are considered
to be defined mod $k$).  In this case, $c_1(\RR_s)=0$, so
\begin{equation}\label{zbe}\int {\ch}_4(\V_\tau)=\delta_{\tau 0}.\end{equation}
This determines the E4-brane charges of fractional D3-branes: it equals 1 for $\tau=0$ and
otherwise vanishes.

Now in (\ref{celme}), $f=c_1(\V_\tau)$ is a cohomology class with
compact support, while $b=c_1(\RR_s)$ is an unrestricted cohomology
class. According to (\ref{zumor}), the pairing of such classes is
obtained by the obvious inner product $\sum_\sigma b_\sigma
f_\sigma$ of the sequences $\{b_\sigma|\sigma=1,\dots, k\}$ and
$\{f_\sigma|\sigma=1,\dots,k\}$ representing them.  As in
(\ref{exon}), $c_1(\RR_s)$ corresponds to the sequence
\begin{equation}\label{texon}b_\sigma=\begin{cases} -1 & {\mathrm {if}} ~ y_\sigma<s\\
  0 & {\mathrm {if}} ~y_\sigma>s. \end{cases}\end{equation}
Given this, we find to obey (\ref{celme}) that the sequence
corresponding to $c_1(\V_\tau)$ is
\begin{equation}\label{hexon}c^{(\tau)}_\sigma=\delta_{\sigma,\,\tau}-
\delta_{\sigma,\,\tau+1}.\end{equation}
This determines the E2-brane charges of fractional D3-branes.

From the definition of linking number, a fractional D3-brane of type $\B_\tau$, stretching between the
$\tau^{th}$ and $\tau+1^{th}$ NS5-brane,  contributes 1
to the linking number $\ell_\tau$, $-1$ to $\ell_{\tau+1}$, and
0 to $\ell_\sigma$ for other values of $\sigma$. But this is the same as the contribution of the
same brane to the
E2-brane charges in (\ref{hexon}). In other words,
the contribution of such a brane to the linking number sequence is
precisely the same as its contribution to the sequence representing
the first Chern class.  This is what we aimed to show:  even after including fractional D3-branes,
the first Chern class
of the sheaf over $\TN_\k$ corresponding to a brane configuration is represented by the linking number sequence.

The formula (\ref{hexon}) for ${\mathrm {ch}}_2(\V_\sigma)$
is invariant under cyclic permutations of all NS5-branes.  However, the
formula (\ref{zbe}) for ${\mathrm {ch}}_4(\V_\sigma)$ lacks this
symmetry.  This is expected, because a $B$-field gauge transformation, as noted above,
can modify $\ch_4(\V_\tau)$ but not $\ch_2(\V_\tau)$.

The results (\ref{hexon}) and (\ref{zbe}) confirm the expectation
that a complete D3-brane contributes zero to the E2-brane charge and
1 to the E4-brane charge.  Indeed, (\ref{hexon}) vanishes if summed
over $\tau$, while (\ref{zbe}) sums to 1.

\subsubsection{The E0-Brane Charge}\label{eob}

To compute $\ch_4(V)$, where $V$ is the instanton bundle derived
from a NS5-D5-D3 system, we must add to (\ref{zbe}) the D5
contribution.  The Chern character of a line bundle $\L$ is
$\ch(\L)=\exp(c_1(\L))$, so we have
$\int\ch_4(\L)=\frac{1}{2}\int_{\TN_\k} c_1(\L)^2$.  We have to be careful in using this formula,
however, since $\TN_\k$ is not compact.  The line bundles of interest are tensor products of
line bundles $\L_\rho$ and $\L_*^t$ studied in section \ref{lb}; their connection forms are given in eqns. (\ref{pij}) and
(\ref{elfk}).

If $\L=\otimes_\rho \L_\rho^{n_\rho}$ with $\sum_\rho n_\rho=0$, then $c_1(\L)$ is a cohomology class of compact
support, represented by the sequence $\{n_\rho|\rho=1,\dots,k\}$.   In this case, we can compute $\int c_1(\L)^2$ using
(\ref{zumor}), with the result $\int c_1(\L)^2=\sum_\rho n_\rho^2$.

Now let us consider the opposite case of a topologically trivial line bundle $\L_*^t$.  Here, the first Chern class
vanishes in cohomology.  However, the curvature form is $F=\d\Lambda_t=t\d\Lambda$, where $\Lambda$ is the globally-defined one-form (\ref{elfk}).  So $\int_{\TN_\k}F\wedge F/4\pi^2=kt^2\int_{\partial \TN_\k}\Lambda\d\Lambda/4\pi^2
= kt^2$, where $\partial\TN_\k$, which is an $S^1$ bundle over $S^2$, is the ``boundary'' of $\TN_\k$.  We have used
these facts: (i) $\int_{S^1}\Lambda=2\pi$; (ii) $\d\Lambda$ is a pullback from $S^2$; (iii) $\int_{S^2}\d\Lambda=2\pi k$.

For a line bundle $\L=\L_*^t \otimes (\otimes_\sigma \L_\sigma^{n_\sigma})$, with $\sum_\sigma n_\sigma=0$,
we combine these facts and get
\begin{equation}\label{horsefly}
\int_{\TN_\k}c_1(\L)^2=kt^2+\sum_\sigma n_\sigma^2.\end{equation}  Finally, if $\sum_\sigma n_\sigma
\not=0$, we set $n_*=\sum_\sigma n_\sigma$ and $\tilde n_\sigma=n_\sigma-n_*/k$, and write formally $\L=\L_*^{t+n_*/k}\otimes(\otimes_\sigma \L_\sigma^{\tilde n_\sigma})$.  Even though the $\tilde n_\sigma$ may not be integers, we can still use (\ref{horsefly}),
with the result that
\begin{equation}\label{orseful}\int_{\TN_\k}{c_1(\L)}^2=kt^2+2tn_*+\sum_\sigma n_\sigma^2.\end{equation}   One
way to  justify this use of (\ref{horsefly}) is to replace $\L$ with $\L^k$, using the fact that $\int c_1(\L)^2=\int c_1(\L^k)^2/k^2$.
The last expression can be evaluated using (\ref{horsefly}) without running into fractional exponents.

Finally, we apply this to the line bundle $\RR_s$ associated with a D5-brane at $y=s$, where $y_\rho<s<y_{\rho+1}$.  According to
(\ref{incmon}), this is the case that $t=s/2\pi R$, and that the nonzero $n_\sigma$ are $n_\sigma=-1$ for $\sigma\leq \rho$.
So we get
\begin{equation}\label{seful}\int_{\TN_\k}\ch_4(\RR_s)=\frac{1}{2}\int_{\TN_\k}c_1(\RR_s)^2=\frac{1}{2}\left(kt^2-2t\rho+\rho\right).\end{equation}

\subsection{The Monodromy At Infinity And The ALE Limit}\label{all}

\def\tell{\tilde\ell}
\def\p{\mathrm p}
We have interpreted the linking numbers $\ell_\sigma$ of NS5-branes in terms of the first Chern class of
an instanton bundle over $\TN_\k$.  The question now arises of finding a similarly interesting
interpretation of the linking numbers $\tilde\ell_\lambda$  of D5-branes.

One easy answer is that we could apply an $S$-duality transformation
that exchanges NS5-branes with D5-branes.  This then gives a new
configuration with $p$ NS5-branes and $k$ D5-branes. $T$-duality on
$S^1$ will lead to $U(k)$ instantons on $\TN_\p$ (rather than $U(p)$
instantons on $\TN_\k$, as studied so far).  In this description,
the first Chern class of the instanton bundle is given by the
sequence $\{\tilde\ell_\lambda|\lambda=1,\dots,p\}$.   From the
point of a reduction to gauge theory on $\R^3$ (recall section
\ref{reduction}), the two descriptions differ by three-dimensional
mirror symmetry.  Three-dimensional mirror symmetry was introduced
in \cite{IS} and interpreted via $S$-duality of brane configurations
in \cite{G,HW}.

However, we would like to interpret the $\tell_\lambda$ in terms of the ``original''  instanton
bundle  $V\to\TN_\k$.  After arranging the D5-branes as in section \ref{reduction}, so that no
D3-brane ends on a D5-brane, the linking numbers $\tell_\lambda$ are determined by the
D5-brane positions $s_\lambda$: if $y_\rho<s_\lambda<y_{\rho}+1$ then $\tell_\lambda=\rho$.

The D5-brane positions have a simple interpretation in terms of the
instanton bundle $V$. In section \ref{lineb}, we explained that the
line bundle $\RR_s$ has monodromy $\exp(is/R)$ over the circle
$\tilde S^1$ at infinity  ($R$ is the radius of the original circle
$S^1$).  The behavior at infinity is not affected by D3-branes,
fractional or otherwise, as they are localized in $\vec X$. Hence
the monodromy at infinity of an instanton bundle, which we will
denote as $U_\infty$, can be read off from the positions of the
D5-branes. If the D5-branes are located at positions $s_1,
\dots,s_p$ along $S^1$, then
\begin{equation}\label{oklo}U_\infty={\mathrm {diag}}(\exp(is_1/R),\exp(is_2/R),\dots,\exp(is_p/R)).
\end{equation}

This explains what the D5-brane positions mean in terms of the instanton bundle over $\TN_\k$.
However, the linking numbers $\tell_\lambda$ contain less information than the D5-brane positions:
they depend only on the positions of D5-branes relative to NS5-branes.  There is a simple
variant of what we have just said in which one sees only the $\tell_\lambda$.

We simply take the limit $R\to 0$ (keeping fixed the angular
positions of the fivebranes). In the limit, which was discussed in
section \ref{el}, the dual circle at infinity in $\TN_\k$
decompactifies and the ALF space $\TN_\k$ becomes an ALE space which
is the resolution of an ${\mathrm A}_{k-1}$ singularity. In the ALE
limit, the fundamental group at infinity is $\ZZ_k$, so all
monodromies at infinity are of order $k$.  As explained in section
\ref{el}, in this limit
 the line bundles $\L_\rho$ all have monodromy at infinity
equal to $\exp(-2\pi i/k)$.

In the same limit, the line bundle $\RR_s$ is
$\otimes_{\sigma=1}^\rho \L_\sigma^{-1}$, where
$y_\rho<s<y_{\rho+1}$.  This follows from (\ref{genve}) and the fact
that the connection on $\L_*$ is trivial in the ALE limit.  Hence the monodromy at
infinity of $\RR_s$ is $\exp(2\pi i\rho/k)$.  But a D5-brane located
at $y_\rho<s<y_{\rho+1}$ has linking number $\rho$.  So the linking
number corresponds directly to the exponent of monodromy.

More generally, if we are given any set of D5-branes with linking
numbers $\tell_1,\dots,\tell_p$, then the monodromy at infinity in
the ALE limit is conjugate to
\begin{equation}\label{zolp}U_{\infty}={\mathrm {diag}}(\exp(2\pi i\tell_1/k),\exp(2\pi i\tell_2/k),\dots,
\exp(2\pi i\tell_p/k)).\end{equation}

This result can be stated in another way.  For $\sigma=1,\dots,k$,
let $a_{\sigma,\infty}$ be the number of eigenvalues of
$U_\infty$ that equal $\exp(2\pi i \sigma/k)$.  The D5-branes that contribute this eigenvalue of the
monodromy are those with $\tell=\sigma$, and since the number of those is $m_\sigma$, we have
\begin{equation}\label{helme} a_{\sigma,\infty}=m_\sigma.\end{equation}

We can use these results to clarify what sort of configurations on $\TN_\k$ can be represented
by a canonical NS5-D5-D3 configuration in which D3-branes end on NS5-branes only.  If the monodromy at infinity
of an instanton bundle on $\TN_\k$ is given, we read off from (\ref{oklo}) what the positions of the D5-branes
must be.  The first Chern class of the instanton bundle determines the D5 linking numbers.  If we want D3-branes
to end on NS5-branes only, the ordering of the NS5-branes on $S^1$, relative to the D5-branes, are determined by
the D5 linking numbers.  This constrains the $B$-field on $\TN_\k$, since the $B$-field is determined by the NS5-brane positions
on $S^1$.  (We are still free to vary the positions of the NS5-branes in $\vec X$, so we can vary the geometrical
moduli of $\TN_\k$.)  A general choice of $B$-field on $\TN_\k$ forces us to arrrange the NS5-branes on $S^1$ in a more
general fashion, so that D3-branes will end on D5-branes as well as NS5-branes.  A configuration of this more general
type has a less simple description in gauge theory on $\R^3\times S^1$, as explained in \cite{Gaiotto:2008sa}.

\bigskip\noindent{\it First Chern Class At Infinity}

A similar interpretation of the  $\tell_\lambda$ can be given
without taking the ALE limit. An instanton bundle $V\to \TN_\k$
associated to an NS5-D5-D3 configuration (in which D3-branes end on
NS5-branes only) is generically not a simple sum of line bundles.
But near infinity, it always naturally decomposes as a direct sum of
line bundles $\L_{s_\lambda}$, corresponding to the D5-brane
positions $s_\lambda$. The region near infinity in $\TN_\k$ is
homotopic to $S^3/\ZZ_k$, and $H^2(S^3/\ZZ_k,\ZZ)\cong \ZZ_k$. So
when restricted to the region near infinity, each $\L_{s_\lambda}$
has a first Chern class valued in $\ZZ_k$.  This is equal  precisely
to its linking number $\tell_\lambda$.  This statement is equivalent
to the previous statement in terms of holonomy at infinity in the
ALE limit, since in that limit, the $\L_{s_\lambda}$ become flat,
and a flat line bundle over $S^3/\ZZ_k$ with holonomy $\exp(2\pi i
r/k)$ has first Chern class $r$.

\subsubsection{NS5-Brane Linking Numbers And Monodromy}\label{lmon}

So far, we have interpreted the NS5-brane linking numbers in terms of the first Chern class
and the D5-brane linking numbers in terms of monodromy at infinity in the ALE limit.
However, it is also possible to interpret the NS5-brane linking numbers in terms of monodromy,
in a certain limit.

The space $\TN_\k$ develops an ${\mathrm A}_{k-1}$ singularity,
looking locally like $\R^4/\ZZ_k$, when all NS5-branes are at the
same location in $\R^3_\X$ (though possibly at different points in
$S^1$), or in other words when the parameters $\vec x_\sigma$ in
(\ref{gomely}) are all equal.  This singularity develops even before
taking the ALE limit $\lambda\to\infty$, and for the moment, we do
not consider that limit.

In general, how does one define gauge theory in the presence of a
singularity of the form $\R^4/\ZZ_k$? It is defined as
$\ZZ_k$-invariant gauge theory on the covering space $\R^4$.
However, there is some freedom in picking the action of $\ZZ_k$.  In
general, we pick an element $U_0$ of the gauge group $G$, such that
$U_0^k=1$, and we consider gauge fields on $\R^4$ that are invariant
under the action of $\ZZ_k$ on $\R^4$ together with a gauge
transformation $U_0$.  We can think of $U_0$ as the monodromy around
a small circle at the origin. In the present context, the gauge
group is $G=U(p)$, with $p$ the number  of D5-branes, so $U_0$ has
$p$ eigenvalues which are all $k^{th}$ roots of unity.

Let us write $a_{\rho,0}$ for the number of eigenvalues of $U_0$
that equal $\exp(2\pi i\rho/k)$. Standard results about quiver gauge
theories  \cite{KN} can be used to show that $a_{\rho,0}$ is equal
to the relative linking number
$\Delta\ell_\rho=\ell_{\rho+1}-\ell_\rho$, or equivalently,
according to eqn. (\ref{betterref}),
\begin{equation}\label{zello}a_{\rho,0}=n_{\rho+1}+n_{\rho-1}-2n_\rho+m_\rho.\end{equation}

We postpone to Appendix B the general derivation of this result
using quiver gauge theory. For now, we explain only the special case
that there are no D3-branes, so that all $n_\tau$ vanish.
  In this case,
the instanton bundle is simply a direct sum of line bundles $\RR_s$.
Using the familiar result (\ref{genve})  describing these line
bundles in terms of $\L_\rho$ and $\L_*$, we can compute the
monodromy of $\RR_s$ near the ${\mathrm A}_{k-1}$ singularity.
According to (\ref{pij}), if we set all $\vec x_\sigma$ equal (to generate the singularity), then
the connection form of the line bundle $\L_\rho$ behaves for $\rr\to
\vec x_\sigma$ as $\Lambda_\rho\to -\d\tilde\chi/k+\dots$, where the
omitted terms are less singular.  So the monodromy of $\L_\rho$ near
the singularity is $\exp(-2\pi i/k)$, independent of $\rho$.

This is the same as the monodromy of $\L_\rho$ at infinity in the
ALE limit $\lambda\to\infty$.  That fact has a simple explanation.
If we take all $\vec x_\sigma$ equal and also take
$\lambda\to\infty$, then $\TN_\k$ reduces to $\R^4/\ZZ_k$, as one
can deduce from (\ref{gomely}), and according to (\ref{pij}),
$\Lambda_\rho$ equals $-\d\tilde\chi/k$ exactly.  The last formula implies
that $\L_\rho$ is flat when $\TN_\k=\R^4/\ZZ_k$, so it has the same
monodromy at the singularity and at infinity.

The monodromy at the singularity of $\L_*=\otimes_{\tau=1}^k\L_\tau$
is therefore trivial. Hence it follows from (\ref{genve}) that the
monodromy at the singularity of $\RR_s$ is $\exp(2\pi  i\rho/k)$,
where $y_\rho<s<y_{\rho+1}$.

Finally, taking a direct sum of line bundles $\RR_s$ with different
values of $s$, the number of eigenvalues $\exp(2\pi i\rho/k)$ of the
monodromy at the singularity is equal to $m_\rho$, the number  of
D5-branes between $y_\rho$ and $y_{\rho+1}$.  This verifies the
claim (\ref{zello}) for the case that there are no D3-branes.  The
general case is deferred to the appendix.

Because a choice of gauge for the $B$-field entered the definition of the line bundles $\L_\rho$,
the description (\ref{zello}) depends on this choice of gauge, just as (\ref{helme}) does.  In the limit
that $\TN_\k$ reduces to $\R^4/\ZZ_k$, a change of gauge for the $B$-field multiplies both $U_0$ and
$U_\infty$ by the same $k^{th}$ root of unity.

\subsubsection{Duality Of Young Diagrams}\label{duality}

\begin{figure}
  \begin{center}
    \includegraphics[width=5in]{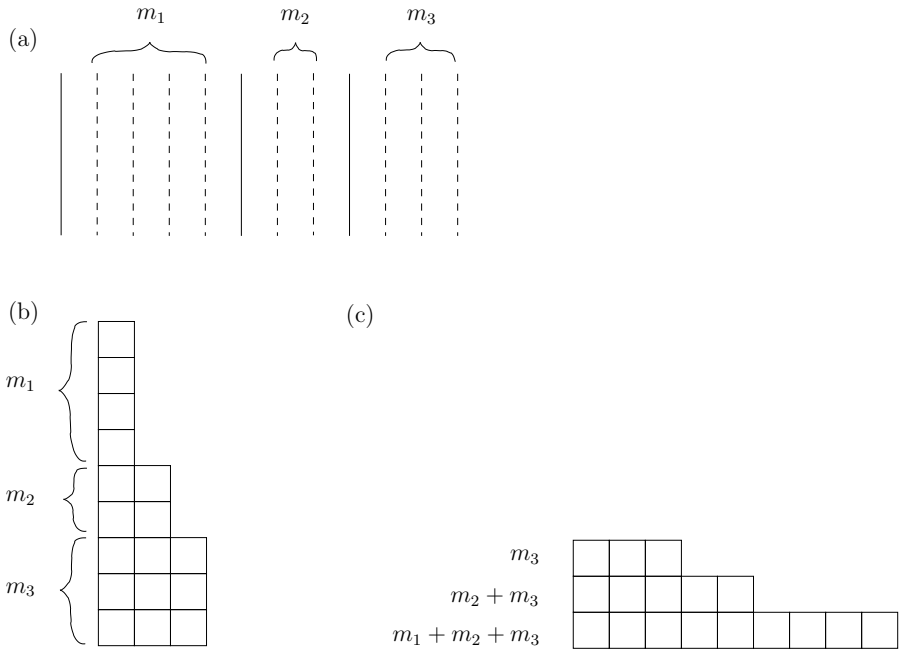}
  \end{center}

\caption{\small
(a) A configuration of three NS5-branes together with D5-branes.  We have selected an example without
D3-branes.  (b) The D5-brane
linking numbers can be arranged to make this Young diagram, by a recipe described in the text.   (c) The NS5-brane linking numbers can be arranged
to make this Young diagram.  (When there are no D3-branes, we have $\Delta\ell_\sigma=m_\sigma$, and the
recipe described in the text gives the result shown here.) In general, the two Young diagrams are independent, but when there are no D3-branes, they are mapped to each
other by a ``flip'' that exchanges the vertical and horizontal axes. }
  \label{octo}\end{figure}

We have used the nondecreasing sequence
$\tell_1,\tell_2,\dots,\tell_p$ of  D5-brane linking numbers  to
determine a collection of $k^{th}$ roots of unity, which we
interpret as eigenvalues of a $U(p)$-valued monodromy at infinity in
the ALE limit.    The $\tell_\lambda$ are integers that change by
$\pm k$ when a D5-brane is transported around the circle. For the
present discussion, it is convenient to add multiples of $k$ so that
they take values in the set $\{1,2,3,\dots,k\}$.  After doing this,
we arrange the D5-branes so that the $\tell_\lambda$ are
nondecreasing.

We have also used the sequence of linking number differences
$\Delta\ell_1,\Delta\ell_2,\dots,\Delta\ell_k$ of NS5-branes to determine
another collection of $k^{th}$ roots of unity, which we interpret as
a $U(p)$-valued monodromy near an ${\mathrm A}_{k-1}$ singularity.
After adding suitable multiples of $p$, we consider the
$\Delta\ell_\sigma$ to be valued in the set $\{0,1,\dots,p-1\}$.

The sequences have different lengths and take values in different
sets.  But a sequence of either type exactly suffices to determine a
conjugacy class in $U(p)$ that is of order $k$. There must,
therefore, be a direct map between sequences of the two types. The
relevant map is a duality of Young diagrams that appeared for rather
similar reasons (in comparing sequences of NS5-brane and D5-brane linking numbers)
in section 3.3.2 of \cite{Gaiotto:2008ak}.

We arrange the D5-brane linking numbers into a Young diagram as follows.
Given a nondecreasing sequence of $p$ elements $\tell_\lambda$ of the set $\{1,2,\dots,k\}$,
we arrange them from top to bottom as successive rows of a Young diagram, as in fig. \ref{octo}(b). This Young diagram
fits in a $p\times k$ rectangle. (There are always $p$ nonzero rows, so the height of the
diagram is strictly $p$.  The number of columns may be less than $k$ if
no $\tell_\lambda$ is equal to $k$.)

Alternatively, we can make a Young diagram that fits in a $k\times
p$ rectangle by using the linking numbers of the NS5-branes.   In
doing this, we define the linking number of the first NS5-brane to
be $\ell_1=0$ (we recall the additive ambiguity in the linking
numbers), and, recalling the definition $\Delta\ell_\sigma=
\ell_{\sigma+1}-\ell_\sigma$, we then have $\ell_2=\Delta\ell_1$,
$\ell_3=\Delta\ell_1+\Delta\ell_2$, and in general
$\ell_\sigma=\Delta\ell_1+\dots+\Delta\ell_{ \sigma-1}$.    Roughly
speaking, we now want to make a Young diagram from the $\ell_\sigma$
just as in the last paragraph we made a Young diagram from the
$\tell_\lambda$.  However, it turns out that the duality we are
about to state is slightly more elegant if we include a minus sign
in defining the Young diagram. (Alternatively, we could omit the
minus sign in defining the Young diagram and include it in stating
the duality.) Thus, we take the {\it negatives} of the
$\ell_\sigma$, add multiples of $p$ to shift them into the range
$\{1,2,\dots,p\}$, and then  arrange them in nondecreasing order as
the rows of a Young diagram.  As $\sum_\tau \Delta\ell_\tau=p$, we
have $-\ell_\sigma+p=
\Delta\ell_\sigma+\Delta\ell_{\sigma+1}+\dots+\Delta\ell_k$, and
these numbers are in the set $\{1,2,\dots,p\}$. Arranging these
numbers in ascending order, the recipe is to form a Young diagram in
which the first row is of length $\Delta\ell_k$, the second of
length $\Delta\ell_{k-1}+\Delta\ell_k$, and in general the $q^{th}$
row, for $q=1,\dots,k$, is of length
$\Delta\ell_{k-q+1}+\Delta\ell_{k-q+2}+\dots +\Delta\ell_k$.  This
Young diagram fits in a $k\times p$ rectangle. (Its $k^{th}$  row is
precisely of length $p$, so the width is precisely $p$, but the
height is less than $k$ if $\Delta\ell_k=0$.) As in the fig.
\ref{octo}, a ``flip'' that exchanges the horizontal and vertical
axes maps a $p\times k$ rectangle to a $k\times p$ rectangle, and
maps a Young diagram of one type to a Young diagram of the other
type.

In general, the Young diagram obtained from the $\tell_\lambda$'s
encodes the monodromy  at infinity in the ALE limit, while the
flipped version of the Young diagram obtained from the
$\Delta\ell_\sigma$'s encodes, in a suitable limit, the monodromy at
an ${\mathrm A}_{k-1}$ singularity. If there are no D3-branes, then
the two monodromies are equal, as we learned in section \ref{lmon}.
That one diagram is equivalent to the flipped version of the other,
when there are no D3-branes, is illustrated in the figure.

 When
D3-branes are present, the two Young diagrams are almost
independent. The only relation between them is that they have the
same number of boxes, consistent with the fact that the $U(1)$
bundle $\det V$ must have the same holonomy at the origin as at
infinity.

\subsubsection{$M$-Theory Interpretation}\label{mth}

We began our analysis in section \ref{concrete} with an NS5-D5-D3 configuration in Type IIB superstring theory on
$M=\R^3\times S^1\times \R^3_\X\times \R^3_\Y$.  The paper has been based on the equivalence of this, by $T$-duality on $S^1$, to a D2-D6 configuration in Type IIA superstring theory on
$\R^3\times \TN_\k\times \R^3_\Y$.  There are $p$ D6-branes, wrapped on $\R^3\times \TN_\k$.
In one branch of the moduli space of vacua, the D2-branes dissolve into instantons on $\TN_\k$.

\def\p{{\mathbf p}}
This configuration has a more symmetrical description based on a lift to $M$-theory.  For any
seven-manifold $N_7$,
Type IIA  on $N_7\times \R^3$  with a D6-brane wrapped on $N_7$
lifts to $M$-theory on $N_7\times \TN$ \cite{Townsend}.  With $p$ wrapped D6-branes, the lift
is to $M$-theory on $N_7\times \TN_\p$.  In our case, $N_7=\R^3\times \TN_\k$,
so the model that we have been studying is equivalent to $M$-theory on $\R^3\times \TN_\k\times \TN_\p$.  In this $M$-theory interpretation, the D2-branes turn into M2-branes.

At a generic point in the supersymmetric moduli space of the system,
$\TN_\k$ and $\TN_\p$ are smooth, and the M2-brane charge is carried
by M2-branes and by the flux $G=\d C$ of the three-form field $C$.
However, either or both of $\TN_\k$ or $\TN_\p$ may develop an
orbifold singularity.\footnote{In this paper, until the present
point, we have considered the moduli of $\TN_\k$, but not the moduli
of $\TN_\p$.  In the Type IIB description, those correspond to the
positions of the D5-branes in $\Y$.  They have  been set to zero
throughout our analysis, which means from the $M$-theory point of
view that we have been sitting on the ${\mathrm A}_{p-1}$
singularity.} An ${\mathrm A}_{p-1}$ singularity of $\TN_\p$ leads
to an $SU(p)$ gauge symmetry on $\R^3\times \TN_\k$, which is
enhanced to $U(p)$ by the effects of the $C$-field. Conversely, an
${\mathrm A}_{k-1}$ singularity of $\TN_\k$ leads to an $SU(k)$
gauge symmetry on $\R^3\times \TN_\p$, enhanced to $U(k)$ by the
$C$-field.   In either case, the relevant modes with $G$-flux
collapse at the singularity, but instantons appear that can carry
the M2-brane charge.

A particularly interesting case is that we vary the moduli so that $\TN_\k$ reduces to $\R^4/\ZZ_k$
and $\TN_\p$ reduces to $\R^4/\ZZ_p$, giving us $M$-theory on $\R^3\times \R^4/\ZZ_k\times\R^4/\ZZ_p$.  We refer to such an intersection of singularities as an $\A_{k-1}\times \A_{p-1}$ singularity.
To specify an $M$-theory model associated with such a singularity, we must choose (a) the monodromy
at infinity of the $U(p)$ gauge fields on $\R^4/\ZZ_\k$, (b) the monodromy at infinity of the
$U(k)$ gauge fields on $\R^4/\ZZ_\p$, and (c) the M2-brane charge.    The monodromies at the
origin are not selected as part of the specification of the model, because they can change in a dynamical
process, as we explain momentarily.  We write $U_\infty$ and $\tilde U_\infty$ for the two monodromies at infinity.

The process in which the monodromy at the origin changes is easily
described from a field theory point of view.  On $\R^4$, an
instanton can shrink to a point, with the result that the instanton
number (of the smooth part of the gauge field) changes by an
integer.  Letting $\ZZ_k$ or $\ZZ_p$ act on $\R^4$ with an isolated
fixed point at the origin, it is possible for instantons to shrink
to the origin in a $\ZZ_k$ or $\ZZ_p$-invariant way, such that the
change in instanton number is not divisible by $k$ or $p$. Such a
process can be interpreted on the quotient $\R^4/\ZZ_k$ or
$\R^4/\ZZ_p$ as the collapse to the origin of a fractional number of
instantons; in this process the monodromy around the singularity
changes.\footnote{The most basic case arises if we consider a
one-instanton solution on $\R^4$, centered at the origin.  Such a
solution is automatically $\ZZ_k$ or $\ZZ_p$ invariant, and when it
shrinks to a point, the instanton number on $\R^4$ is reduced by 1,
and the instanton number on $\R^4/\ZZ_k$ or $\R^4/\ZZ_p$ is reduced
by $1/k$ or $1/p$.  In such a process, which we call the shrinking
of fractional instantons, the Chern-Simons invariant of the
monodromy around the singularity changes by an amount equal mod 1 to
minus the change in the instanton number.}

The map from Type IIB data to $M$-theory data is clear from the results that we have described.
The monodromy at infinity on $\R^4/\ZZ_k$ is determined by
the D5-brane linking numbers.  Similarly, the monodromy at infinity on $\R^4/\ZZ_p$ is determined
by the NS5-brane linking numbers.  Finally, in a branch of vacua in which all M2-brane charge
is carried by instantons on $\TN_\k$, the M2-brane charge corresponds in our analysis
to ${\mathrm {ch}}_4(V)$ and was computed in sections \ref{threec}, \ref{eob}.  In general, one must sum the contributions from instantons
on one side or the other plus free M2-branes.

So we know what the models are.  Now we would like to understand how they behave.
In any component of the moduli space of vacua of an $M$-theory model on $\R^3\times\R^4/\ZZ_k\times
\R^4/\ZZ_p$, the M2-brane charge is carried
by a mixture of different components: free M2-branes and instantons on one branch or the other.  Let us examine
the possible branches more precisely.

From a low energy field theory point of view, to define instantons on $\R^4/\ZZ_k$ requires
specifying a monodromy at the origin, which we call $U_0$.  Similarly, to define
instantons on $\R^4/\ZZ_p$ requires specifying a second monodromy at the origin, which
we call $\tilde U_0$.  In low energy field theory, in addition to these two local monodromies, one can also label the
$\A_{k-1}\times \A_{p-1}$ singularity by the M2-brane charge that is supported
at the intersection of the two singularities.
 However, low energy field theory is not powerful enough to tell
us what values these invariants can have.  For this, we require some input from $M$-theory.

If we are given an allowed set of invariants, corresponding to an $M$-theory state that looks
like an $\A_{k-1}\times \A_{p-1}$ singularity from a field theory point of view, we can always
make new $M$-theory singularities, with greater M2-brane charge and possibly
 with different $U_0$ or $\tilde U_0$,
by letting M2-branes approach the singularity or by letting instantons or fractional instantons shrink
on one side or the other.   Any $\A_{k-1}\times\A_{p-1}$ singularity that is generated in
this way from  some other $\A_{k-1}\times\A_{p-1}$ singularity will necessarily
have moduli -- namely the moduli of the M2-branes or instantons.  It also has a greater M2-brane charge
than the singularity we started with.

Low energy field theory is thus powerful enough to predict the
existence of infinitely many possible $\A_{k-1}\times \A_{p-1}$
singularities in $M$-theory, given the existence of any one such
singularity.   But  the existence or nature of ``irreducible''
$M$-theory singularities that  {\it cannot} be obtained in this way
from more primitive ones is  beyond the reach of low energy field
theory. It is reasonable to expect that the irreducible
singularities are precisely the rigid ones, that is, the ones
without moduli.

From the relation to an NS5-D5-D3 configuration in Type IIB, we can easily identify the
irreducible  $\A_{k-1}\times \A_{p-1}$
singularities in $M$-theory.  We put any NS5-D5-D3 configuration in the form of
section \ref{reduction}, arranging the branes so that  D3-branes, if any, end on NS5-branes only.
In this form, the condition that there are no moduli is simply that there are no D3-branes.  D3-branes
(whether full or fractional) are free to move in the $\X$ direction, a process dual in $M$-theory
to moving full or fractional instantons away from the singularity on $\R^4/\ZZ_k$.

So the irreducible $M$-theory singularities correspond to arrangements of NS5- and D5-branes on a circle, with no D3-branes.
The linking numbers $\tell_\lambda$ of the D5-branes can be specified at will; this
determines how to arrange the branes, up to isomorphism, so the linking numbers
$\ell_\sigma$ of the NS5-branes are then uniquely determined.  Conversely, we may specify
the $\ell_\sigma$ arbitrarily and then the $\tell_\lambda$ are determined.  The two sets of
linking numbers are related by the duality of Young diagrams explained in section \ref{duality}.

Thus, in $M$-theory, rigid $\A_{k-1}\times \A_{p-1}$ singularities can be classified either by
the monodromy $U_0$ around the $\A_{k-1}$ singularity or by the monodromy $\tilde U_0$ around
the $\A_{p-1}$ singularity.  Either one of these monodromies determines the other.

From the Type IIB point of view, a $B$-field gauge transformation can multiply $U_0$ or $\tilde U_0$
by a root of unity.  This will correspond in $M$-theory to a $C$-field gauge transformation.   There is undoubtedly
more to say about the role of the $C$-field in this problem.

\vskip 2 cm

I wish to thank S. Cherkis and G. Moore for careful reading of the manuscript and helpful comments.

\appendix
\section{Complex Structures And Nahm's Equations}

\def\n{{ n}}
Here we will review the application \cite{Cherkis:2008ip} of Nahm's
equations to instanton moduli spaces on $\TN_\k$ associated with canonical NS5-D5-D3 configurations
(in which D3-branes end on NS5-branes only).
The purpose is to show that, as a complex symplectic manifold in any one complex structure, the moduli space of
instantons on $\TN_\k$  associated to  such a configuration is independent of the radius of the circle
at infinity as well as the monodromy at infinity. As
explained in section \ref{morep}, the argument immediately extends
to show that an instanton bundle $V\to\TN_\k$, viewed as a
holomorphic bundle in one of the complex structures of $\TN_\k$, is
independent of the same parameters.

Nahm's equations for the pair $(\vec X,A)$ have been written in eqn.
(\ref{cuffy}):
\begin{equation}\label{zur}\frac{D\vec X}{Dy}+\vec X\times \vec
X=0.\end{equation} The inclusion of an NS5-brane at, say,  $y=y_0$
leads to a modification explained in eqn. (\ref{zelf}): the fields
$(\vec X,A)$ and the  gauge transformations acting on them become
possibly discontinuous at $y=y_0$. Thus, locally there are separate
gauge theories on the regions $y\leq y_0$ and $y\geq y_0$. In
general, the gauge group may be $U(n_-)$ for $y\leq y_0$ and
$U(n_+)$ for $y\geq y_0$. At $y=y_0$, there is supported a
bifundamental hypermultiplet, which transforms as $(\n_-,\overline
\n_+)\oplus (\overline \n_-,\n_+)$ under $U(n_-)\times U(n_+)$. (We
denote the fundamental representation of $U(n)$ and its dual as $n$
and $\bar n$; their extensions to representations of $GL(n,\C)$ are
denoted below as $n$ and $n^\vee$.)  If we are on a circle and there
is only one NS5-brane, then the two gauge theories are connected by
going around the circle and therefore $n_-=n_+$. Writing $\vec X^-$
and $\vec X^+$ for the limits of $\vec X$ as $y\to 0$ from the left
or the right, $\vec X$ obeys boundary conditions that were written
in eqn. (\ref{zelf}):
\begin{equation}\label{gelf}\vec X^-(y_0)=\vec \mu^-,~~~\vec
X^+(y_0)=-\vec\mu^+.\end{equation} Here $\vec \mu^-$ and $\vec\mu^+$
are the hyper-Kahler moment maps for the actions of $U(n_-)$ and $U(n_+)$ on the bifundamental hypermultiplet.

In addition, we include D5-branes at points $y=s_\lambda$, $\lambda=1,\dots,p$,
which we take to be distinct from $y_0$.  At each of these points, there is
a fundamental hypermultiplet of $U(n)$, which arises from
quantization of the D3-D5 strings.  We write $\vec\nu_\lambda$ for the
moment map of the hypermultiplet that is supported at $y=s_\lambda$. The
D5-branes contribute source terms to Nahm's equations, which become
\begin{equation}\label{gommo}\frac{D\vec X}{Dy}+\vec X\times \vec
X+\sum_{\lambda=1}^p\delta(y-s_\lambda)\vec\nu_\lambda=0.\end{equation}

Instantons on $\TN_\k$ correspond to solutions of Nahm's equations
with the sources indicated in (\ref{gommo}) and the discontinuities
described in (\ref{gelf}), modulo gauge transformations.  The
eigenvalues of the monodromy at infinity are $\exp(is_\lambda/R)$,
according to eqn. (\ref{oklo}), and the Chern classes of the instanton
bundle are described in section \ref{moveb}.

\def\X{{\mathcal X}}
\def\A{{\mathcal A}}
\def\D{{\mathcal D}}
Let us first consider the case $k=1$; thus, we study instantons on
the basic Taub-NUT manifold $\TN$.  We choose $y_0=0$. Since the regions to the left and right of $y=0$ are
connected by going around the circle, we set $n_+=n_-=n$. We
``unwrap'' the circle to the interval $I:0\leq y\leq 2\pi R$. The
boundary conditions become
\begin{equation}\label{ovoc}\vec X(0)=-\vec\mu^+,~~\vec X(2\pi
R)=\vec\mu^-.\end{equation}

Nahm's equations and the boundary conditions become much more
tractable if we consider the moduli space $\M$ of its solutions not as
a hyper-Kahler manifold, but only as a complex symplectic manifold in one of
its complex structures.  To do this, we introduce the complex fields $\X=X_1+iX_2$, $\A=A-iX_3$, which are
holomorphic in one of the complex structures.  Ignoring the sources and
boundary conditions for a moment, Nahm's equations reduce to a
complex equation
\begin{equation}\label{omy}\frac{\D\X}{\D\A}=0,~~{\D}=\frac{d}{dy}+[\A,\,\cdot\,]\end{equation}
together with a real equation.  The complex equation is invariant
under $GL(n,\C)$-valued gauge transformations acting on $\X$ and
$\A$ in the usual way.  The real equation is, in effect, a gauge-fixing condition that
reduces $GL(n,\C)$ gauge-invariance to the usual $U(n)$ gauge-invariance. A standard argument shows that, to
understand $\M$ as a complex symplectic manifold in the chosen complex
structure, instead of imposing all of Nahm's equations and dividing
by $U(n)$-valued gauge transformations, it is equivalent
to impose only the complex Nahm equation and divide by
$GL(n,\C)$-valued gauge transformations.

This alternative procedure leads to many simplifications.  First of all, on the
interval $I$, there is no gauge-invariant information in the gauge
field $\A$.  So we can use $GL(n,\C)$-valued gauge transformations
to set $\A=0$.  This condition is invariant under constant
$GL(n,\C)$-valued gauge transformations, which we consider later.
With $\A=0$, the complex Nahm equation (\ref{omy}) reduces to
$\d \X /\d y=0$.

In one of its complex structures, a hypermultiplet in the
fundamental representation of $U(n)$ is just a pair $(B,C)$
transforming as $\n\oplus \n^\vee$ under $GL(n,\C)$. A certain
bilinear expression in $B$ and $C$ transforms in the adjoint
representation of $GL(n,\C)$; we write this expression simply as
$BC$ (in other words, we view $B$ as a column vector, and $C$ as a row vector,
so $BC$ is an $n\times n$ square matrix).

Now we want to include in the complex Nahm equations the delta
function source terms that appear in (\ref{gommo}).  The
hypermultiplet at $y=s_\lambda$ is a pair $(B_\lambda,C_\lambda)$ as above, and in our
chosen complex structure, the complex moment map is simply $B_\lambda C_\lambda$.
Nahm's equations, in the gauge $\A=0$, therefore take the form
\begin{equation}\label{turkey}\frac{\d\X}{\d
y}+\sum_{\lambda=1}^p\delta(y-s_\lambda)B_\lambda C_\lambda=0.\end{equation} We can solve
this immediately, and learn that
\begin{equation}\label{urkey}\X(2\pi
R)=\X(0)-\sum_{\lambda=1}^sB_\lambda C_\lambda.\end{equation} The parameters $s_\lambda$ and
$R$ have disappeared, so it is already clear that,  in any one complex structure, $\M$ is independent of the radius
and monodromy at infinity.

An analogous simplification occurs for the boundary conditions.  The bifundamental hypermultiplet of $U(N)\times U(N)$ that is
supported at $y=0$ is equivalent, in one complex structure, to a pair of $n\times n$ matrices $S,T$, transforming
respectively as $(n,n^\vee)$ and $(n^\vee,n)$.  The complex moment maps are $\mu_\C^-=ST$, $\mu_\C^+=TS$,
and the boundary conditions (\ref{gelf}) become
\begin{equation}\label{omelf}\X(0)=ST,~~\X(2\pi R)=TS.\end{equation}
Combining (\ref{urkey})   with the boundary conditions, we can eliminate $\X$ and get an equation
for hypermultiplets only:
\begin{equation}\label{zog}[T,S]+\sum_{\lambda=1}^p B_\lambda C_\lambda = 0.\end{equation}
As a complex manifold, the moduli space $\M$ of instantons on $\TN$ with gauge group $U(p)$ and instanton number $n$
is the space of solutions of this equation modulo the action of $GL(n,\C)$.

The equations (\ref{zog}) have a simple interpretation.  They are the ADHM equations for instantons on $\R^4$,
adapted to one complex structure; consequently \cite{Cherkis:2008ip},  in any one complex structure,
instanton moduli space on $\TN$ is equivalent to instanton moduli space on $\R^4$.  One might anticipate this result,
because as a complex symplectic manifold in one complex structure, both $\TN$ and $\R^4$ are equivalent to $\C^2$.
Hence  one might hope that instanton moduli space on both $\TN$ and $\R^4$ would be equivalent to a moduli space
of holomorphic bundles on $\C^2$.   However, $\TN$ and $\R^4$ are not compact, and their metrics behave quite
differently at infinity.  It is therefore not so clear {\it a priori} in what sense instantons on $\TN$ or $\R^4$, understood as anti-selfdual Yang-Mills with square integrable curvature, are equivalent to holomorphic bundles on $\C^2$ or to each other.  (Bearing in
mind that the definition of instantons on $\TN$ involves a choice of monodromy at infinity, this may seem even less clear.)
Hence the simple comparison between instanton moduli spaces on $\TN$ and $\R^4$ may come as a pleasant surprise.
As explained in section \ref{morep}, the equivalence of instanton moduli spaces on $\TN$ and $\R^4$ also extends
to an equivalence of the holomorphic bundles that correspond to the instantons.

\def\ad{{\mathrm {ad}}}
It may help to recall how the equations (\ref{zog}) arise in the
context of instantons on $\R^4$.  $U(p)$ instantons on $\R^4$ with
instanton number $n$ can be described by a system consisting of $n$
E0-branes supported at a point in $\R^4$, and $p$ E4-branes of
world-volume $\R^4$.  The E0-E0 strings are a hypermultiplet $H_\ad$
in the adjoint representation of $U(N)$, and the E0-E4 strings are
$p$ hypermultiplets $H_\lambda$, $\lambda=1,\dots,p$ in the
fundamental representation.  The hypermultiplets parametrize a space
$\R^{4n^2+4np}$, and the hyper-Kahler quotient
$\R^{4n^2+4np}/\negthinspace/\negthinspace/U(n)$ is the moduli space
${\mathcal N}$ of supersymmetric states of this system, or in other
words the moduli space of instantons on $\R^4$. This gives the ADHM
construction of instanton moduli space on $\R^4$. To describe
$\mathcal N$ as a complex manifold in one complex structure, one can
replace the hyper-Kahler quotient by a complex symplectic quotient,
in which one sets to zero a complex moment map and divides by
$GL(n,\C)$.  From the point of view of one complex structure,
$H_\ad$ is a pair $S,T$ of matrices in the adjoint representation of
$GL(n,\C)$, the $H_\lambda$ correspond to pairs
$B_\lambda,\,C_\lambda$ in the fundamental representation and its
dual, and the complex moment map is $\mu_\C=[T,S]+\sum_\lambda
B_\lambda C_\lambda$.  Eqn. (\ref{zog}) is thus equivalent to
$\mu_\C=0$, and when we impose this condition and divide by
$GL(n,\C)$, we get the moduli space $\mathcal N$, viewed as a
complex symplectic manifold in one complex structure.

The foregoing can be adapted rather directly for canonical configurations with more than one NS5-brane
related to instantons on a more general ALF space $\TN_\k$.  For this,
we introduce NS5-branes at points $y=y_\sigma$, $\sigma=1,\dots,k$.  Between
$y_\sigma$ and $y_{\sigma+1}$, we place
$n_\sigma$ D3-branes.  In that interval, we also place $m_\sigma$ D5-branes, supported at points
$s_{\sigma,\lambda}$,
$\lambda=1,\dots,m_\sigma$.  In each interval $I_\sigma=[y_\sigma,y_{\sigma+1}]$, there is a
$U(n_\sigma)$ gauge
theory, interacting with hypermultiplets in the fundamental representation that are supported
at the points $y_{\sigma,\lambda}$
and with bifundamentals of $U(n_{\sigma-1})\times U(n_\sigma)$  and of $U(n_\sigma)\times U(n_{\sigma+1})$
at the left
and right ends of the intervals.  The corresponding component  $\M$ of the moduli space of instantons on $\TN_\k$ is the space of solutions of Nahm's equations
with delta function sources as in (\ref{turkey}) and jumping conditions at $y=y_\sigma$  analogous to (\ref{ovoc}).  To describe
$\M$ as a complex manifold in one complex structure, we introduce complex fields $\X,\,\A$ in each interval $I_\sigma$.
After going to the gauge $\A=0$, and replacing the fundamental hypermultiplet at $s_{\sigma,\lambda}$
by a pair
$B_{\sigma\lambda},\,C_{\sigma\lambda}$, the complex Nahm equation becomes
\begin{equation}\label{compn}\frac{\d\X}{\d y}+\sum_{\lambda=1}^{m_\sigma}\delta(y-s_{\sigma,\lambda})B_{\sigma\lambda}
C_{\sigma\lambda}=0,\end{equation} implying that
\begin{equation}\label{tome}\X^-(y_{\sigma+1})-\X^+(y_\sigma)+\sum_{\lambda=1}^{m_\sigma}B_{\sigma\lambda}C_{\sigma
\lambda}=0.  \end{equation}  (The D5-brane positions have dropped
out, so again $\M$ is independent of the asymptotic radius of the
circle as well as the asymptotic monodromy.) The bifundamental
hypermultiplet at $y=y_\sigma$ is equivalent to a pair of complex
fields $S_\sigma,T_\sigma$ transforming as $(n_{\sigma-1},
n_\sigma^\vee)\oplus (n_{\sigma-1}^\vee,n_\sigma)$ of
$GL(n_{\sigma-1})\times GL(n_\sigma)$, and the jumping conditions
that generalize (\ref{gelf}) become\footnote{\label{noct} It is also
possible to add constants $x_\sigma$ to the moment maps, in which
case the boundary conditions become  $\X^+(\sigma)=S_\sigma
T_\sigma+x_\sigma, $  $\X^-(\sigma)=T_\sigma S_\sigma+x_\sigma$. The
constants $x_\sigma$ become the moduli of $\TN_\k$ (in one complex
structure), as in eqn.  (\ref{zocon}). For brevity, we omit this in
the text, meaning that we write all formulas for the case that
$\TN_\k$ has an ${\mathrm A}_{k-1}$ singularity.}
 \begin{equation}\label{jumping}\X^+(\sigma)=S_\sigma T_\sigma, ~X^-(\sigma)=T_\sigma S_\sigma.\end{equation}
Together with (\ref{tome}), the boundary conditions enable us to eliminate $\X$ to get
\begin{equation}\label{ohem} T_{\sigma+1}S_{\sigma+1}-S_\sigma T_\sigma  +\sum_{\lambda=1}^{m_\sigma}
B_{\sigma\lambda}C_{\sigma\lambda}=0.\end{equation}
The moduli space $\M$ is obtained by solving these equations for $\sigma=1,\dots,k$ and then dividing by
$\prod_{\sigma=1}^k GL(n_\sigma)$.

On the other hand, let $\N$ be the Higgs branch associated with the ${\mathrm A}_{k-1}$ quiver of fig. \ref{quiver}.  The gauge group of
this quiver is $H=\prod_{\sigma=1}^k U(n_\sigma)$, and the matter fields are $m_\sigma$ fundamental hypermultiplets of each $U(n_\sigma)$ group
and the usual bifundamental hypermultiplets of $U(n_\sigma)\times U(n_{\sigma+1})$.  Let $Z$ be the space
parametrized by the hypermultiplets.  $\N$ can be described
as a hyper-Kahler manifold by taking the hyper-Kahler quotient $Z/\negthinspace/\negthinspace/ H$.  It can be described much more simply as a complex
symplectic manifold in one complex structure by taking the complex symplectic quotient $Z/\negthinspace/H_\C$,
with $H_\C=\prod_{\sigma=1}^k GL(n_\C)$ the complexification of $H$.  Equations (\ref{ohem}) are precisely the equations
for vanishing of the complex moment map of the quiver, so we conclude that the moduli space $\M$ of instantons on
$\TN_\k$ is equivalent as a complex symplectic manifold to $\N$, the Higgs branch of the ${\mathrm A}_{k-1}$ quiver.

\def\diag{{\mathrm{diag}}}
This quiver is usually studied \cite{KN,Douglas:1996sw} as a way to describe instantons on $\R^4/\ZZ_k$, or its hyper-Kahler resolution.   So as a complex symplectic manifold, instanton moduli space on a space $\TN_\k$ with an ${\mathrm A}_{k-1}$
singularity coincides with instanton moduli space on $\R^4/\ZZ_k$.  This equivalence persists when one deforms
away the ${\mathrm A}_{k-1}$ singularity on each side; the deformation parameters $x_\sigma$ on the $\TN_\k$ side
(see footnote \ref{noct})
map to the usual deformation parameters in the quiver description of instantons on a deformed $\R^4/\ZZ_k$.

Let us recall how the quiver theory arises in studying instantons on $\R^4/\ZZ_k$.  One uses the fact that instantons on $\R^4/\ZZ_k$ are $\ZZ_k$-invariant
instantons on $\R^4$; in turn, instantons on $\R^4$ are described by the ADHM construction, or equivalently, as explained
above, by a configuration of $n$ E0-branes and $p$ E4-branes.  To go to $\R^4/\ZZ_k$, first we pick an action of $\ZZ_k$ on
$\R^4$ such
that the generator 1 of $\ZZ_k$ acts on $\R^4\cong \C^2$ as $w=\diag(\omega,\omega^{-1})$, with $\omega=\exp(2\pi i /k)$.
We also pick a $\ZZ_k$ action on the Chan-Paton space $V\cong \C^n$ of the E0-branes, generated by an element
$g\in U(n)$ that obeys $g^k=1$.  $g$ can be diagonalized with eigenvalues of the form $\omega^\sigma$, $\sigma=1,\dots,k$;
we choose $g$ to have eigenvalues $\omega^\sigma$ with multiplicity $n_\sigma$.  Similarly, we pick $\ZZ_k$ to
act on the Chan-Paton space $W\cong \C^p$ of the E4-branes by an element $u\in U(p)$ that has eigenvalues
$\omega^\sigma$ with multiplicity $m_\sigma$.

Instantons on $\R^4/\ZZ_k$ are now described simply by taking
the $\ZZ_k$-invariant subtheory of the original E0-E4 theory.   This subtheory is obtained by replacing $U(n)$ by its
subgroup that commutes with $\ZZ_k$ and keeping only the $\ZZ_k$-invariant hypermultiplets.
The subgroup of $U(n)$ that commutes with
$\ZZ_k$ is $H=\prod_{\sigma=1}^n U(n_\sigma)$.
$\ZZ_k$-invariant hypermultiplets come from the $\ZZ_k$-invariant part of the original
system of $p$ fundamental hypermultiplets of $U(n)$ and a hypermultiplet in the adjoint representation.
The contribution from the  fundamental hypermultiplets
 is the direct sum of $m_\sigma$ fundamental hypermultiplets of $U(n_\sigma)$; these are represented by the boxes of the quiver
diagram of  fig. \ref{quiver}.  Slightly more subtle is the analysis of the adjoint hypermultiplet of $U(n)$.  It parametrizes
$\C^2\otimes V\otimes V^\vee$, where $V^\vee$ is the dual of $V$.  $\ZZ_k$ acts on $\C^2$ via $w$ (one learns this
by quantizing the E0-E0 strings to get these hypermultiplets) and on $V$ and $V^\vee$
via $g$.  Taking this into account, $\ZZ_k$ invariance reduces the adjoint hypermultiplet of $U(n)$ to a direct sum of
bifundamental hypermultiplets of $U(n_\sigma)\times U(n_{\sigma+1})$, $\sigma=1,\dots,k$; these are represented by
the links of the quiver diagram.  The upshot is that the
quiver theory is simply the $\ZZ_k$-invariant part of the theory describing instantons on $\R^4$, so it describes instantons
on $\R^4/\ZZ_k$.

\section{The Monodromy At The Origin}\label{monorg}

When an ALF space $\TN_\k$ develops an ${\mathrm A}_{k-1}$ singularity, the definition of gauge theory on this
space requires a choice of a monodromy $U_0$ at the singularity, as discussed in section \ref{lmon}.  In eqn. (\ref{zello}), we have given a formula
for $U_0$ in terms of NS5-brane linking numbers in a dual Type IIB description.  This formula (which in section \ref{lmon}
was justified only in the absence of D3-branes) will be derived here.  For this purpose, it suffices to consider
canonical NS5-D5-D3 configurations and to view the
instanton bundle in just one complex structure.   Under these circumstances, as we have learned in Appendix A,
 we can replace a
$\TN_\k$ space that has an ${\mathrm A}_{k-1}$ singularity by the corresponding $\R^4/\ZZ_k$.  (\ref{zello}) is then a standard consequence of the quiver description of instantons on $\R^4/\ZZ_k$, as we will now
explain.

The ADHM description of an instanton bundle on $\R^4$, regarded as a holomorphic bundle over $\C^2$, is as follows.
First of all, we parametrize $\C^2$ with complex coordinates $s,t$, on which $\ZZ_k$ acts by $s\to \omega s$,
$t\to \omega^{-1}t$, with $\omega=\exp(2\pi i/k)$.  In the notation of Appendix A, we can regard  $S,T$ as maps $V\to V$, and similarly  $C$ as a map $V\to W$, and $B$ as a map $W\to V$.
We define linear maps
\begin{equation}\label{plp} V\overset{f}{\rightarrow}\C^2\otimes V\oplus W\overset{k}{\rightarrow}V \end{equation}
by
\begin{align}\label{fuss} f & = (S-s)\oplus (T-t)\oplus C\\
                        \notag k & =(T-t)\oplus - (S-s)\oplus B.\end{align}
Here we regard $\C^2\otimes V$ as $V\oplus  V$; $f$ maps $v\in V$ to $(S-s)v\oplus (T-t)v\oplus Cv$,
and $k$ maps $a\oplus b\oplus c$ to $(T-t)a-(S-s)b+Bc$.  The complex version of the ADHM
equations, namely $[T,S]+BC=0$,  is equivalent to $kf=0$. (The sum over $\lambda$ in (\ref{zog}) is now absorbed
in regarding $C$ and $B$ as  maps to and from $W$.)  Hence, when the ADHM equations are obeyed, one can define ``cohomology groups''
\begin{align}\label{guss} H^0&={\mathrm {ker}}\, f  \\
 \notag    H^1&={\mathrm {ker}}\,k/{\mathrm {im}}\,f \\
 \notag H^2& = V/{\mathrm {im}}\,k.\end{align}
For $q=0,1,2$, we write $h^q={\mathrm {dim}}\,H^q$.  The ``Euler characteristic'' $h^0-h^1+h^2$ is independent of $f$
and $k$ and hence can be evaluated at $f=k=0$, giving
\begin{equation}\label{eul}h^0-h^1+h^2=-{\mathrm {dim}}\,W=-p.\end{equation}
According to the ADHM construction, a smooth and irreducible instanton solution on $\R^4$ is associated with a solution of
the ADHM equations with\footnote{In general, $H^0$ generates the Lie algebra of gauge symmetries of a given instanton
solution, and $h^2=h^0$ because of   a kind of duality.} $h^0=h^2=0$.    Hence for each $s,t$, $E_{s,t}=H^1$ is a vector space of dimension $p$. ($H^1$ depends on $s$ and $t$, even though this is not shown in the notation, because $s$ and $t$ appear in the
definition of $f$ and $k$.)
As $s$ and $t$ vary, $E_{s,t}$ varies as the fiber of a holomorphic vector bundle $E\to \C^2$.  This is the instanton
bundle associated with the given solution of the ADHM equations, understood as a holomorphic bundle in a particular complex structure.

Now, we consider a $\ZZ_k$-invariant solution of the ADHM equations, with  $\ZZ_k$ acting  on the spaces $\C^2$,
$V$, and $W$ as multiplication by $w$, $g$, and $u$, as in Appendix A.  Then in particular, $\ZZ_k$ acts on $E_{0,0}$, the fiber of $E$ at
$s=t=0$.  We want to describe this action in terms of the data $n_\sigma$ and $m_\sigma$ of the ${\mathrm A}_{k-1}$
quiver diagram.

First of all, $w$ is completely determined by the geometrical action of $\ZZ_k$ on $\C^2$;
$w$ acts on $\C^2$ with eigenvalues $\omega$ and $\omega^{-1}$.  (We can assume that the eigenvectors correspond
to $s$ and $t$.)  The action of $g$ and $u$ determines
the integers $n_\sigma$ and $m_\sigma$ appearing in a particular ${\mathrm A}_{k-1}$ quiver (fig. \ref{quiver}).   Specifically, we decompose
\begin{equation}\label{tired}V=\oplus_{\sigma=1}^kV_\sigma,~~ W=\oplus_{\sigma=1}^k W_\sigma,\end{equation}
where the element $1\in \ZZ_k$ acts on $V_\sigma$ and on $W_\sigma$ as multiplication by $\omega^\sigma$.
Then the quiver data are $n_\sigma={\mathrm{dim}}\,V_\sigma$, $m_\sigma={\mathrm{dim}}\,W_\sigma$, as is clear from the derivation of
the quiver diagram in Appendix A.

To understand how $\ZZ_k$ acts on the fiber $E_{0,0}$ at the origin,
 we first set $s=t=0$.  This ensures that $\ZZ_k$ commutes with $f$ and $k$ and hence acts on the $H^q$. So we can
   make the decomposition
\begin{equation}\label{deco} H^q=\oplus_{\sigma=0}^{k-1} H_\sigma^q,\end{equation}
where again the element $1\in \ZZ_k$ acts on $H^q_\sigma$ as multiplication by $\omega^\sigma$.
We set $h^q_\sigma={\mathrm{dim}}\,H^q_\sigma$.  For each value of $\sigma$, the generalized Euler characteristic $h^0_\sigma-h^1_\sigma+h^2_\sigma$
is independent of $f$ and $k$, as long as they commute with the chosen action of $\ZZ_k$.  Hence the Euler
characteristic can again be evaluated at  $f=k=0$, and is
\begin{equation}\label{horm} h^0_\sigma-h^1_\sigma+h^2_\sigma = 2n_\sigma-n_{\sigma+1}-n_{\sigma-1}-m_\sigma.
\end{equation}
Now again, in the case of a solution of the ADHM equations that
corresponds to a smooth and irreducible instanton bundle, we have
$H^0=H^2=0$.  So all $H^0_\sigma$ and $H^2_\sigma$ vanish, and
$h^0_\sigma=h^2_\sigma=0$. Hence for a smooth $\ZZ_k$-invariant
instanton solution, we get
\begin{equation}\label{zorm} h^1_\sigma =n_{\sigma+1}+n_{\sigma-1}-2n_\sigma+m_\sigma.\end{equation}

The object $H^1$ that appears in this computation is the fiber at the origin of the instanton bundle $E$,
so  $h^1_\sigma$ is the dimension of the subspace of this fiber that transforms as $\omega^\sigma$.  Hence
$h^1_\sigma$ coincides with the object $a_{\sigma,0}$
defined in section \ref{lmon}, and (\ref{zorm}) is equivalent to the claimed result (\ref{zello}).

\begin{footnotesize}

\end{footnotesize}
\end{document}